\documentclass{article}
\usepackage[top=3cm,right=2.5cm,bottom=3cm,left=2.5cm]{geometry}

\usepackage[scr=euler]{mathalfa}

\usepackage{graphicx} 
\usepackage{tikz}
\usetikzlibrary{matrix,positioning,arrows.meta,shapes}
\usepackage{caption}
\usepackage[caption = false]{subfig}

\usepackage{diagbox}
\usepackage{booktabs}
\usepackage{enumitem}

\usepackage{amsmath, amssymb}
\usepackage{resizegather} 
\usepackage{dsfont}

\usepackage{bm}

\usepackage{natbib}
\usepackage{multirow}
\usepackage{url}

\usepackage{algorithm}
\usepackage{algpseudocode}

\usepackage{float}

\title{Modeling temporal dependence in a sequence of spatial random partitions driven by spanning tree: an application to mosquito-borne diseases}

\author{Jessica Pavani\textsuperscript{1}, Rosangela Helena Loschi\textsuperscript{2} \& Fernando Andr{\'e}s Quintana\textsuperscript{3} \\
\textsuperscript{1}\small Department of Mathematics and Statistics, University of Calgary, Calgary, Canada \\
\textsuperscript{2}\small Departmento de Estat{\'i}stica, Universidade Federal de Minas Gerais, Belo Horizonte, Brazil \\
\textsuperscript{3}\small Departamento de Estad{\'i}stica, Pontificia Universidad Cat{\'o}lica de Chile, Santiago, Chile
}

\date{}

\begin{document}

\maketitle

\begin{abstract}
Time-dependent regionalization, or spatially restricted grouping, is a significant area of research focused on understanding the evolution of spatial clusters over time. In this study, we adopt a probabilistic approach to regionalization, conceptualizing it as a random partition of geographic space at each time point, with the sequence of spatial partitions exhibiting time dependency. This methodology facilitates inference regarding the temporal dynamics of clusters. We employ a product partition prior for the random partitions at each time point, introducing temporal correlation among partitions through the temporal structure associated with prior cohesions. To explore partition search space effectively and ensure spatially constrained clustering, we utilize random spanning trees. This research is motivated by a pertinent applied problem: the identification of spatial and temporal patterns associated with mosquito-borne diseases. Given the overdispersion inherent in this type of data, we propose a spatio-temporal Poisson mixture model in which both mean and dispersion parameters vary according to spatio-temporal covariates. We apply the proposed model to analyze weekly reported cases of dengue from 2018 to 2023 in the Southeast region of Brazil. Additionally, we assess modeling performance using simulated data. Results indicate that our model is competitive in analyzing the temporal evolution of spatial clustering.

\vspace{0.1cm}

\end{abstract}


\section{Introduction} \label{sec:intro}

The proposed spatio-temporal clustering model is motivated by a significant public health challenge that primarily affects countries in tropical regions: the control of vector-borne diseases transmitted by mosquitoes, such as dengue, chikungunya, Zika, and yellow fever. A comprehensive discussion of the problem to be addressed is presented in Section~\ref{sec:motivation}. The primary objective of this research is to identify and cluster neighboring regions exhibiting similar infection dynamics while concurrently examining the temporal evolution of these spatial patterns. The identification and grouping of neighboring regions with analogous infection trajectories, along with the investigation of temporal changes in these patterns, are critical for the formulation of improved control policies and the evaluation of previously implemented disease management strategies. This inquiry represents a classic problem of time-dependent regionalization.

Regionalization, or spatially restricted clustering, presents significant challenges due to the need to explore a vast number of potential partitions within the search space. These challenges are further exacerbated in the context of spatio-temporal clustering, where a primary objective is to evaluate the evolution of spatial clusters over time. In response to these difficulties, this study addresses the regionalization problem from a probabilistic perspective \citep{Hegarty2008, Teixeira2015, Page2016}, treating it as a random partition model for the map at each time point. Given that areal clusters can evolve over time, we specifically propose a time-dependent sequence of product partition models for spatial partitions at each time instance. 

Contrary to the assumptions made in previous studies \citep{Teixeira2019, Page2022}, we introduce temporal correlation between partitions through a time-dependent structure for prior cohesions. Specifically, we integrate spanning trees \citep{Jungnickel2013} with product partition models \citep[PPM,][]{Hartigan1990} to account for temporal dependence in spatial random partitions. Under this framework, the map is represented as an undirected graph, with spanning trees generated randomly and independently from this graph at each time $t$. A partition of the map at time $t$ is derived by pruning the corresponding spanning tree. Consequently, the prior cohesions emerge as a function of the probabilities of edge removal at time $t$. We hypothesize that the probabilities of edge removal may vary over time and are modeled using an autoregressive structure \citep{Jara2013}. To develop a more flexible class of models, we employ Poisson mixture models to address the issue of overdispersion, which is commonly encountered in count data \citep{Saraiva2022}. Specifically, we utilize a Poisson-inverse Gaussian (PIG) model, which has demonstrated superior performance compared to alternative methods for modeling overdispersed and heavy-tailed data \citep{BarretoSouza2015, Perrakis2015}. Our formulation incorporates a spatio-temporal dispersion component and posits that, conditional on a positive latent variable representing heterogeneity, the response variables are independent and identically distributed (iid) Poisson realizations. The rate of these Poisson realizations is determined by a random effect that captures heterogeneity, as well as a mean component linked to explanatory variables via a log-link function (see Section~\ref{s:PIG}). 

\begin{figure} \begin{center}
\includegraphics[width=\textwidth]{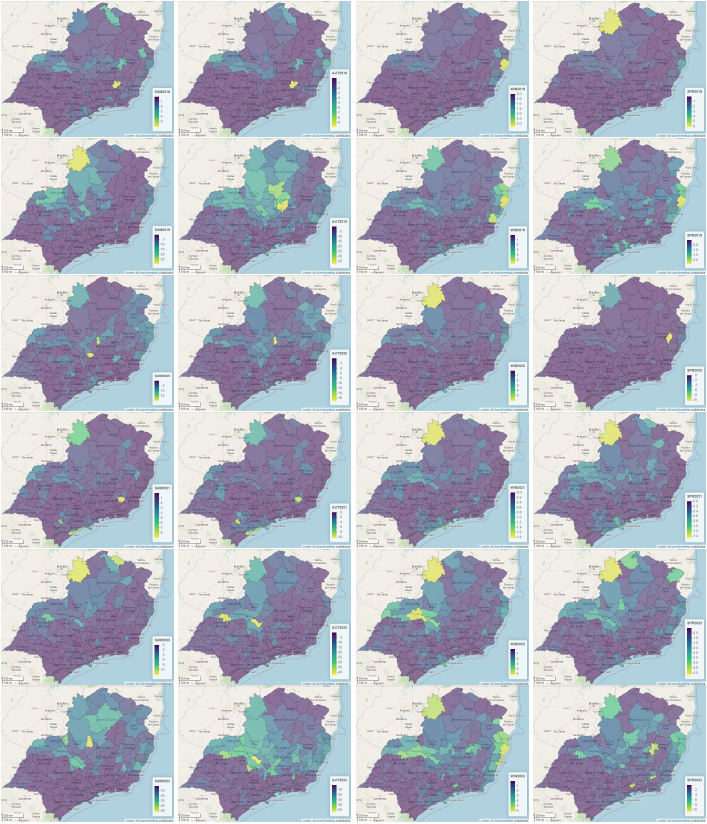}
\end{center}
\caption{ Spatial distribution of standardized incidence ratios of dengue across 145 microregions in the Brazilian Southeast region from 2018 (top) to 2023 (bottom) by seasons - summer (1st column), autumn (2nd column), winter (3rd column), and spring (4th column). } \label{fig:sp_response}
\end{figure}

The combination of spanning trees and product partition models was initially proposed by \cite{Teixeira2015} as an alternative solution to the regionalization problem. In this context, spanning trees serve as a mechanism to reduce the search space of spatial partitions and ensure spatially constrained clustering. An extension addressing the time-dependent regionalization problem was introduced by \cite{Teixeira2019}, which considered a spatio-temporal graph while maintaining a consistent probability of edge removal over time. Despite its innovative approach, applying this strategy to spatio-temporal data by constructing a spatio-temporal graph results in an exceedingly large search space, rendering it impractical for data analysis over extended periods. Another limitation of this methodology is the absence of a clear mechanism for quantifying the temporal relationships of partitions using standard time series methods. Addressing these limitations was a primary focus of our research. The model proposed by \cite{Page2022} also presents an interesting alternative for time-dependent partitioning. However, it is not well-suited for regionalization since it permits the formation of clusters comprised of non-contiguous areas. The approaches discussed in \cite{Cremaschi2023} and \cite{Pavani2025} similarly exhibit limitations with respect to regionalization. Furthermore, in the context of our motivating application, Figure~\ref{fig:sp_response} illustrates a pronounced seasonal effect on the clustering of areas, indicating that neighboring regions tend to exhibit similar incidence rates. Both of these features must be duly considered in the proposed model.

In summary, this study tackles the challenge of creating a prior distribution for sequences of random spatial partitions that can adapt to various spatio-temporal clustering patterns. Our aim is to uncover spatial and temporal trends in mosquito-borne diseases. Key contributions include: (i) integrating spanning trees into a hierarchical Bayesian model for a new spatio-temporal product partition model; (ii) introducing a new strategy to incorporate temporal dependence in spatial random partitions to assess temporal autocorrelation; (iii) employing an overdispersed Poisson mixture model with parameters influenced by spatial, temporal, and spatio-temporal covariates; and (iv) identifying spatio-temporal clusters to aid policymakers in disease prevention and control strategies. While some of the model choices are specifically tailored to the motivating application, the novel prior for time-evolving sequences of random partitions can be employed more generally. The remainder of this manuscript is organized as follows: Section~\ref{sec:works} presents a literature review of related work. Section~\ref{sec:motivation} discusses the motivating data. Section~\ref{sec:model} outlines the hierarchical model for overdispersed count data and introduces a spatio-temporal PPM driven by a spanning tree, including guidance for prior elicitation. In Section~\ref{sec:simulations}, we detail simulation studies that assess the model's performance. Section~\ref{sec:dataapp} presents results from applying the model to mosquito-borne disease data. The paper concludes with a discussion in Section~\ref{sec:conclusion}. Computation details, simulations, and additional results are provided in the Appendix.

\section{Related works} \label{sec:works}

The PPM-based clustering is a flexible strategy for modeling heterogeneous data and has been used for various purposes. The main feature of PPM is to express the prior distribution of a partition ${\bm \pi} = \{ \mathscr{C}_{1}, \ldots, \mathscr{C}_{k} \}$ of $n$ areal units into $k$ clusters in a product form such that $\mathds{P} \big[ {\bm \pi} = \{ \mathscr{C}_{1}, \ldots, \mathscr{C}_{k} \} \big] \propto \prod_{j=1}^{k} C(\mathscr{C}_{j})$, where the cohesion function $C(\mathscr{C}_{j})$ is any nonnegative function measuring how likely the elements of $\mathscr{C}_{j}$ are to co-cluster. From a PPM perspective, the number of clusters is not fixed \textit{a priori}, and its behavior is derived from the distribution of random partitions. More details will be provided later in Section~\ref{s:SPPPM}, and a general review can be found in \cite{Quintana2018}. In the spatial context, \cite{Hegarty2008} introduced a prior for the random partitioning of areal data based on the number of neighbors. This strategy was later applied to the spatio-temporal context \citep{Cremaschi2023, Pavani2025}. In terms of geo-referenced data, \cite{Page2016} developed a prior distribution that considers the distance between areas. 

Although clustering strategies based on PPM have been gaining traction in the literature, the large number of possible partitions comprising the search space makes computation challenging. To facilitate the exploration of the partition space, \cite{Teixeira2015} incorporated random spanning trees into PPM for areal clustering, which also ensures that the clustering remains spatially constrained. The spanning tree is a well-established method for effective regionalization in machine learning \citep{Assuncao2006} and continues to garner interest for new advancements \citep{Duan2023, Tam2025}. In clustering analysis, it reduces the search space by creating a connected subgraph that includes all nodes without cycles. This contrasts with the original PPM, which requires examining all possible partitions of the full graph; spanning trees limit partitions to those compatible with the tree, enabling contiguous clustering and the detection of irregular shapes. By pruning trees, the spatial constraints is inherently respected, ensuring connected clusters. Studies such as \cite{Teixeira2015}, \cite{Luo2021}, \cite{Criscuolo2023}, and \cite{Luo2023} further highlight the advantages of spanning trees in spatial clustering and regression settings. Additionally, \cite{Teixeira2019} extended the strategy presented in \cite{Teixeira2015} to the spatio-temporal context by building a graph that links each region to its neighboring areas at time $t$, as well as to itself and its neighbors at the subsequent time $t+1$. Thus, each tree takes spatial and temporal information into account. When pruning these trees, spatio-temporal partitions are automatically obtained, with dependence inherited from the graph structure.

Other spatio-temporal clustering methods are documented in the literature. \cite{Napier2018} proposed a spatio-temporal mixture model that categorizes regions based on their temporal patterns using probabilities from a Dirichlet distribution. Similarly, \cite{Zhong2024} introduced a model that groups adjacent areas with similar disease spread using spanning trees. Both models classify temporal trends. Other studies focus on the temporal correlation of random probability measures. \cite{Gutierrez2016}, \cite{Jo2017}, and \cite{DeIorio2023} integrated dependence through stick-breaking representations, and \cite{Caron2017} utilized a generalized P{\'{o}}lya urn scheme. \cite{Page2022} introduced a method for modeling dependence in random partition sequences, using a PPM prior combined with an auxiliary variable for partition similarity. Recently, multiview clustering approaches have emerged. \cite{Franzolini2023} applied a conditional partially exchangeable model to link clustering arrangements across features, while \cite{Dombowsky2024} used a nonparametric prior to create dependent random distributions centered around a random product measure. Additionally, \cite{Giampino2024} employed a state-space modeling framework to capture dependence between partitions, highlighting temporal evolution.

\section{Motivating dataset} \label{sec:motivation}

Brazil has the highest number of mosquito-borne disease cases in the Americas, particularly dengue fever, which the World Health Organization (WHO) has flagged as a potential global epidemic. Dengue poses serious health risks and is a leading cause of child mortality in parts of Asia and Latin America. The first dengue cases in Brazil appeared in the early 1980s in Roraima, and the disease has persisted across the nation since then. After peaking in 2023 with over three million cases, Brazil is confronting an even larger outbreak in 2024, with reported cases during the first five weeks of the year up by 378\% compared to the same timeframe in 2023. Our study focuses on the Southeast region, which includes the states of Esp{\'i}rito Santo, Minas Gerais, Rio de Janeiro, and S{\~a}o Paulo, home to nearly 90 million people across 145 microregions. Cases were tracked weekly over 313 epidemiological weeks, from January 2018 to December 2023, with a total of 5,309,984 cases (see in Appendix Figure~\ref{fig:time_response}). Additional figures can be found in the Appendix.

Analyzing data on mosquito-borne diseases like dengue is challenging, particularly due to its geographical distribution. Figure~\ref{fig:sp_response} illustrates the spatial distribution of standardized incidence ratios (SIR) for 2018 and 2023. For each area $i = 1, \ldots, 145$, SIR is defined as the ratio of observed counts to expected counts, i.e., SIR$_{i} = Y_{i}/E_{i}$, where $E_{i}$ denotes the total number of cases that one would expect if the population of the $i$-area behaved the way the standard population behaves \citep[Chapter 5]{Moraga2019}. It indicates whether the disease risk is lower (SIR $< 1$), equal (SIR $= 1$), or higher (SIR $> 1$) than expected from the standard population. Dengue SIR patterns are similar in neighboring areas, forming large clusters at times, but they evolve over time and vary with seasons. For instance, during the summer of 2018, SIR was consistent across most regions with few hotspots, whereas in summer 2023, the southern zones showed significantly lower SIR than the northern ones. A comparison of winter SIR from 2018 to 2023 highlights the impact of climate change, as 2023 exhibits higher SIR levels and more clusters. This data pattern poses both practical modeling challenges for mosquito-borne diseases and theoretical challenges in developing models that account for the temporal evolution of spatial clusters.

In addition to geographic challenges, we must consider the temporal patterns of mosquito-borne diseases, which are influenced by changing weather and climate conditions. These diseases typically follow seasonal trends and exhibit yearly variations linked to meteorological factors \citep{Franklinos2019}. Temperature significantly affects the spread of vector-borne illnesses, with extreme temperatures threatening vector survival, as optimal growth occurs between 22 and 32$^{\circ}$C \citep{Marinho2016}. Humidity also plays a crucial role, as low humidity increases mosquito desiccation risk. Therefore, we included minimum temperature and minimum humidity as spatio-temporal covariates to explain the mean and dispersion of the data (see Section~\ref{s:PIG} for details). Their temporal trends during the study period are shown in Appendix Figure~\ref{fig:time_cov}, while their spatial distributions are depicted in Appendix Figures~\ref{fig:sp_humidity} and \ref{fig:sp_temperature}. From now on, for simplicity, we will refer to minimum temperature and minimum humidity as temperature and humidity, respectively. Socio-demographic factors also influence mosquito-borne diseases, prompting us to include the Human Development Index (HDI) as a spatial variable to analyze mean case numbers. The HDI, which integrates life expectancy, education, and income, is a key indicator of human development. We retrieved HDI values from the 2010 Demographic Census for each regional unit. Appendix Figure~\ref{fig:HDI} depicts the HDI's spatial distribution, revealing a distinctive pattern.

\section{Model specification} \label{sec:model}

Consider a map with $n$ areas that remain constant over time, with $i = 1, \ldots, n$. Our model incorporates temporal structure of the data through two time units: seasons, denoted by $s = 1, \ldots, S$, and epidemiological weeks, denoted by $t = 1, \ldots, T$. Seasons refer to summer, autumn, winter and spring, listed in that order, and each season lasts approximately 13 weeks. Therefore, $T = 13\times S$. Cluster labels in season $s$ are denoted by $j_{s} = 1, \ldots, k_{s}$, where $k_{s}$ is the total number of clusters in season $s$. 

\subsection{Modeling overdispersed count data} \label{s:PIG} 

Denote by $y_{it}$ the number of cases registered in area $i$ during week $t$ and by $z_{is}$ the random effect accounting for heterogeneity in area $i$ and season $s$. The proposed Poisson-inverse Gaussian regression model is specified as
\allowdisplaybreaks \begin{align}
    (y_{it} \mid O_{it}, \lambda_{it}, z_{is}) & \overset{ind}{\sim} \text{Poi}(O_{it} \lambda_{it} z_{is}),\quad 13\times (s-1)+1\le t \le 13\times s, \label{eq:Poi} \\
    (z_{is} \mid \psi_{is}) & \overset{ind}{\sim} \text{IG}(1, \psi_{is}), \label{eq:IG}
\end{align} 
where $O_{it}$ and $\lambda_{it} > 0$ respectively denote the offset and the component accounting for the mean in area $i$ and week $t$, and $\psi_{is} > 0$ is the shape parameter of the inverse-Gaussian (IG) distribution in area $i$ and season $s$. The distribution in \eqref{eq:IG} imposes $\mathds{E}(Z_{is}) = 1$ and $\mathds{V}(Z_{is}) = \psi_{is}^{-1}$. By assuming \eqref{eq:Poi}-\eqref{eq:IG}, the marginal distribution of the response variable is $(y_{it} \mid \lambda_{it}, \psi_{is}) \overset{ind}{\sim} \text{PIG}(O_{it} \lambda_{it}, \psi_{is})$, so that the marginal mean and variance are given by $\mathds{E}(Y_{it} \mid \lambda_{it}, \psi_{is}) = O_{it} \lambda_{it}$ and $\mathds{V}(Y_{it} \mid \lambda_{it}, \psi_{is}) = O_{it} \lambda_{it} + (O_{it}\lambda_{it})^{2} \psi_{is}^{-1}$, respectively \citep[Chapter~6]{Hilbe2014}. Thus, for any positive random variable $Z_{is}$, we obtain an overdispersed distribution, where $\psi_{is}$ is the dispersion parameter \citep{BarretoSouza2015}. 

As commonly assumed, we consider a regression structure for $\lambda_{it}$ using the log-link function. To account for seasonal effects in the mean, we consider a spatio-temporal random intercept that is specific to spatial clusters and seasons. Let $\theta_{is}$ be the spatio-temporal parameter for area $i$ and season $s$. To define the random clustering structure on $\theta_{is}$ at each season $s$, we consider a graph $\mathcal{G}$ of neighboring areas. Let ${\bm \pi}_{s} = \{\mathscr{C}_{1}^{s}, \ldots, \mathscr{C}_{k_{s}}^{s}\}$ be a random partition in season $s$ obtained by removing some specific edges in $\mathcal{G}$ (see Section~\ref{s:SPPPM} for a discussion). Given ${\bm \pi}_{s}$, we assume that the parameters $\theta_{is}$ are identical across all areas within each cluster, that is,
$\theta_{is} = \theta_{j_{s}s}^{\star}$ for all $i \in \mathscr{C}_{j_{s}}^{s}$, where $\theta_{j_{s}s}^{\star}$ is the cluster-season-specific parameter for all areas belonging to cluster $\mathscr{C}_{j_{s}}^{s}$. Thus, $\lambda_{it}$ is given by
\allowdisplaybreaks \begin{align}
    \lambda_{it} &= \exp \{ {\bm X}_{it}^{\top} {\bm \beta} \} \theta_{j_{s}s}^{\star}, \; i \in \mathscr{C}_{j_{s}}^{s}, \label{eq:lambda}\\
    {\bm \beta} &\sim N_{p_{1}}({\bm \mu}_{\beta}, {\bm \Sigma}_{\beta}), \label{eq:beta_prior}
\end{align}
where ${\bm X}_{it}$ is a $p_{1}$-dimensional design vector that considers spatio-temporal predictors for area $i$ at week $t$ and ${\bm \beta} = \{ \beta_{1}, \ldots, \beta_{p_{1}} \}$ denotes their respective coefficients.
We assume that
\begin{equation}
    \theta^{\star}_{j_{s}s} \overset{iid}{\sim} \text{Ga}(a_{\theta}, b_{\theta}), \; a_{\theta}>0, \; b_{\theta}>0. \label{eq:theta_priorTCS} 
\end{equation}

As the vector ${\bm \beta}$ does not include a term for the intercept, $\log({\theta}_{j_{s}s}^{\star})$ works as the random intercept for cluster $j_{s}$ in season $s$. Thus, all areas belonging to the same spatial cluster share the same value for $\theta$ during all weeks of a season. This assumption is driven by the specific characteristics of mosquito-borne diseases. Appendix Figure~\ref{fig:time_response} shows that there is little variation in the number of cases reported from one week to the next, indicating that clustering is unlikely to change on a weekly basis. However, there is a clear seasonal pattern, with an increase in the number of cases occurring between October of one year and May of the following year. Figure~\ref{fig:sp_response} also supports the notion that spatial clustering behavior is seasonal. By incorporating a cluster-season-specific intercept in the model, we account for the overall impact on the Poisson distribution rate. We also allow for individual behavior in each area during each season. Areas are grouped together in each season based on the similarity of the $\theta_{i}$'s. Different groupings for each season are assumed to follow the structure detailed in Section~\ref{s:SPPPM}.

Although the dispersion parameter in \eqref{eq:IG} is usually assumed to be constant, it may be more reasonable to allow it to vary over time and space. Following \cite{BarretoSouza2015}, we consider a log-linear structure for $\psi_{is}$ letting
\begin{equation}
    \psi_{is} = \exp \{ {\bm V}_{is}^{\top} {\bm \delta} \}, \label{eq:psi} 
\end{equation}
where ${\bm V}_{is}$ is a $p_{2}$-dimensional design vector considering spatio-temporal predictors that might affect the dispersion, and ${\bm \delta} = \{ \delta_{0}, \ldots, \delta_{p_{2}} \}$ represents their effects. We also assume
\begin{equation}
    {\bm \delta} \sim N_{p_{2}}({\bm \mu}_{\delta}, {\bm \Sigma}_{\delta}). \label{eq:delta_prior}
\end{equation}

Unlike the mean, which is assumed to be different for each area $i$ and week $t$, dispersion is area-specific but remains constant within each season $s$. This assumption is justified by the characteristics of the context that motivates this study. The heterogeneity of mosquito-borne diseases is greater in summer than in winter (Figure~\ref{fig:sp_response}). This occurs because the climatic characteristics of summer are more favorable for the reproduction and spread of mosquitoes, affecting regions differently. The regression structures for both mean and dispersion can be easily modified to accommodate spatial, temporal, and/or spatio-temporal covariates, depending on the context in which the model is applied. For mosquito-borne disease data, the design matrix $\bm X$ comprises covariates measured by area and week, while the design matrix $\bm V$ contains information measured by area and season.

To complete the model specification, we need to specify hyperparameters for the prior distributions of the regression coefficients ${\bm \beta}$ and ${\bm \delta}$, as well as for the cluster-specific parameters $\theta_{j_{s}s}^{\star}$. Although the use of vague prior distributions is usually preferred for making posterior inferences when dealing with complex models, it is convenient to use informative priors, at least for some parameters. We consider vague priors for ${\bm \beta}$ and ${\bm \delta}$ by assigning them normal distributions with a mean of zero and a large variance. However, the prior distribution for the cluster-specific parameters requires more attention. As defined in \eqref{eq:lambda}, $\log( \theta_{j_{s}s}^{\star} )$ works as a random spatio-temporal cluster-specific intercept. Therefore, its hyperparameters should be specified according to the expected relative risk of the disease. One way to define these values would be to consider the simplest case of the model, where data overdispersion is not observed and the covariates have no effect, i.e., ${\bm z} = {\bm 1}$ and ${\bm \beta} = {\bm 0}$. In this case, $\theta_{is}$ is the only random effect explaining the outcome in area $i$ and season $s$. When ${\bm \theta} = {\bm 1}$, the outcome is entirely determined by the offset. Given this, it seems natural to set both the prior mean and variance of $\theta_{is}$ equal to one.

\subsection{Spatio-temporal PPM driven by spanning tree} \label{s:SPPPM}

One of our key contributions is developing a model for the temporal evolution of spatial partitions formed by contiguous clusters. Considering the seasonal impact of climate on mosquito-borne diseases, it is reasonable to expect greater fluctuations in case numbers during certain times of the year, especially when conditions favor the spread of these diseases, leading to an increase in clusters. Inspired by the approach of \cite{Teixeira2015}, we propose a spatio-temporal model using spanning trees and a product partition prior for random partitions. For a deeper understanding of spanning trees, please refer to Appendix Section~\ref{app:back}. Our model expands on the original framework by integrating the changes in partitions over time. To account for this time evolution on the partitions, we consider that the probabilities of removing edges may vary over seasons.
	
As our map is constant over time, the neighborhood structure is represented by a graph $\mathcal{G}$ that is common to all seasons. To simplify the sampling process of the posterior distribution of ${\bm \pi}_{s}$, we introduce a minimum spanning tree $\mathcal{T}_{s}$ that is randomly generated from $\mathcal{G}$ for each season $s$. The edges are conditionally independent and randomly removed from $\mathcal{T}_{s}$ with a probability of $\rho_{s} \in [0,1]$. Thus, a spatial partition of the areas in season $s$ is generated. The probability $\rho_{s}$ is common to all edges linking spatial vertices in season $s$ and influences the estimated number of clusters. Large values of $\rho_{s}$ induce a high expected number of clusters, while the opposite holds for small $\rho_{s}$. The cohesion function of cluster $\mathscr{C}_{j_{s}}^{s}$ is formulated as
\begin{align*}
C(\mathscr{C}_{j_{s}}^{s}) = \begin{cases}
\rho_{s} (1 - \rho_{s})^{|e_{j_{s}}|}, \quad & \text{if} \, \, j_{s} < k_{s}, \\
(1 - \rho_{s})^{|e_{j_{s}}|}, & \text{if} \, \, j_{s} = k_{s}, 
\end{cases} 
\end{align*}
where $|e_{j_{s}}|$ is the total number of edges not removed in $\mathscr{C}_{j_{s}}^{s}$, and $k_{s}$ is the number of clusters in season $s$. Assuming this function, the prior probability of partition given the tree is
\begin{align}
    \mathds{P}[{\bm \pi}_{s} = \{\mathscr{C}_{1}^{s}, \ldots, \mathscr{C}_{k_{s}}^{s}\} \mid \mathcal{T}_{s}, \rho_{s}] &= \begin{cases}
    \rho_{s}^{k_{s} - 1} (1 - \rho_{s})^{n - k_{s}}, \quad & \text{if} \, {\bm \pi}_{s} \prec \mathcal{T}_{s}, \\
    0, & \text{otherwise}. \label{eq:partition_prior} \end{cases}
\end{align}
In \eqref{eq:partition_prior}, ${\bm \pi}_{s} \prec \mathcal{T}_{s}$ denotes that partition and tree are compatible, i.e., ${\bm \pi}_{s}$ can be obtained by pruning $k_{s} - 1$ edges from $\mathcal{T}_{s}$. Note that this effectively reduces the collection of partitions we consider in each season to those compatible with the corresponding tree.

Like partitions, trees also vary across seasons. To ensure compatibility between trees and partitions, we assume independence of trees over seasons. Then, we assume independent uniform distributions so that for each season $s$, all possible trees of $\mathcal{G}$ are equally probable, that is,
\begin{equation}
    \mathds{P}[\{ \mathcal{T} \}_{s}] = \prod_{s = 1}^{S} \mathds{P}[\mathcal{T}_{s}], \quad \quad
		\mathds{P}[\mathcal{T}_{s}] \propto 1. \label{eq:tree_prior}
\end{equation}

While straightforward, the previous approach has the disadvantage of creating an independent sequence of partitions. To alleviate this problem while still keeping tractability, we consider an indirect correlation between partitions, by introducing a time-dependent structure in the vector ${\bm \rho} = (\rho_{1}, \ldots, \rho_{S})$. Specifically, we model ${\bm \rho}$ with an autoregressive time series model that is constrained to the unit interval. To achieve this, we adopt the hierarchical framework introduced by \cite{Jara2013}, where a sequence of beta random variables is linked through a set of exchangeable latent indicators built in an autoregressive way. For this purpose, let $\{ u_{s} \}$ and $\{ c_{s} \}$ be sequences of non-negative integer-valued latent variables. The prior distribution for $\rho_{s}$, given $\{ u_{s} \}$ and $\{ c_{s} \}$, is given by
\allowdisplaybreaks \begin{align}
(\rho_{s} \mid u_{s}, \ldots, u_{s-q}, c_{s}, \ldots, c_{s-q}, \upsilon, \kappa) & \overset{ind}{\sim} \text{Be} \left( \upsilon + \sum_{l = 0}^{q} u_{s-l}, \kappa + \sum_{l = 0}^{q} (c_{s-l} - u_{s-l}) \right), \label{eq:rho_prior} \\
(u_{s} \mid c_{s}, w) & \overset{ind}{\sim} \text{Bin}(c_{s}, w), \label{eq:u_prior}\\
(c_{s} \mid \zeta) & \overset{iid}{\sim} \text{Poi}(\zeta), \label{eq:c_prior} \\
(w \mid \upsilon, \kappa) & \sim \text{Be}(\upsilon, \kappa), \label{eq:w_prior}
\end{align}
with $\upsilon > 0$, $\kappa > 0$, and $\zeta > 0$. Latent variables $u_{s}$ and $c_{s}$ are defined to be zero for $s \leq 0$. In this formulation, the role of $\{ u_{s} \}$ is to establish a link between the $\rho_{s}$'s. The hyperparameter $w$ is the common success probability that determines the overall series level, and $q \geq 1$ represents the order of the autoregressive process.
 
The advantages of using this approach are numerous. By defining the temporal dependence through a sequence of latent variables instead of directly on the original probability vector, we ensure a beta marginal distribution for $\rho_{s}$, specifically $\rho_{s} \sim \text{Be}(\upsilon, \kappa)$ \citep{Jara2013}. This property is helpful for obtaining marginal distributions for partitions and the number of clusters in each season \citep{Teixeira2015, Teixeira2019}. By identifying how the number of clusters is distributed {\it a priori}, we can calculate its mean and variance as
\begin{equation}
    \mathds{E}(k_{s} \mid \cdot) = (n - 1) \frac{\upsilon}{\upsilon + \kappa} + 1 \quad \text{and} \quad
    \mathds{V}(k_{s} \mid \cdot) = (n - 1) \frac{\upsilon \kappa (\upsilon + \kappa + n - 1)}{(\upsilon + \kappa)^{2} (\upsilon + \kappa + 1)}, \label{eq:exp_var_ncl}
\end{equation}
respectively. This is a valuable result for prior elicitation purposes, since it is clear from \eqref{eq:exp_var_ncl}, that $\rho_{s}$ directly impacts the number of clusters through its hyperparameters. If $\upsilon = \kappa$, the average marginal probability of removing the edge is 50\%, implying an expected number of clusters around $n/2$. If $\upsilon \rightarrow 0$ and/or $\kappa \rightarrow \infty$, then $\{ \rho_{s} \} \rightarrow 0$ and all areas are grouped into the same cluster. Conversely, if $\upsilon \rightarrow \infty$ and/or $\kappa \rightarrow 0$, then $\{ \rho_{s} \} \rightarrow 1$ creating $n$ clusters.

Another advantage of this approach lies in the temporal autocorrelation of $\{ \rho_{s} \}$ and consequently of $\{ {\bm \pi}_{s} \}$. The proposed model assumes: (i) trees are independent for each season; (ii) given the tree and the probability of removing edges, partitions are independent for each season; and (iii) probabilities of removing edges are autocorrelated over time. Thus, partitions exhibit autocorrelation due to the underlying process that drives them, something similar to the construction established in Hidden Markov Models. Within seasons, the correlation between observations occurs indirectly since they share the common parameters ${\bm \theta}$ and ${\bm z}$ (see in Appendix Section~\ref{ss:rho}). For model \eqref{eq:rho_prior}--\eqref{eq:w_prior}, \cite{Jara2013} showed that the autocorrelation function of $\{ \rho_{s} \}$ can be computed in closed form as
\begin{equation}
    \text{corr}(\rho_{s}, \rho_{s+l}) = \frac{(\upsilon + \kappa) \sum\limits_{h=0}^{q-l} c_{s-h} + \left(\sum\limits_{h=0}^{q} c_{s-h}\right) \left(\sum\limits_{h=0}^{q} c_{s+l-h}\right)}{\left( \upsilon + \kappa + \sum\limits_{h=0}^{q} c_{s-h}\right) \left( \upsilon + \kappa + \sum\limits_{h=0}^{q} c_{s+l-h}\right)},
\label{eq:cor_rho}\end{equation}
for $s, l \geq 1$. By varying the values of $\upsilon$, $\kappa$, and $\{c_{s}\}$, we can obtain different degrees of autocorrelation. For all $s$, $l$, and ${\bm c}$, if $( \upsilon + \kappa ) \rightarrow 0$, then corr$(\rho_{s}, \rho_{s+l}) \rightarrow 1$, whereas, if $( \upsilon + \kappa ) \rightarrow \infty$, then corr$(\rho_{s}, \rho_{s+l}) \rightarrow 0$. Regarding ${\bm c}$, the first term of the numerator in \eqref{eq:cor_rho} considers shared values of $\{ c_{s} \}$, while the other terms consider sets of $\{ c_{s} \}$ that define $\rho_{s}$ and $\rho_{s+l}$. Hence, when $\rho_{s}$ and $\rho_{s+l}$ do not share any $\{ c_{s} \}$, then $\sum_{h=0}^{q-l} c_{s-h} = 0$, and the autocorrelation function is driven by the values of $\upsilon$ and $\kappa$. Furthermore, when $\{c_{s}\} \rightarrow 0$, then corr$(\rho_{s}, \rho_{s+l}) \rightarrow 0$ for all $s$, $l$, $\upsilon$, and $\kappa$.

The full Bayesian hierarchical structure is achieved by assuming prior distributions for the remaining hyperparameters, which we choose as
\begin{equation}
\upsilon \sim \text{Ga}(a_{\upsilon}, b_{\upsilon}), \quad 
\kappa \sim \text{Ga}(a_{\kappa}, b_{\kappa}), \quad 
\zeta \sim \text{Ga}(a_{\zeta}, b_{\zeta}). \label{eq:priors}
 \end{equation}
The choice of values for $a_{\upsilon}$, $b_{\upsilon}$, $ a_{\kappa}$, and $b_{\kappa}$ in \eqref{eq:priors} must be guided by the expected number of clusters in the first season, while the values for $a_{\zeta}$ and $b_{\zeta}$ impact the autocorrelation function. See in Appendix Sections~\ref{ss:ncl_prior}, \ref{ss:rho}, and \ref{app:prior_spec} for further details.

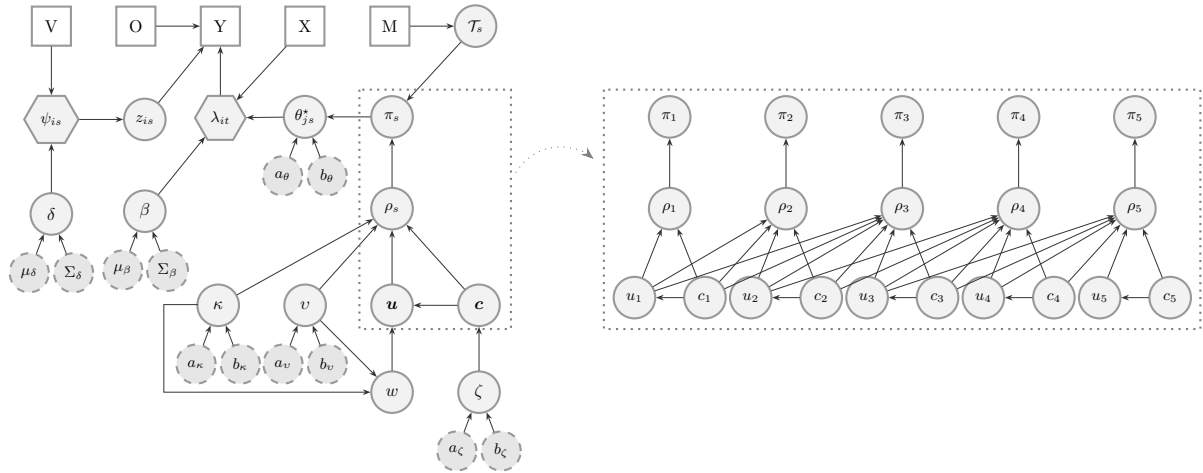
\begin{figure} [hbt!] \centering
\resizebox{\columnwidth}{!}{%
\begin{tikzpicture}[scale=0.65, transform shape,
squarednode/.style={rectangle, draw=black!40, thick, minimum size=8mm},
hexagonode/.style={regular polygon,regular polygon sides=6, draw=black!40, fill=black!5, thick, minimum size=9.5mm},
roundnode/.style={circle, draw=black!40, fill=black!5, thick, minimum size=8.5mm},
roundnode2/.style={circle, dashed, draw=black!40, fill=black!10, thick, minimum size=8mm}]

    \node[squarednode] (v) {V};
    \node[squarednode, right=9mm of v] (o) {O};
    \node[squarednode, right=9mm of o] (y) {Y};
    \node[squarednode, right=9mm of y] (x) {X};
    \node[squarednode, right=9.1mm of x] (g) {M};
    
    \node[hexagonode, below=10mm of v] (psi) {\small $\psi_{is}$};
    \node[roundnode, below=10mm of psi] (delta) {$\delta$};
    \node[roundnode2, below left=6mm and -1.5mm of delta] (md) {\small $\mu_{\delta}$};
    \node[roundnode2, below right=6mm and -1.5mm of delta] (sd) {\small $\Sigma_{\delta}$};

    \node[roundnode, right=9mm of psi] (z) {\small $z_{is}$};
    \node[hexagonode, below=10mm of y] (lambda) {\small $\lambda_{it}$};
    \node[roundnode, below=10mm of x] (theta) {\small $\theta^{\star}_{js}$};

    \node[roundnode, below=10mm of z] (beta) {$\beta$};  
    \node[roundnode2, below left=6mm and -1.5mm of beta] (mb) {\small $\mu_{\beta}$};
    \node[roundnode2, below right=6mm and -1.5mm of beta] (sb) {\small $\Sigma_{\beta}$};
    
    \node[roundnode, right=9mm of theta] (pi) {\small $\pi_{s}$};
    \node[roundnode, right=9mm of g] (t) {\small $\mathcal{T}_{s}$};
    \node[roundnode, below=10mm of pi] (rho) {\small $\rho_{s}$}; 
    
    \node[roundnode2, below left=6mm and -1.5mm of theta] (at) {\small $a_{\theta}$};    
    \node[roundnode2, below right=6mm and -1.5mm of theta] (bt) {\small $b_{\theta}$};

    \node[roundnode, below=11mm of rho] (u) {$\bm u$};
    \node[roundnode, left=9mm of u] (up) {$\upsilon$};
    
    \node[roundnode, right=9mm of u] (c) {$\bm c$};
    \node[roundnode, below=9mm of c] (zeta) {$\zeta$};
    \node[roundnode2, below left=6mm and -1.5mm of zeta] (az) {\small $a_{\zeta}$};
    \node[roundnode2, below right=6mm and -1.5mm of zeta] (bz) {\small $b_{\zeta}$};
    
    \node[roundnode, below=9mm of u] (w) {$w$};
    \node[roundnode, left=9mm of up] (k) {$\kappa$};
    
    \node[roundnode2, below left=6mm and -1.5mm of k] (ak) {\small $a_{\kappa}$};
    \node[roundnode2, below right=6mm and -1.5mm of k] (bk) {\small $b_{\kappa}$};
    \node[roundnode2, below left=6mm and -1.5mm of up] (au) {\small $a_{\upsilon}$};
    \node[roundnode2, below right=6mm and -1.5mm of up] (bu) {\small $b_{\upsilon}$};

    \node[roundnode, right=48mm of pi] (pi1) {\small$\pi_{1}$};
    \node[roundnode, right=15mm of pi1] (pi2) {\small$\pi_{2}$};
    \node[roundnode, right=15mm of pi2] (pi3) {\small$\pi_{3}$};
    \node[roundnode, right=15mm of pi3] (pi4) {\small$\pi_{4}$};
    \node[roundnode, right=15mm of pi4] (pi5) {\small$\pi_{5}$};

    \node[roundnode, below=10mm of pi1] (rho1) {\small$\rho_{1}$};
    \node[roundnode, below=10mm of pi2] (rho2) {\small$\rho_{2}$};
    \node[roundnode, below=10mm of pi3] (rho3) {\small$\rho_{3}$};
    \node[roundnode, below=10mm of pi4] (rho4) {\small$\rho_{4}$};
    \node[roundnode, below=10mm of pi5] (rho5) {\small$\rho_{5}$};
    
    \node[roundnode, below left=12mm and 1mm of rho1] (u1) {\small$u_{1}$};
    \node[roundnode, below left=12mm and 1mm of rho2] (u2) {\small$u_{2}$};
    \node[roundnode, below left=12mm and 1mm of rho3] (u3) {\small$u_{3}$};
    \node[roundnode, below left=12mm and 1mm of rho4] (u4) {\small$u_{4}$};
    \node[roundnode, below left=12mm and 1mm of rho5] (u5) {\small$u_{5}$};

    \node[roundnode, below right=12mm and 1mm of rho1] (c1) {\small$c_{1}$};
    \node[roundnode, below right=12mm and 1mm of rho2] (c2) {\small$c_{2}$};
    \node[roundnode, below right=12mm and 1mm of rho3] (c3) {\small$c_{3}$};   
    \node[roundnode, below right=12mm and 1mm of rho4] (c4) {\small$c_{4}$};
    \node[roundnode, below right=12mm and 1mm of rho5] (c5) {\small$c_{5}$};

    \path[-{Stealth[]}, line width=0.0mm, black!80]
    (md) edge (delta) (sd) edge (delta) (delta) edge (psi) (v) edge (psi) (psi) edge (z) 
    (z) edge (y) (o) edge (y) (x) edge (lambda) (lambda) edge (y) (beta) edge (lambda)
    (theta) edge (lambda) (at) edge (theta) (bt) edge (theta) (mb) edge (beta) (sb) edge (beta)
    (pi) edge (theta) (t) edge (pi) (rho) edge (pi) (g) edge (t) (c) edge (rho) (up) edge (rho)
    (u) edge (rho) (k) edge (rho) (ak) edge (k) (bk) edge (k) (w) edge (u) (up) edge (w) 
    (zeta) edge (c) (az) edge (zeta) (bz) edge (zeta) (au) edge (up) (bu) edge (up) (c) edge (u); 

    \draw[-{Stealth[]}, line width=0.0mm, black!80] (3,-5.7) -| +(-0.7,-0.5) |- (w);    

    \path[-{Stealth[]}, line width=0.0mm, black!80]
    (rho1) edge (pi1) (rho2) edge (pi2) (rho3) edge (pi3) (rho4) edge (pi4) (rho5) edge (pi5) 
    (u1) edge (rho1) (u2) edge (rho2) (u3) edge (rho3) (u4) edge (rho4) (u5) edge (rho5) 
    (u1) edge (rho2) (u2) edge (rho3) (u3) edge (rho4) (u4) edge (rho5) 
    (u1) edge (rho3) (u2) edge (rho4) (u3) edge (rho5)
    (c1) edge (u1) (c2) edge (u2) (c3) edge (u3) (c4) edge (u4) (c5) edge (u5)
    (c1) edge (rho1) (c2) edge (rho2) (c3) edge (rho3) (c4) edge (rho4) (c5) edge (rho5)
    (c1) edge (rho2) (c2) edge (rho3) (c3) edge (rho4) (c4) edge (rho5) 
    (c1) edge (rho3) (c2) edge (rho4) (c3) edge (rho5); 

    \draw[black!50,thick,dotted] (6.3,-1.3) -- (6.3,-6.2) -- (9.4,-6.2) -- (9.4,-1.3) -- (6.3,-1.3); 
    \draw[black!50,thick,dotted] (11.3,-1.3) -- (11.3,-6.2) -- (23.5,-6.2) -- (23.5,-1.3) -- (11.3,-1.3);
    \draw [-{Stealth[]}, black!50, dotted] (9.5,-3) arc (150:50:30pt);
    
\end{tikzpicture} %
}
\caption{Graphical representation of the model highlighting an autoregressive process of order $q = 2$. Square nodes represent inputs -- $M$: adjacency matrix, $X$ and $V$: design matrices, $O$: offset, and $Y$: outcome. Hexagon nodes represent generated quantities. Circle nodes represent unknown (solid lines) and known (dashed lines) parameters.} \label{fig:model}
\end{figure}

Figure~\ref{fig:model} provides a graphical representation of our proposed model, highlighting the fact that the time dependence of probabilities $\bm \rho$ is induced by the sequences of latent variables $\{u_{s}\}$ and $\{c_{s}\}$. Even more importantly, the time dependence of the sequence of partitions $\{ {\bm \pi}_{s} \}$ is induced by the time dependence of $\bm \rho$.

\section{Simulation study} \label{sec:simulations}

The objectives of this section are twofold: (i) to compare the performance of Poisson and PIG models in both overdispersed and equidispersed spatio-temporal scenarios, and (ii) to evaluate the impact of the order of temporal dependence on clustering. In both studies, we generated synthetic datasets using adaptations of the model defined in \eqref{eq:Poi}--\eqref{eq:delta_prior}. To create realistic scenarios, we utilized the 70 microregions of Minas Gerais, Brazil, where two regions are deemed neighbors if they share a geographic boundary. Additionally, we incorporated observed data as offsets and covariates in our analysis. Population size (per 100,000 people) served as the offset, while temperature and humidity were included as covariates, forming ${\bm X}$ and ${\bm V}$ (data details are in Section~\ref{sec:motivation}). Unlike ${\bm X},$ which contains weekly measurements, ${\bm V}$ consists of average values for each season and includes a column of $1$'s for the intercept. These settings applied to all simulated datasets, with specific study details in Appendix Section~\ref{app:sim}. After data generation, we implemented the MCMC algorithm (described in Appendix Section~\ref{app:MCMC}), saving 1,000 samples from 10,000 iterations while discarding the first 70\% as burn-in and thinning by 3 to reduce correlation. Convergence was monitored graphically. Prior specifications are detailed in Appendix Section~\ref{app:sim}. We used the Watanabe-Akaike information criterion \citep[WAIC,][]{Gelman2014} to evaluate model fit and the {\tt salso} R package \citep{Dahl2020} along with the variation information (VI) loss function for estimating partitions. We then calculated the Rand Index \citep[RI,][]{Hubert1985} to assess the similarity between true and estimated partitions. Key characteristics and results for each study are summarized below, with further details in Appendix Section~\ref{app:sim}.

\subsection{Simulation 1 - Comparing PIG and Poisson models when fitting equidispersed and overdispersed data} \label{s:sim1}

The data generation process consists of two stages. First, we establish the spatial partitions, followed by the generation of the data itself. Rather than simulating partitions based on the model structure defined in \eqref{eq:partition_prior}--\eqref{eq:w_prior}, we adopt an arbitrary approach to determine the clustering structure. This approach encompasses three distinct scenarios. Scenario 1 consists of a partition formed by 4 clusters that remain constant over time. In Scenario 2 each season has a different partition remaining the same over the years (i.e., summer's partition is the same for all years, autumn's partition is the same for all years, and so on). In this case, partitions are made up of 4, 3, 2, and 2 clusters for summer, autumn, winter, and spring, respectively. Finally, we consider a most complex case, Scenario 3, where partitions are different for each season and year resulting in 12 different cluster formations with 2 to 7 components. Maps showing each partition structure considered in this simulation study are displayed in Appendix Figure~\ref{fig:map_part_sim1}. Once the partition structure is defined, we utilize \eqref{eq:Poi}--\eqref{eq:delta_prior} to generate the data. A total of 156 time points were created to represent the epidemiological weeks (i.e., $t = 1, \ldots, 156$). Given that a year comprises 52 weeks, this corresponds to a duration of three years, resulting in 12 distinct seasons, each lasting 13 weeks (i.e., $s = 1, \ldots, 12$). The difference in producing equidispersed and overdispersed data lies in the way the component $z_{is}$ accounting for data heterogeneity is defined. To produce equidispersed data, it is sufficient to assume ${\bm z} = {\bm 1}$; thus, the mean and variance will be equal. In the other case, $z_{is}$ is randomly generated from the IG density as presented in \eqref{eq:IG}. The values assigned to the parameters $\bm \beta$, $\bm \theta$, and $\bm \delta$ can be found in the Appendix Section~\ref{app:sim1}. Once we had the dataset, we fitted PIG and Poisson models with $q = 1$.

We first evaluated the models' ability to recover regression coefficients (Table~\ref{tab:sim1}). The coverage associated with $\bm \beta$ was similar for all models. This result was expected, as the regression structure used to explain the average is the same for all datasets. In the case of $\bm \delta$ coefficients, we observed that when the PIG model was applied to overdispersed data, it was able to recover these parameters. However, when we fitted the PIG model to data where the mean and variance are equal, the model tended to overestimate these coefficients, especially the intercept. This occurs because, when estimating high values for $\bm \delta$, we obtain high values for $\bm \psi$, which implies that $\mathds{V}(z_{is}) \rightarrow 0$, resulting in $z_{is} \approx 1$. Indeed, the coverage associated with $\bm z$ is very high for all scenarios. In cases where data are equidispersed, the model correctly estimates $z_{is} = 1$, which is equivalent to a Poisson model. Another characteristic explored in this study concerns the goodness of fit (Table~\ref{tab:sim1}). For equidispersed data, models presented very similar performance. When considering overdispersed data, PIG model proved to be advantageous, exhibiting lower WAIC values that were consistent across all scenarios. RI for PIG and Poisson models when the data were equidispersed had similar performance. In this case, the accuracy of the partition estimates obtained was high, which agrees with our previous discussion. For overdispersed data, Poisson model almost completely lost its clustering capacity, while PIG model provided averaged RI values exceeding 60\%. The conclusion drawn from this simulation study is that a PIG model can be applied to both equidispersed and overdispersed data. When overdispersion is not present, PIG model assumes its particular Poisson model case. Therefore, this model can simultaneously deal with estimation and clustering. The opposite does not hold true; when a Poisson model is applied to overdispersed data, the model's clustering capacity decreases drastically. In this case, $\bm \theta$ attempts to explain the interaction between the cluster-specific intercept and the dispersion parameter, making it difficult for areas to co-cluster. Additional results can be found in Appendix Section~\ref{app:sim1}.
\begin{center} \begin{table} [!t]
 \caption{Model fit performance metrics used to compare Poisson-inverse Gaussian and Poisson models applied to equidispersed and overdispersed data. Values are averaged over the 100 generated datasets. Coverage rate was calculated based on 95\% credible intervals. Lower WAIC values indicate better fit. Higher RI value indicate higher accuracy of partition estimates.} \label{tab:sim1}
\resizebox{\columnwidth}{!}{ %
\begin{tabular}{ccccccccc}
\toprule
\multicolumn{2}{@{}c}{\multirow{2}{*}{\bf Equidispersed}} & \multicolumn{5}{@{}c}{Coverage rate $|$ posterior mean (standard deviation)} & \multirow{2}{*}{\small WAIC} & \multirow{2}{*}{\small RI}  \\ 
\cmidrule{3-7}
 & & $\beta_{1}$ & $\beta_{2}$ & $\delta_{0}$ & $\delta_{1}$ & $\delta_{2}$ &  & \\ 
\cmidrule{1-1} \cmidrule{2-2} \cmidrule{3-7} \cmidrule{8-9}
 \multirow{2}{*}{Sce 1} & {\small PIG} & 0.98 $|$ 0.40 (0.01) & 0.99 $|$ 0.10 (0.01) & 0.00 $|$ 9.51 (0.59) & 0.83 $|$ -0.01 (0.57) & 0.80 $|$ -0.36 (0.59) & 53085 & 0.99 \\
                             & Poisson & 0.98 $|$ 0.40 (0.01) & 0.99 $|$ 0.10 (0.01) & -- & -- & -- & 53086 & 0.99 \\
 \multirow{2}{*}{Sce 2} & {\small PIG} & 0.97 $|$ 0.40 (0.01) & 0.68 $|$ 0.11 (0.01) & 0.00 $|$ 9.62 (0.54) & 0.83 $|$ 0.13 (0.59) & 0.65 $|$ -0.52 (0.57) & 51416 & 0.96 \\
                             & Poisson & 0.98 $|$ 0.40 (0.01) & 0.68 $|$ 0.11 (0.01) & -- & -- & -- & 51414 & 0.96 \\
 \multirow{2}{*}{Sce 3} & {\small PIG} & 0.95 $|$ 0.40 (0.01) & 0.90 $|$ 0.11 (0.01) & 0.00 $|$ 8.64 (0.52) & 0.82 $|$ -0.04 (0.53) & 0.85 $|$ -0.33 (0.56) & 51373 & 0.98 \\
                             & Poisson & 0.98 $|$ 0.40 (0.01) & 0.87 $|$ 0.11 (0.01) & -- & -- & -- & 51368 & 0.98 \\
 \midrule
 \multicolumn{2}{@{}c}{\bf Overdispersed} & & & & & \\
\cmidrule{1-2}
 \multirow{2}{*}{Sce 1} & {\small PIG} & 0.98 $|$ 0.40 (0.00) & 1.00 $|$ 0.10 (0.00) & 0.72 $|$ -0.16 (0.06) & 0.96 $|$ 0.13 (0.08) & 0.82 $|$ -0.22 (0.09) & 48756 & 0.65 \\
                             & Poisson & 0.98 $|$ 0.40 (0.00) & 1.00 $|$ 0.10 (0.00) & -- & -- & -- & 49238 & 0.04 \\
 \multirow{2}{*}{Sce 2} & {\small PIG} & 0.98 $|$ 0.40 (0.00) & 0.88 $|$ 0.11 (0.00) & 0.86 $|$ -0.19 (0.07) & 0.96 $|$ 0.13 (0.08) & 0.92 $|$ -0.32 (0.09) & 47182 & 0.61 \\
                             & Poisson & 1.00 $|$ 0.40 (0.00) & 0.84 $|$ 0.11 (0.00) & -- & -- & -- & 47577 & 0.02 \\
 \multirow{2}{*}{Sce 3} & {\small PIG} & 1.00 $|$ 0.40 (0.00) & 0.76 $|$ 0.11 (0.00) & 0.86 $|$ -0.21 (0.07) & 0.98 $|$ 0.16 (0.08) & 0.94 $|$ -0.31 (0.09) & 47108 & 0.65 \\
                             & Poisson & 0.98 $|$ 0.40 (0.00) & 0.60 $|$ 0.11 (0.00) & -- & -- & -- & 47491 & 0.03 \\
\bottomrule
\end{tabular} %
} 
\end{table} \end{center}

\subsection{Simulation 2 - Exploring the order of temporal dependence and its impact on clustering} \label{s:sim2}

In accordance with the data generation process employed in the preceding simulation study, we first arbitrarily established spatial partitions, then, we generated data utilizing part of the proposed model. This study investigates a clustering scenario in which each season is associated with a distinct partition that is consistently repeated over multiple years. We examined two distinct situations: in Scenarios 1 and 2, the number of clusters remains relatively stable, comprising 4, 3, 1, and 2 clusters for summer, autumn, winter, and spring, respectively. Scenarios 3 and 4 introduce greater variability in the number of clusters, consisting of 5, 10, 4, and 2 clusters for the respective seasons. Appendix Figure~\ref{fig:map_part_sim2} illustrates the maps depicting each partition structure considered in this simulation study.  Once the partitions were produced, model \eqref{eq:Poi}--\eqref{eq:delta_prior} was considered as a data-generating mechanism to create 100 synthetic datasets with 260 time points representing the epidemiological weeks (i.e., $t = 1, \ldots, 260$). This is equivalent to 5 years, resulting in 20 seasons, each lasting 13 weeks (i.e., $s = 1, \ldots, 20$). Data in  Scenarios 1 and 2 where generated assuming different values for the cluster-specific parameters. The same strategy was assumed in Scenarios 3 and 4. The cluster-specific parameters were set as
\begin{equation*} \resizebox{\columnwidth}{!}{ \begin{tabular}{ll} 
			\multicolumn{1}{c}{\bf Scenario 1} & \multicolumn{1}{c}{\bf Scenario 2} \\
			${\bm \theta}^{\star}_{s} = \begin{cases}
				(1, 3, 6, 9) & \rm{for} \; s = 1, 5, 9, 13, 17 \\
				(1, 5, 9) & \rm{for} \; s = 2, 6, 10, 14, 18 \\
				1 &  \rm{for} \; s = 3, 7, 11, 15, 19 \\
				(1, 4) &  \rm{for} \; s = 4, 8, 12, 16, 20 \\
			\end{cases}$ & ${\bm \theta}^{\star}_{s} = \begin{cases}
				(1, 10, 25, 45) &  \rm{for} \; s = 1, 5, 9, 13, 17 \\
				(1, 10, 25) &  \rm{for} \; s = 2, 6, 10, 14, 18 \\
				1 &  \rm{for} \; s = 3, 7, 11, 15, 19 \\
				(1, 10) &  \rm{for} \; s = 4, 8, 12, 16, 20 \\
			\end{cases}$ \\
		\\
			\multicolumn{1}{c}{\bf Scenario 3} & \multicolumn{1}{c}{\bf Scenario 4} \\
			${\bm \theta}^{\star}_{s} = \begin{cases}
				(1, 3, 6, 9, 13) &  \rm{for} \; s = 1, 5, 9, 13, 17 \\
				(1, 3, 6, 9, 12, 16, 20, 25, 30, 35) &  \rm{for} \; s = 2, 6, 10, 14, 18 \\
				(1, 3, 6, 9) &  \rm{for} \; s = 3, 7, 11, 15, 19 \\
				(1, 4) &  \rm{for} \; s = 4, 8, 12, 16, 20 \\
			\end{cases}$ & ${\bm \theta}^{\star}_{s} = \begin{cases}
				(1, 10, 25, 45, 70) &  \rm{for} \; s = 1, 5, 9, 13, 17 \\
				(1, 9, 20, 35, 55, 75, 100, 120, 150, 180) &  \rm{for} \; s = 2, 6, 10, 14, 18 \\
				(1, 10, 25, 40) &  \rm{for} \; s = 3, 7, 11, 15, 19 \\
				(1, 15) &  \rm{for} \; s = 4, 8, 12, 16, 20 \\
			\end{cases}$ \end{tabular} } \end{equation*}
These configurations allows us to compare the ability of the proposed model to estimate the partition when the clusters are similar (Scenarios 1 and 3), and when there are more pronounced differences among them (Scenarios 2 and 4). The values assigned to the other parameters can be found in the Appendix Section~\ref{app:sim2}.

\begin{figure} [!t] \centering
 \includegraphics[width=\textwidth]{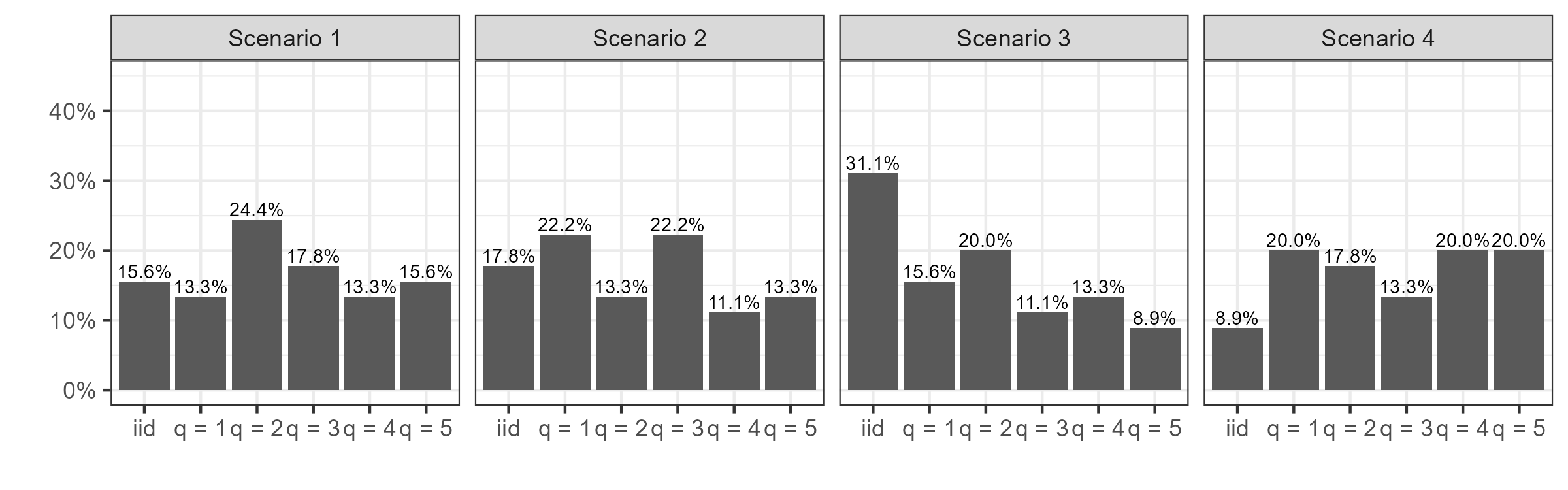}
\caption{Barplot indicating how frequently each model obtained the lowest WAIC value over 100 datasets. In case of a tie, the simplest model was credited.} \label{fig:sim2}
\end{figure}

To analyze the simulated data, we fitted PIG models with $q = 1, 2, \ldots, 5$. Additionally, we fitted  a particular case of the proposed model in which partitions are time-independent. Note that to do this, it is sufficient to fix ${\bm u} = {\bm c} = {\bm 0}$, thus $\rho_{s}$ is independent and identically distributed as Be$(\upsilon, \kappa)$. To evaluate the goodness of fit, we calculated the WAIC for each fitted model. Then we computed the frequency of times each model obtained the lowest WAIC value over the 100 datasets, which is displayed in Figure~\ref{fig:sim2}. Overall, adding a temporal structure to the partition prior tends to enhance the fit performance compared to using the independent version. There was only one scenario in which the model with independent partitions outperformed the others. However, it was unclear which order of dependence yielded the best fit. Looking at the averaged WAIC values (Table~\ref{tab:sim2_1} in the Appendix), the difference between models seems imperceptible, and this occurred similarly in all scenarios. Appendix Figure~\ref{fig:sim2_1} and Table~\ref{tab:sim2_1} show that the proposed model provided good accuracy in partition estimation, with RI values averaging over 90\% for all scenarios. Nonetheless, similar to the WAIC, it is difficult to identify the best order of dependence. On average, the RI values are comparable among the models (Appendix Figure~\ref{fig:sim2_1}). Indeed, all parameters used in the temporal structure were estimated similarly across the models, as observed in Appendix Table~\ref{tab:sim2_2}.

\section{An application to dengue data} \label{sec:dataapp}

We applied the proposed model to analyze data described in Section~\ref{sec:motivation} related to dengue cases reported weekly in the Southeast region of Brazil between 2018 and 2023. The response variable $y_{it}$ represents the number of dengue cases; $O_{it}$ is the population size (per 100,000 people); and ${\bm X}_{it}$ is the design vector composed of temperature, humidity, and HDI, all measured per area $i = 1, \ldots, 145$ and epidemiological week $t = 1, \ldots, 313$. The design vector ${\bm V}_{is}$ is composed of temperature and humidity averaged per season (spring, summer, autumn, and winter) in each area, with $s = 1, \ldots, 24$. We fitted the proposed model \eqref{eq:Poi}--\eqref{eq:priors} by fixing the prior parameters as follows: for regression coefficients, ${\bm \mu}_{\beta} = {\bm 0}$ and ${\bm \Sigma}_{\beta} = 10 {\bm I}$; for the cluster- and season-specific parameter, $a_{\theta} = b_{\theta} = 1$; we kept $a_{\upsilon} = 10$, $b_{\upsilon} = 1$, $a_{\kappa} = 100$, and $b_{\kappa} = 1$ so that the {\it a priori} expected number of clusters in the first season is around 10\% of the total number of areas; and $a_{\zeta} = b_{\zeta} = 1$ so that the {\it a priori} temporal correlation of the sequence of probabilities is close to zero. For the MCMC, we saved a sample of size 1,000 by running 15,000 iterations, discarding the first two-thirds as burn-in, and thinning by 5. Convergence was monitored graphically. Prior elicitation and the MCMC algorithm are available in Appendix Sections~\ref{app:prior_spec} and \ref{app:MCMC}, respectively.

To account for temporal trends, we varied the dependence order parameter $q$ from 1 to 12, corresponding to a time span of three years. This approach enabled us to assess the impact of different $q$ values on the posterior inference of $\bm \rho$ and the resulting partition. Additionally, we fitted a simpler model that assumed independent partitions. Model selection was conducted using the WAIC, as detailed in Appendix Table~\ref{tab:app_1}. Our findings indicate that a first-order autoregressive structure provides the best fit for the dengue data from Southeast Brazil. All subsequent results presented in this section were derived from fitting the proposed model with a dependence order of $q = 1$.
\begin{figure} \begin{center}
\includegraphics[width=\textwidth]{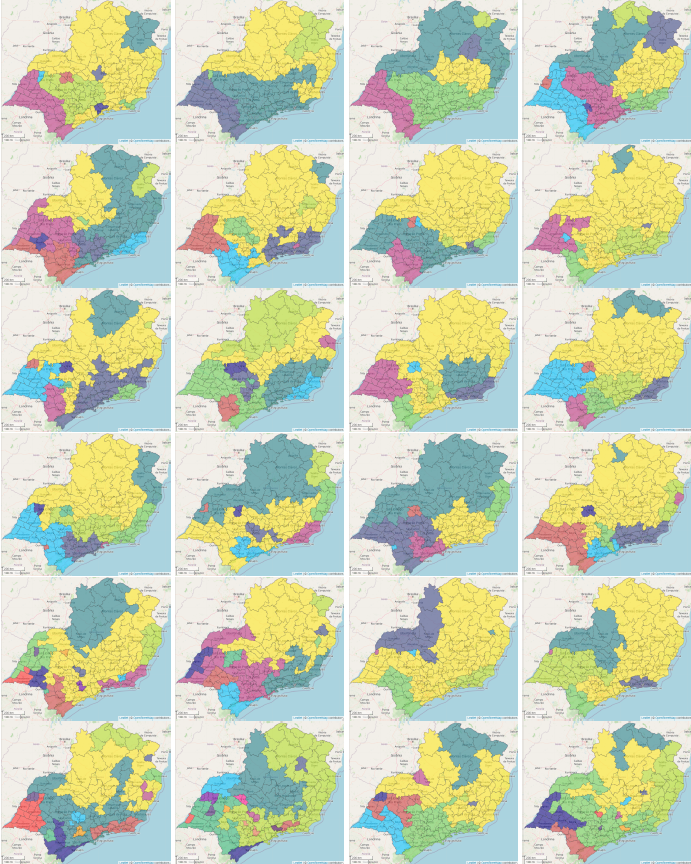}
\end{center}
\caption{Posterior estimate of the random partition for dengue cases on the Brazilian Southeast region from 2018 (top) to 2023 (bottom) by seasons - summer (1st column), autumn (2nd column), winter (3rd column), and spring (4th column).} \label{fig:time_est_part}
\end{figure} 

Figure~\ref{fig:time_est_part} displays estimated partitions for each season, where colors differentiate clusters within each partition. Although colors are repeated over time, they do not necessarily represent the same clusters or even relate them. Overall, our model estimated between five and eighteen clusters. Winter was the season with the lowest estimated number of clusters (6 to 9), followed by spring (6 to 13), summer (9 to 14), and autumn, when the largest number of clusters was estimated (5 to 18). This pattern corroborates the temporal trend of the observed data. In Brazil, dengue cases usually begin rising in December and reach their peak between April and May, which corresponds to summer and autumn. In this region, the time from summer to early autumn is characterized by warm and rainy weather, creating suitable conditions for the survival of vectors. The opposite occurs in winter and extends until spring. Additionally, we should consider the mosquito's lifespan and the time of virus incubation, which may delay the emergence of cases. In general, we observed a greater number of singletons in the partitions estimated for 2023, mainly in autumn and spring. This may be due to the significant increase in the number, scale, and simultaneous occurrence of multiple outbreaks observed in 2023, which spread into regions that were previously unaffected.

Apart from the temporal trend observed in the number of clusters, there is a smooth change in the formation of clusters over time, with no significant variations in their configuration. A spatial pattern of clusters can also be observed when we examine partitions of the same season over years, which may suggest a seasonal dependence. To measure similarity between estimated partitions over time, we calculated the lagged RI values using the {\tt salso} R package, as shown in Appendix Figure~\ref{fig:corr_plot}(A). Although the partitions are similar, with an averaged RI of 69\%, it is difficult to detect a temporal or even seasonal pattern from the similarity matrix. On the other hand, when examining the estimated values of the probability of removing edges in each season (Appendix Figure~\ref{fig:corr_plot}(B)), we observed a certain temporal trend. A pattern repeats itself over the first four years, with a rising trend between spring and autumn, followed by a rising trend between autumn and spring. Additionally, Appendix Figure~\ref{fig:corr_plot}(C) shows the autocorrelation function. Recall that we chose the model with $q = 1$, so we show the correlation between $\rho_{s}$ and $\rho_{s-1}$.
\begin{figure} \begin{center}
\includegraphics[width=\textwidth]{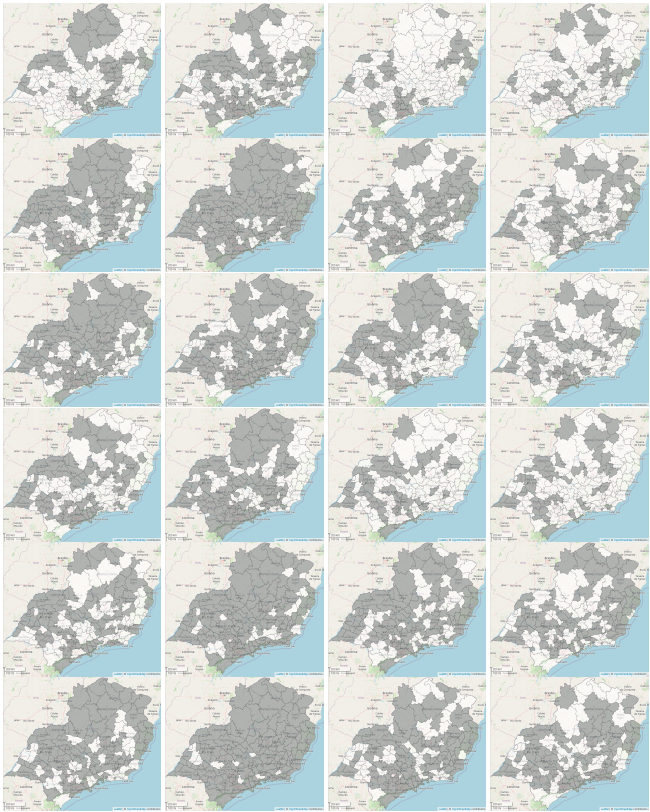}
\end{center}
\caption{Dispersion indicators by areas constructed from the posterior distribution of $\bm z$ from 2018 (top) to 2023 (bottom) by seasons - summer (1st column), autumn (2nd column), winter (3rd column), and spring (4th column). $z \approx 1$ (white); $z \neq 1$ (gray).} \label{fig:time_est_disp}
\end{figure} 

\begin{figure} \begin{center}
 \includegraphics[width=\textwidth]{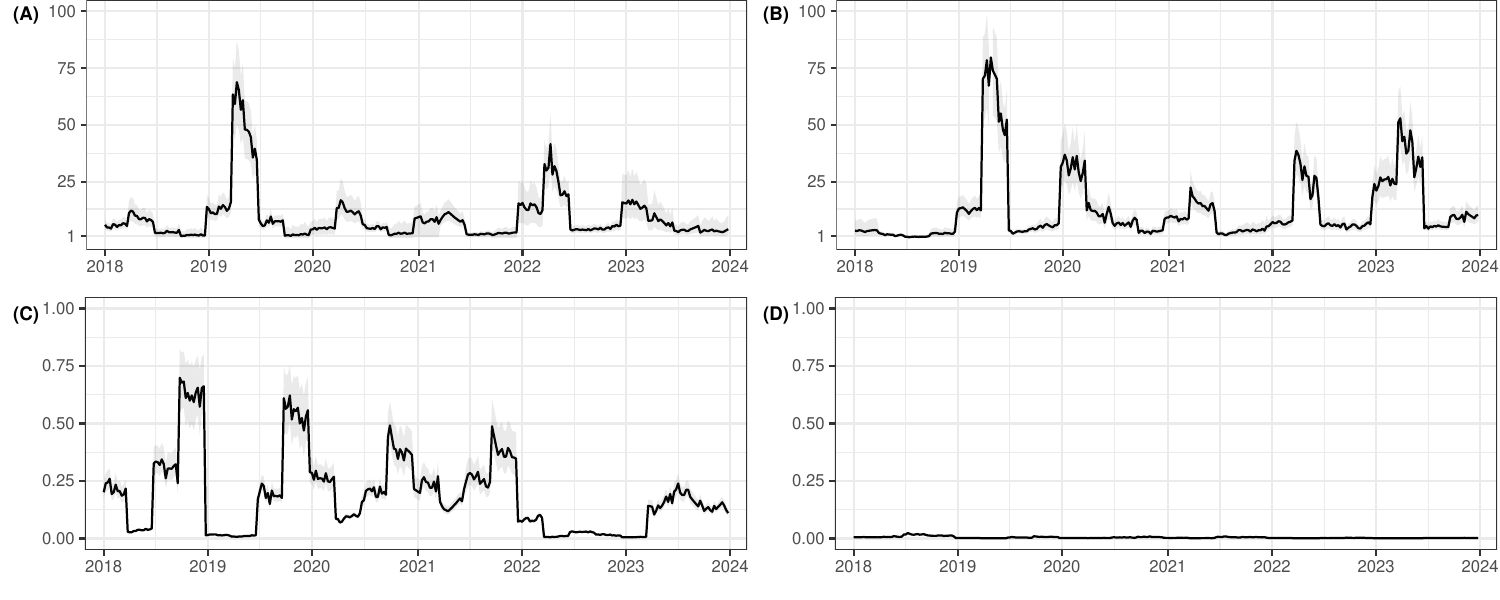} 
\end{center}
\caption{Posterior mean and 95\% credible interval of $\bm \lambda$ (A)-(B) and the ratio $\mathds{E}(Y_{it} \mid \lambda_{it}, \psi_{is}) / \mathds{V}(Y_{it} \mid \lambda_{it}, \psi_{is})$ (C)-(D) over time for two selected area: Janu{\'a}ria - MG ((A)-(C)) and Campinas - SP ((B)-(D)).} \label{fig:time_z_lambda}
\end{figure}

Another important result obtained by fitting our model is the estimated space-time dispersion. To identify which areas and seasons exhibit overdispersed data, we construct dispersion indicators from the posterior distribution of $\bm z$ and represent them using maps; see Figure~\ref{fig:time_est_disp}. Such indicators point out whether one lies within the 95\% credible interval of $z$ or not. We observed more overdispersion during periods of higher prevalence of dengue. Autumn was the season with the highest level of overdispersion, followed by summer, winter, and spring. For instance, in the autumn of 2023, nearly all areas of Minas Gerais exhibited overdispersion. In contrast, during the winter and spring of 2018, most areas displayed equidispersion. It is important to note that $z=1$ is not the only evidence of equidispersion. In the PIG distribution, we have that $\mathds{E}(Y_{it} \mid \lambda_{it}, \psi_{is}) = \mathds{V}(Y_{it} \mid \lambda_{it}, \psi_{is})$ whenever $\mathds{V}(Z_{is}) = 0$ (see Section~\ref{s:PIG}). To further analyze how dispersion changes over time, we selected two areas and calculated the ratio $\mathds{E}(Y_{it} \mid \lambda_{it}, \psi_{is}) / \mathds{V}(Y_{it} \mid \lambda_{it}, \psi_{is})$. The first area, Janu{\'a}ria, is located in the northern part of the state of Minas Gerais, and its corresponding ratio is shown in Figure~\ref{fig:time_z_lambda}(C). The second area, Campinas, a significant region in the northwest of the state of S{\~a}o Paulo, is depicted in Figure~\ref{fig:time_z_lambda}(D). In both regions, we observed that $ \mathds{V}(Y_{it} \mid \lambda_{it}, \psi_{is}) > \mathds{E}(Y_{it} \mid \lambda_{it}, \psi_{is})$, which confirms the overdispersion evident in the data. This overdispersion is particularly pronounced in Campinas. Refer to the Appendix for maps of dispersion indicators for all seasons, as well as the temporal evolution of $\mathds{E}(Y_{it} \mid \lambda_{it}, \psi_{is}) / \mathds{V}(Y_{it} \mid \lambda_{it}, \psi_{is})$ for additional areas.

\begin{figure} [!h] \centering
	\includegraphics[width=\textwidth]{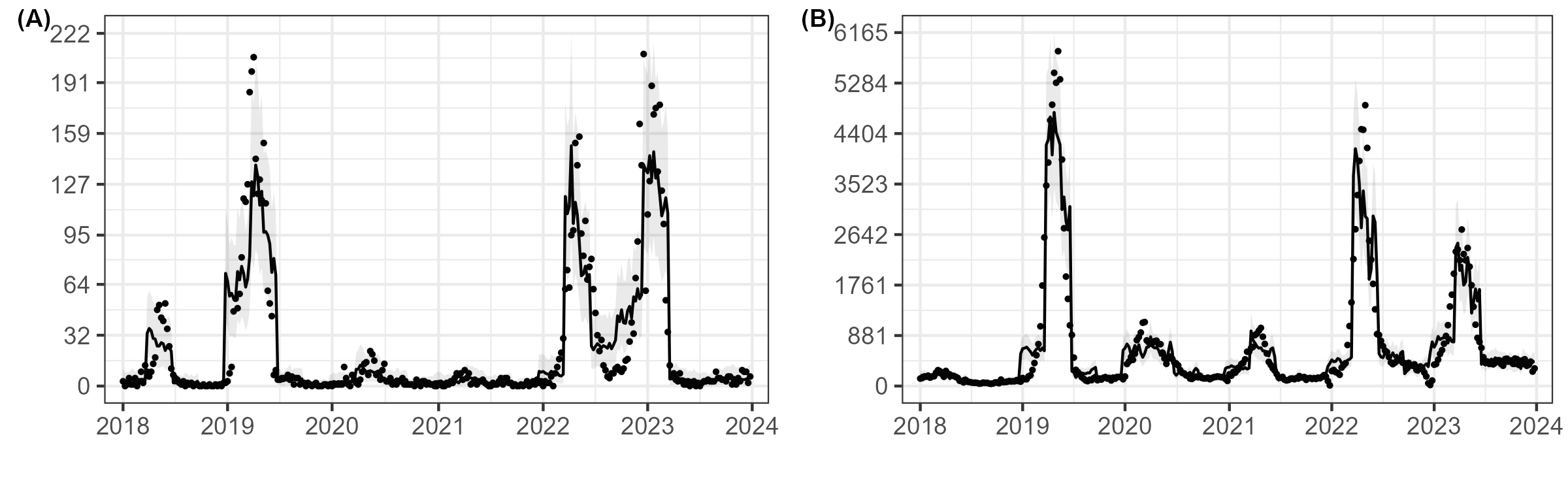}
	\caption{Observed number of cases (points) and estimated number of cases (solid lines) accompanied by its 95\% credible intervals for the two selected areas, Janu{\'a}ria - MG (A) and  Campinas - SP (B).} \label{fig:rate_plot}
\end{figure}

As presented in \eqref{eq:IG}, the PIG regression model considers two parameters. The spatial-temporal effect of $\bm z$, which accounts for heterogeneity, has already been explored in Figures~\ref{fig:time_est_disp} and \ref{fig:time_z_lambda}(C)-(D). Now, we focus on the temporal evolution of $\bm \lambda$. In this case, $\bm \lambda$ depends on temperature and humidity, both of which vary in space and time, and on HDI, which varies only in space. To explore the temporal trend of $\bm \lambda$, we selected two areas, Janu{\'a}ria and Campinas, which have the lowest (0.60) and highest (0.77) observed HDI, respectively. For both regions, $\bm \lambda$ reached its highest peak in 2019, as shown in Figure~\ref{fig:time_z_lambda}(A)-(B). Although with less intensity, annual peaks can still be observed. It is worth noting that the cluster-specific parameter $\bm \theta$ also contributes to explaining $\bm \lambda$. Janu{\'a}ria and Campinas were estimated to be in the same cluster during four seasons: summer 2019, winter 2020 and 2022, and spring 2022, as illustrated in Figure~\ref{fig:time_est_part}. A graphic representation of the product $\bm{O z \lambda}$ for these areas is shown in  Figure~\ref{fig:rate_plot}.

We conclude our analysis by evaluating the regression coefficients considered to explain both the mean and data dispersion. Temperature has a positive effect on both the mean and dispersion, with values of $0.30 [0.299, 0.302]$ and $0.30 [0.198, 0.387]$, respectively. This indicates that a warmer climate contributes to an increase in dengue cases, as well as their heterogeneity. In contrast, humidity has an opposite effect on the mean and dispersion, with values of $0.13 [0.124, 0.127]$ and $-0.47 [-0.469, -0.206]$, respectively. While higher humidity positively affects the increase in cases, it negatively impacts the dispersion; that is, the increase in cases is more homogeneous over time and space in a humid climate. The HDI was used only to explain the average number of cases, where it has a positive impact of $0.08 [0.071, 0.083]$. Appendix Figure~\ref{fig:reg_coef} displays the posterior distribution of the regression coefficients.

\section{Concluding remarks} \label{sec:conclusion}

Inspired by the scarcity of literature concerning spatial-temporal partition modeling, we have developed a Bayesian model that incorporates temporal dependence in a series of spatial random partitions. Our approach employs a product partition prior for random partitions at each time point, thereby introducing temporal correlation among partitions through the temporal structure associated with prior cohesions. Furthermore, we utilize random spanning trees to effectively explore the partition search space and ensure spatially constrained clustering. The practical applicability of our model is motivated by data related to mosquito-borne diseases. We leveraged the characteristics of this type of data to inform the specification of certain modeling features. For instance, we implemented a Poisson mixture model to address overdispersion, a prevalent issue in this field.

Through simulation studies, we initially demonstrated the necessity of employing a model capable of effectively capturing the overdispersion present in the data. As illustrated in Section~\ref{s:sim1}, the Poisson inverse-Gaussian model exhibits superior performance compared to the Poisson model. Furthermore, in Section~\ref{s:sim2}, we investigated the order of temporal dependence and its influence on clustering. Our findings indicate that incorporating a temporal structure into the partition prior generally enhances fitting performance relative to the independent version. However, the optimal order of temporal dependence is contingent upon the specific characteristics of the data. Additionally, in the Appendix Section~\ref{app:sim3}, we investigate how the model deals with non-contiguous clusters. As mentioned in Section~\ref{sec:works}, the incorporation of random spanning trees within PPM for areal clustering ensures that the resulting clusters maintain spatial contiguity, which has been proven through simulations.

Following an exploration of the characteristics of the proposed model through simulation studies, we applied the model to a dataset of dengue cases in the Southeast region of Brazil from 2018 to 2023. In accordance with the specific characteristics of this context, our model accommodates varying spatial partitions across different seasons. We incorporated climate and socio-demographic factors as risk determinants to quantify their impact on disease transmission. This study enhances understanding of the progression of dengue outbreaks in the region by identifying spatio-temporal clusters, which may assist policymakers in developing effective strategies for disease prevention and control. It is also noteworthy that, while our results from the simulation study were robust concerning the order of dependence, we observed more pronounced differences in the application of the model to real-world data.

The introduction of random spanning trees into PPM for area clustering, combined with the incorporation of a temporal dependence structure in the series of spatial random partitions, is an innovative strategy. Nonetheless, there are still multiple promising opportunities for future research. As discussed in Section~\ref{s:SPPPM}, adopting a sequence of latent variables to define temporal dependence offers benefits, particularly concerning the marginal distribution of $\rho_{s}$, which aids in partition sampling. However, alternative approaches that directly model the original probability vector could be further investigated. Additionally, while our method was developed to model count data, it can be extended to accommodate different types of data by incorporating other distribution families. Another avenue for future research is to explore the less widely studied underdispersion models, which have garnered growing interest in recent years. Finally, the structure could be leveraged more effectively for predictive inference. 

\section*{Acknowledgements}

Fernando A. Quintana gratefully acknowledges support from the Fondo Nacional de Desarrollo Cient{\'i}fico y Tecnol{\'o}gico (FONDECYT) under grant 1220017. Rosangela H. Loschi is partially supported by Conselho Nacional de Desenvolvimento Cient{\'i}fico e Tecnol{\'o}gico (CNPq) under grants 405025/2021-1 and 304268/2021-6, and Funda{\c{c}}{{\~a}}o de Amparo {\`a} Pesquisa do Estado de Minas Gerais (FAPEMIG) under grants APQ00674-24 and APQ01748-24.

\bibliographystyle{apalike}
\renewcommand{\bibname}{References}
\bibliography{Bibliography-MM-MC}

\begin{thebibliography}{}

\bibitem[Assun{\c{c}}{{\~a}}o et~al., 2006]{Assuncao2006}
Assun{\c{c}}{{\~a}}o, R.~M., Neves, M.~C., C\^{a}mara, G., and Freitas, C.~C.
  (2006).
\newblock Efficient regionalization techniques for socio‐economic
  geographical units using minimum spanning trees.
\newblock {\em International Journal of Geographical Information Science},
  20(7):797--811.

\bibitem[Barreto-Souza and Simas, 2015]{BarretoSouza2015}
Barreto-Souza, W. and Simas, A.~B. (2015).
\newblock General mixed {P}oisson regression models with varying dispersion.
\newblock {\em Statistics and Computing}, 26(6):1263--1280.

\bibitem[Barry and Hartigan, 1993]{Barry1993}
Barry, D. and Hartigan, J. (1993).
\newblock A {B}ayesian analysis for change point problems.
\newblock {\em Journal of the American Statistical Association}, 88(421):309.

\bibitem[Caron et~al., 2017]{Caron2017}
Caron, F., Neiswanger, W., Wood, F., Doucet, A., and Davy, M. (2017).
\newblock Generalized {P}{\'{o}}lya rrn for time-varying {P}itman-{Y}or
  processes.
\newblock {\em Journal of Machine Learning Research}, 18:1--32.

\bibitem[Cremaschi et~al., 2023]{Cremaschi2023}
Cremaschi, A., Cadonna, A., Guglielmi, A., and Quintana, F.~A. (2023).
\newblock A change-point random partition model for large spatio-temporal
  datasets.
\newblock {\em arXiv:2312.12396}.

\bibitem[Criscuolo et~al., 2023]{Criscuolo2023}
Criscuolo, T.~L., Assun{\c{c}}{\~a}o, R.~M., Loschi, R.~H., Meira~Jr., W., and
  Cruz-Reyes, D. (2023).
\newblock Handling categorical features with many levels using a product
  partition model.
\newblock {\em The Annals of Applied Statistics}, 17(1):786--814.

\bibitem[Dahl et~al., 2020]{Dahl2020}
Dahl, D., Johnson, D., and M{\"u}ller, P. (2020).
\newblock {\em salso: search algorithms and loss functions for {B}ayesian
  clustering}.
\newblock R package version 0.2.5,
  \url{https://cran.r-project.org/web/packages/salso/index.html}.

\bibitem[De~Iorio et~al., 2023]{DeIorio2023}
De~Iorio, M., Favaro, S., Guglielmi, A., and Ye, L. (2023).
\newblock Bayesian nonparametric mixture modeling for temporal dynamics of
  gender stereotypes.
\newblock {\em The Annals of Applied Statistics}, 17(3):2256--2278.

\bibitem[Dombowsky and Dunson, 2024]{Dombowsky2024}
Dombowsky, A. and Dunson, D.~B. (2024).
\newblock Product centered {D}irichlet processes for dependent clustering.
\newblock {\em arXiv:2312.05365v2}.

\bibitem[Duan and Dunson, 2023]{Duan2023}
Duan, L.~L. and Dunson, D.~B. (2023).
\newblock Bayesian spanning tree: estimating the backbone of the dependence
  graph.
\newblock {\em Journal of Machine Learning Research}, 24(397):1--44.

\bibitem[Franklinos et~al., 2019]{Franklinos2019}
Franklinos, L.~H., Jones, K.~E., Redding, D.~W., and Abubakar, I. (2019).
\newblock The effect of global change on mosquito-borne disease.
\newblock {\em The Lancet Infectious Diseases}, 19(9):e302--e312.

\bibitem[Franzolini et~al., 2024]{Franzolini2023}
Franzolini, B., De~Iorio, M., and Eriksson, J. (2024).
\newblock Conditional partial exchangeability: a probabilistic framework for
  multi-view clustering.
\newblock {\em arXiv:2307.01152v1}.

\bibitem[Gelman et~al., 2014]{Gelman2014}
Gelman, A., Hwang, J., and Vehtari, A. (2014).
\newblock Understanding predictive information criteria for {B}ayesian models.
\newblock {\em Statistics and Computing}, 24:997--1016.

\bibitem[Giampino et~al., 2024]{Giampino2024}
Giampino, A., Guindani, M., Nipoti, B., and Vannucci, M. (2024).
\newblock Local level dynamic random partition models for changepoint
  detection.
\newblock {\em arXiv:2407.20085v1}.

\bibitem[Guti{\'{e}}rrez et~al., 2016]{Gutierrez2016}
Guti{\'{e}}rrez, L., Mena, R.~H., and Ruggiero, M. (2016).
\newblock A time dependent {B}ayesian nonparametric model for air quality
  analysis.
\newblock {\em Computational Statistics {\&} Data Analysis}, 95:161--175.

\bibitem[Hartigan, 1990]{Hartigan1990}
Hartigan, J.~A. (1990).
\newblock Partition models.
\newblock {\em Communications in Statistics - Theory and Methods},
  19(8):2745--2756.

\bibitem[Hegarty and Barry, 2008]{Hegarty2008}
Hegarty, A. and Barry, D. (2008).
\newblock Bayesian disease mapping using product partition models.
\newblock {\em Statistics in Medicine}, 27(19):3868--3893.

\bibitem[Hilbe, 2014]{Hilbe2014}
Hilbe, J.~M. (2014).
\newblock {\em Modeling Count Data}.
\newblock Cambridge University Press, Cambridge, UK.

\bibitem[Hubert and Arabie, 1985]{Hubert1985}
Hubert, L. and Arabie, P. (1985).
\newblock Comparing partitions.
\newblock {\em Journal of Classification}, 2:193--218.

\bibitem[Jara et~al., 2013]{Jara2013}
Jara, A., Nieto-Barajas, L., and Quintana, F.~A. (2013).
\newblock A time series model for responses on the unit interval.
\newblock {\em Bayesian Analysis}, 8(3):723--740.

\bibitem[Jo et~al., 2017]{Jo2017}
Jo, S., Lee, J., M{\"u}ller, P., Quintana, F.~A., and Trippa, L. (2017).
\newblock Dependent species sampling models for spatial density estimation.
\newblock {\em Bayesian Analysis}, 12(2):379--406.

\bibitem[Jungnickel, 2013]{Jungnickel2013}
Jungnickel, D. (2013).
\newblock {\em Spanning trees}, pages 103--134.
\newblock Springer Berlin Heidelberg.

\bibitem[Luo et~al., 2021]{Luo2021}
Luo, Z.~T., Sang, H., and Mallick, B. (2021).
\newblock A {B}ayesian contiguous partitioning method for learning clustered
  latent variables.
\newblock {\em Journal of Machine Learning Research}, 22(37):1--52.

\bibitem[Luo et~al., 2024]{Luo2023}
Luo, Z.~T., Sang, H., and Mallick, B. (2024).
\newblock A nonstationary soft partitioned {G}aussian process model via random
  spanning trees.
\newblock {\em Journal of the American Statistical Association},
  119(547):2105--2116.

\bibitem[Marinho et~al., 2016]{Marinho2016}
Marinho, R.~A., Beserra, E.~B., Bezerra-Gusm{\~a}o, M.~A., Porto, V. d.~S.,
  Olinda, R.~A., and dos Santos, C. A.~C. (2016).
\newblock Effects of temperature on the life cycle, expansion, and dispersion
  of {A}edes aegypti (diptera: Culicidae) in three cities in {P}araiba,
  {B}razil.
\newblock {\em Journal of Vector Ecology}, 41(1):1--10.

\bibitem[Moraga, 2019]{Moraga2019}
Moraga, P. (2019).
\newblock {\em Geospatial health data: modeling and visualization with {R-INLA}
  and {S}hiny}.
\newblock Chapman \& Hall/CRC Biostatistics Series, Boca Raton, US.

\bibitem[Napier et~al., 2018]{Napier2018}
Napier, G., Lee, D., Robertson, C., and Lawson, A. (2018).
\newblock A {B}ayesian space–time model for clustering areal units based on
  their disease trends.
\newblock {\em Biostatistics}, 20(4):681--697.

\bibitem[Page and Quintana, 2016]{Page2016}
Page, G.~L. and Quintana, F.~A. (2016).
\newblock Spatial product partition models.
\newblock {\em Bayesian Analysis}, 11(1):265--298.

\bibitem[Page et~al., 2022]{Page2022}
Page, G.~L., Quintana, F.~A., and Dahl, D.~B. (2022).
\newblock Dependent modeling of temporal sequences of random partitions.
\newblock {\em Journal of Computational and Graphical Statistics},
  31(2):614--627.

\bibitem[Pavani and Quintana, 2025]{Pavani2025}
Pavani, J. and Quintana, F.~A. (2025).
\newblock A {B}ayesian multivariate model with temporal dependence on random
  rartition of areal data for mosquito-borne diseases.
\newblock {\em Statistics in Medicine}, 44(3-4):e10325.

\bibitem[Perrakis et~al., 2015]{Perrakis2015}
Perrakis, K., Karlis, D., Cools, M., and Janssens, D. (2015).
\newblock Bayesian inference for transportation origin–destination matrices:
  the {P}oisson–inverse {G}aussian and other {P}oisson mixtures.
\newblock {\em Journal of the Royal Statistical Society. Series A (Statistics
  in Society)}, 178(1):271--296.

\bibitem[Quintana et~al., 2018]{Quintana2018}
Quintana, F.~A., Loschi, R.~H., and Page, G.~L. (2018).
\newblock Bayesian product partition models.
\newblock {\em Wiley StatsRef: Statistics Reference Online}, pages 1--15.

\bibitem[Saraiva et~al., 2022]{Saraiva2022}
Saraiva, E.~F., Vigas, V.~P., Flesch, M.~V., Gannon, M., and
  de~Bragan{\c{c}}a~Pereira, C.~A. (2022).
\newblock Modeling overdispersed dengue data via {P}oisson inverse {G}aussian
  regression model: a case study in the city of {C}ampo {G}rande, {MS},
  {B}razil.
\newblock {\em Entropy}, 24(9):1256.

\bibitem[Tam et~al., 2025]{Tam2025}
Tam, E., Dunson, D.~B., and Duan, L.~L. (2025).
\newblock Exact sampling of spanning trees via fast-forwarded random walks.
\newblock {\em Biometrika}, asaf031.

\bibitem[Teixeira et~al., 2015]{Teixeira2015}
Teixeira, L.~V., Assun{\c{c}}{\~a}o, R.~M., and Loschi, R.~H. (2015).
\newblock A generative spatial clustering model for random data through
  spanning trees.
\newblock {\em IEEE International Conference on Data Mining}, pages 997--1002.

\bibitem[Teixeira et~al., 2019]{Teixeira2019}
Teixeira, L.~V., Assun{\c{c}}{\~a}o, R.~M., and Loschi, R.~H. (2019).
\newblock Bayesian space-time partitioning by sampling and pruning spanning
  trees.
\newblock {\em Journal of Machine Learning Research}, 20(85):1--35.

\bibitem[van Dyk and Park, 2008]{Dyk2008}
van Dyk, D.~A. and Park, T. (2008).
\newblock Partially collapsed {G}ibbs {S}amplers: theory and methods.
\newblock {\em Journal of the American Statistical Association},
  103(482):790--796.

\bibitem[Zhong et~al., 2024]{Zhong2024}
Zhong, R., Chac{\'o}n-Montalv{\'a}n, E., and Moraga, P. (2024).
\newblock Bayesian spatial functional data clustering: applications in disease
  surveillance.
\newblock {\em arXiv:2407.12633v1}.

\end{thebibliography}

\appendix
\section*{Appendix}

\section{Data description} \label{app:data}

\begin{figure} [!h] \centering
	\includegraphics[width=.45\textwidth]{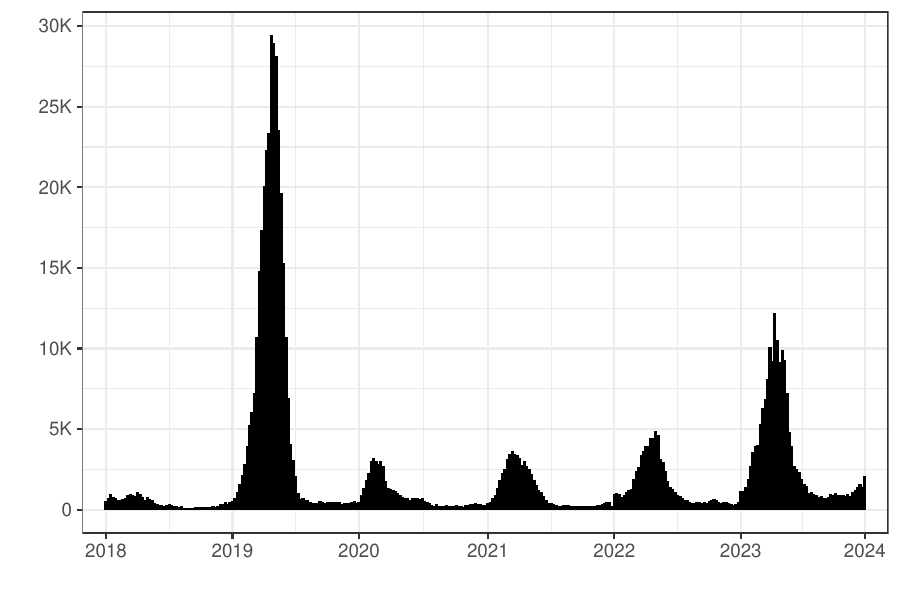}
	\vspace{-0.2cm} \caption{Total number of dengue cases reported per epidemiological week during the years 2018-2023 in the Brazilian Southeast region.} \label{fig:time_response}
\end{figure}

As stated in Section~\ref{sec:motivation} of the original document, this study was inspired by the analysis of data on tropical diseases. In particular, we focused on examining the incidence of dengue in the 145 microregions of the Brazilian Southeast region. Data were collected on a weekly basis from 2018 to 2023, covering a total of 313 epidemiological weeks. Figure~\ref{fig:time_response} illustrates the time series of dengue cases over the weeks. In total, 5,309,984 cases were reported, reaching their peak in 2019.

\begin{figure} [!t] \centering
  \includegraphics[width=\textwidth]{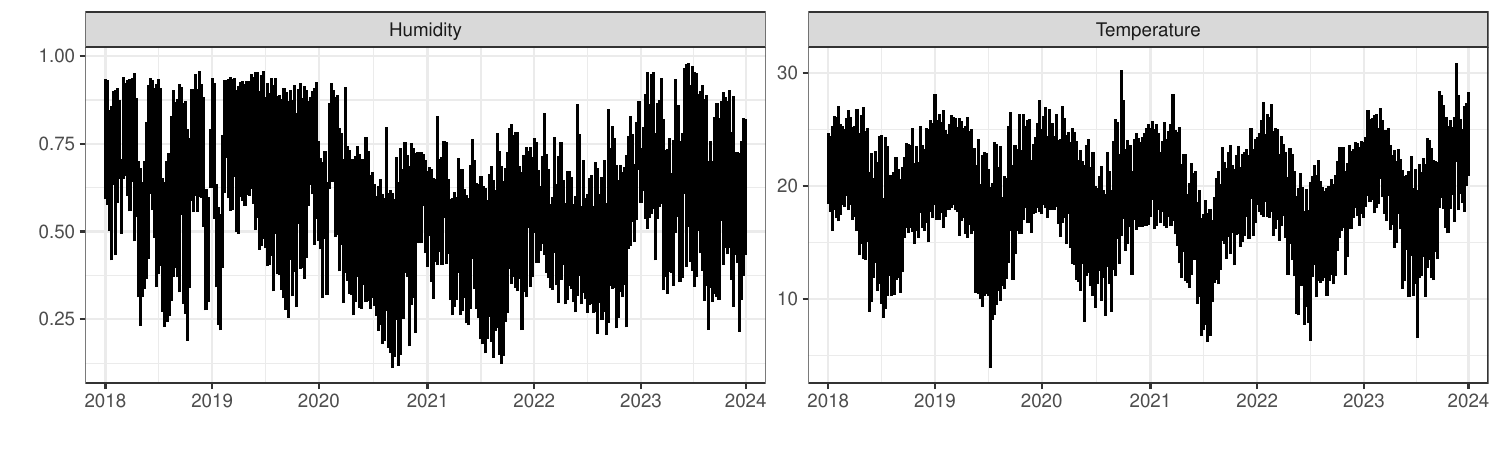}
\caption{ Minimum and maximum humidity (percentage) and temperature (Celsius degree) reported weekly in the Brazilian Southeast region during the years 2018-2023. } \label{fig:time_cov}
\end{figure}

Figure~\ref{fig:time_cov} shows the minimum and maximum values of humidity (percentage) and temperature (Celsius degrees) reported weekly during the years 2018-2023 in the Brazilian Southeast region. Maximum humidity values remained very close to or equal to 100\% over time. Conversely, minimum values varied greatly, in addition to presenting a seasonal pattern. In general, winter was the driest season, reaching the most extreme values in 2020. At this point, it is important to highlight that, although Brazil is located in both hemispheres (93\% of the territory in the Southern Hemisphere and 7\% in the Northern Hemisphere), the region under study is in the Southern Hemisphere. Thus, summer occurs from December to March, followed by autumn from March to June, winter from June to September, and finally, spring from September to December.

\begin{figure}[!t] \centering
\includegraphics[width=.7\textwidth]{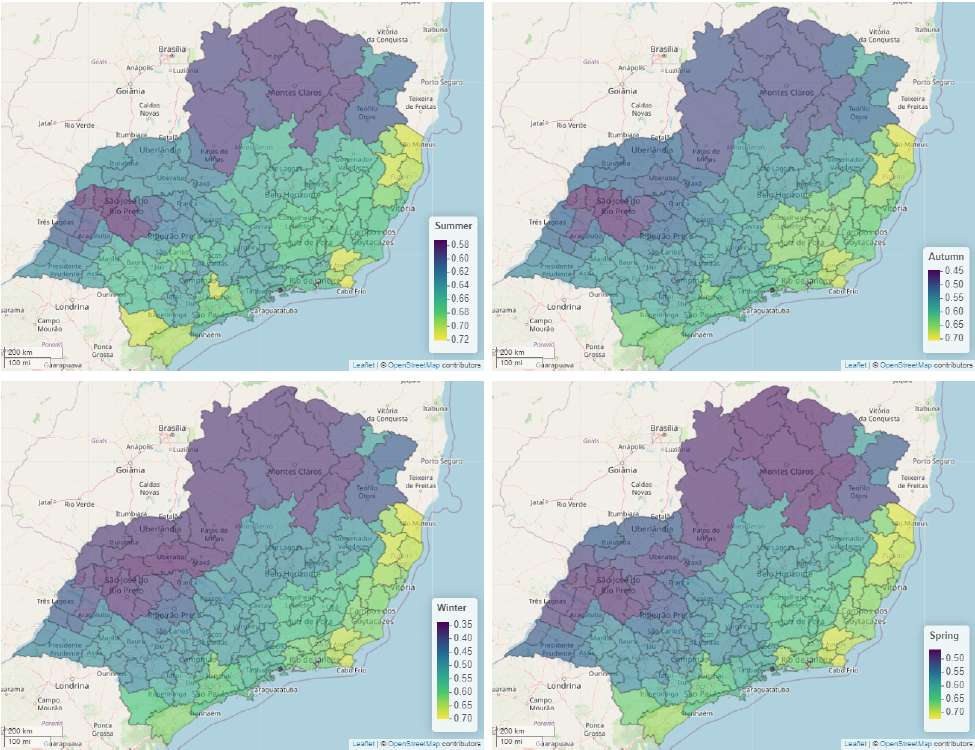}
\caption{ Spatial distribution of minimum humidity over the 145 microregions in the Brazilian Southeast in each season. Average values were obtained by considering the years 2018-2023. } \label{fig:sp_humidity}
\end{figure}

\begin{figure}[!t] \centering
\includegraphics[width=.7\textwidth]{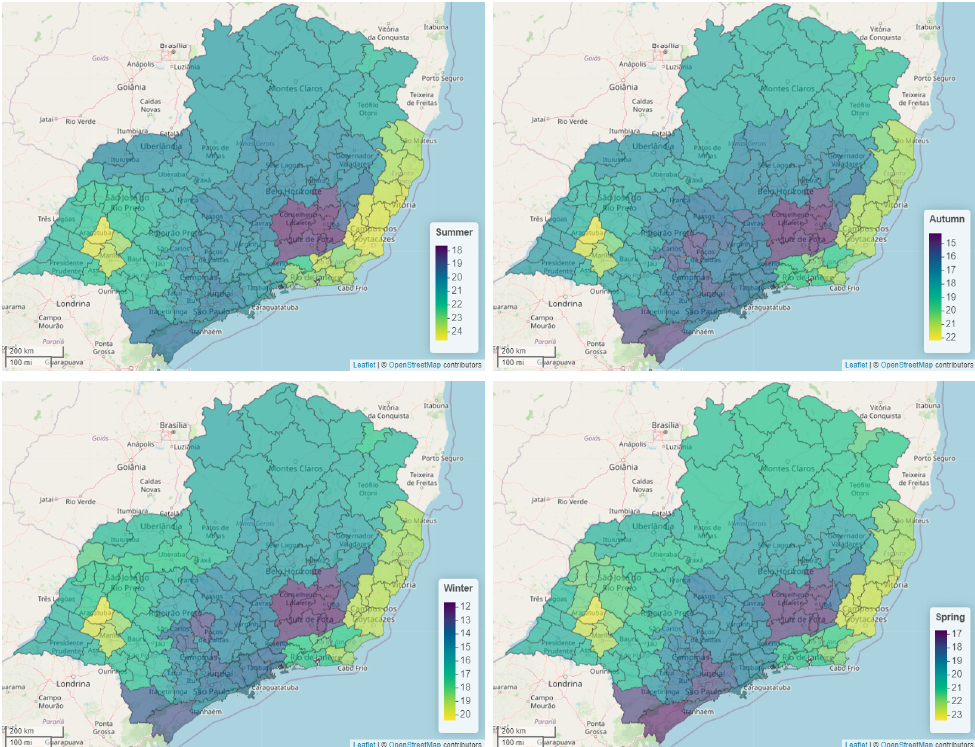}
\caption{ Spatial distribution of minimum temperature across the 145 microregions in the Brazilian Southeast during each season. Average values were obtained by considering the years 2018-2023. } \label{fig:sp_temperature}
\end{figure}

Considering the seasonal pattern, we mapped the distribution of minimum humidity over the 145 microregions in the Brazilian Southeast in each season, as shown in Figure~\ref{fig:sp_humidity}. Areas close to the coast exhibited a wet climate throughout all seasons. The humidity was even more pronounced in the state of Esp{\'i}rito Santo and to the north of Rio de Janeiro. The opposite was observed in the Northwest of S{\~a}o Paulo, where low humidity was noted throughout the year. The North and West of the state of Minas Gerais were also arid regions. Regarding temperature, there was significant temporal variation for both minimum and maximum values, as demonstrated in Figure~\ref{fig:time_cov}. Additionally, a seasonal pattern was clearly observed. As naturally expected, the lowest temperatures occurred in winter, especially in July 2019 when temperatures below 5$^{\circ}$C were recorded. The highest temperatures were recorded in the last week of September and the first week of October 2020, reaching nearly 40$^{\circ}$C. The spatial distribution of minimum temperature over the 145 microregions in the Brazilian Southeast in each season is depicted in Figure~\ref{fig:sp_temperature}. The state of Esp{\'i}rito Santo and Northwestern S{\~a}o Paulo are areas where the highest temperatures were recorded, which remained consistent across all seasons.

\noindent On the other hand, the lowest temperatures were recorded in southeastern Minas Gerais and southern S{\~a}o Paulo. As shown in Figures~\ref{fig:time_cov}, \ref{fig:sp_humidity}, and \ref{fig:sp_temperature}, humidity and temperature were not homogeneous across the region, or even within each state. They varied greatly over time, with the exception of maximum humidity, which remained almost constant. To account for their variability over time, we used minimum temperature and minimum humidity as spatio-temporal covariates in this study.

\begin{figure} [!t] \centering
  \includegraphics[width=.45\textwidth]{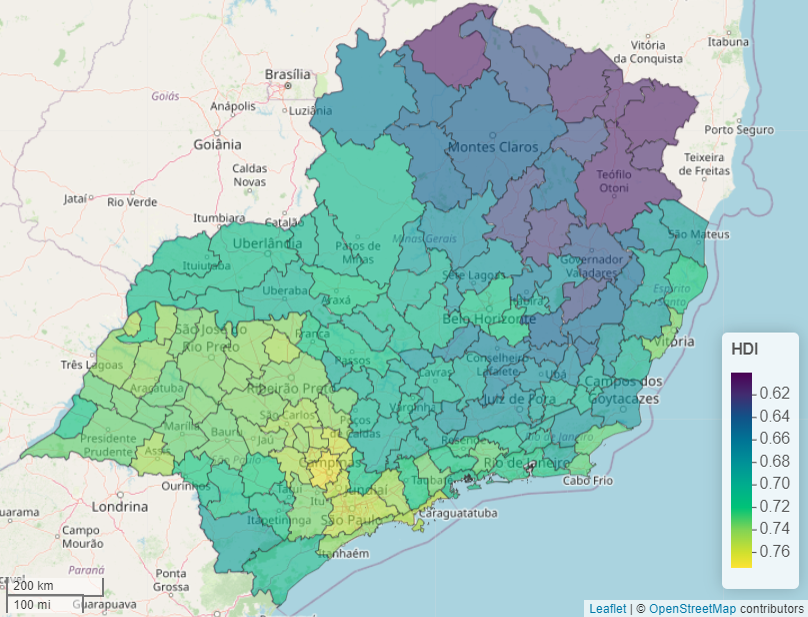}
\caption{ Spatial distribution of the Human Development Index across the 145 microregions in the Brazilian Southeast region. Information is based on the 2010 Demographic Census. } \label{fig:HDI}
\end{figure}

Finally, Figure~\ref{fig:HDI} illustrates the spatial distribution of HDI in the study region, showing a clear pattern. Areas in the northern part of the region, particularly those in Minas Gerais bordering Bahia, have lower HDI values. Conversely, higher HDI values are concentrated in the southern part of the region, particularly in the state of S{\~a}o Paulo. It is worth mentioning that the Southeast region is the most developed in Brazil, exhibiting a smooth variation in HDI values.

\section{Key concepts} \label{app:back}

The proposed model is built assuming that our map containing $n$ areas is represented by an undirected graph. In this section, we provide some preliminary concepts that are assumed throughout the model formulation in the main manuscript.

\subsection{Spanning tree} \label{ss:st}

To improve comprehension of spanning trees (ST), let us review some key concepts from graph theory. A graph is an ordered pair $G = (V, E)$ comprising a set $V$ of vertices or nodes connected by edges in set $E$. An edge, $e = (v_{i}, v_{j})$, connecting vertices $v_{i}$ and $v_{j}$ indicates that those vertices are adjacent to each other or, in a spatial context, that they are neighbors. A graph is said to be connected if, for any pair of nodes $v_i$ and $v_j$, there is at least one path connecting them. It may also contain circuits, i.e., paths that start and end at the same vertex. Graphs can be directed or undirected, depending on whether there is an order among the vertices. Figure~\ref{graph}, left and center, respectively, illustrates these concepts.
\begin{figure} [hbt!] \centering
\subfloat{\begin{tikzpicture}[auto,vertex/.style={draw,circle}]

    \node[vertex] (a) {v$_{1}$};
    \node[vertex, below right=1cm and 1.3cm of a] (b) {v$_{2}$};
    \node[vertex, below left=1cm and 1.3cm of a] (c) {v$_{5}$};
    \node[vertex, below left=1cm and 0.3cm of b] (d) {v$_{3}$};
    \node[vertex, below right=1cm and 0.3cm of c] (e) {v$_{4}$};
    
    \path[-{Stealth[]}]
      (a) edge (b) 
      (a) edge (d)
      (b) edge (d)
      (b) edge (c) 
      (c) edge (d)
      (c) edge (e)
      (d) edge (e);
      
\end{tikzpicture}} \hspace{1cm} 
\subfloat{\begin{tikzpicture}[auto,vertex/.style={draw,circle}]

    \node[vertex] (a) {v$_{1}$};
    \node[vertex, below right=1cm and 1.3cm of a] (b) {v$_{2}$};
    \node[vertex, below left=1cm and 1.3cm of a] (c) {v$_{5}$};
    \node[vertex, below left=1cm and 0.3cm of b] (d) {v$_{3}$};
    \node[vertex, below right=1cm and 0.3cm of c] (e) {v$_{4}$};
    
    \path[-]
      (a) edge (b) 
      (a) edge (d)
      (b) edge (d)
      (b) edge (c) 
      (c) edge (d)
      (c) edge (e)
      (d) edge (e);
      
\end{tikzpicture}} \hspace{1cm} 
\subfloat{\begin{tikzpicture}[auto,vertex/.style={draw,circle}]

    \node[vertex] (a) {v$_{1}$};
    \node[vertex, below right=1cm and 1.3cm of a] (b) {v$_{2}$};
    \node[vertex, below left=1cm and 1.3cm of a] (c) {v$_{5}$};
    \node[vertex, below left=1cm and 0.3cm of b] (d) {v$_{3}$};
    \node[vertex, below right=1cm and 0.3cm of c] (e) {v$_{4}$};
    
    \path[-]
      (a) edge (d)
      (b) edge (d)
      (c) edge (e)
      (d) edge (e);
      
\end{tikzpicture}}
\caption{Different graph types: directed (left), undirected (center), tree (right).} \label{graph}
\end{figure}

A {\it spanning tree} of an undirected connected graph is a subgraph that includes all the nodes of $G$ but only some of its edges, as shown in Figure~\ref{graph} (right). In other words, in the spanning tree, any two nodes in the graph are connected by a unique path. Thus, a tree with $n$ vertices has exactly $n-1$ edges. Many trees can be generated from the same graph; however, in this study, we consider the minimum spanning tree, which is a tree with the minimum possible total edge weight.

A partition of the map is induced by removing some edges from the spanning tree such that the vertices connected by the remaining edges form clusters. Figure~\ref{part} shows a spanning tree and two different partitions induced by removing one and two edges, at the center and left, respectively. It is important to note that there is a significant advantage to clustering from a spanning tree instead of from the original graphs, as the search space of partitions is reduced. Figure~\ref{graph} illustrates this. While the vertices in the undirected graph are connected by seven edges, the tree has only four edges, as shown in Figure~\ref{graph} (center and right, respectively).
\begin{figure} [hbt!]\centering
\subfloat{\begin{tikzpicture}[auto,vertex/.style={draw,circle}]

    \node[vertex] (a) {v$_{1}$};
    \node[vertex, below right=1cm and 1.3cm of a] (b) {v$_{2}$};
    \node[vertex, below left=1cm and 1.3cm of a] (c) {v$_{5}$};
    \node[vertex, below left=1cm and 0.3cm of b] (d) {v$_{3}$};
    \node[vertex, below right=1cm and 0.3cm of c] (e) {v$_{4}$};
    
    \path[-]
      (a) edge (d)
      (b) edge (d)
      (c) edge (e)
      (d) edge (e);
      
\end{tikzpicture}} \hspace{1cm}
\subfloat{\begin{tikzpicture}[auto,vertex/.style={draw,circle}]

    \node[vertex] (a) {v$_{1}$};
    \node[vertex, below right=1cm and 1.3cm of a] (b) {v$_{2}$};
    \node[vertex, below left=1cm and 1.3cm of a] (c) {v$_{5}$};
    \node[vertex, below left=1cm and 0.3cm of b] (d) {v$_{3}$};
    \node[vertex, below right=1cm and 0.3cm of c] (e) {v$_{4}$};
    
    \path[-]
      (a) edge (d)
      (b) edge (d)
      (c) edge (e);
      
\end{tikzpicture}} \hspace{1cm} 
\subfloat{\begin{tikzpicture}[auto,vertex/.style={draw,circle}]

    \node[vertex] (a) {v$_{1}$};
    \node[vertex, below right=1cm and 1.3cm of a] (b) {v$_{2}$};
    \node[vertex, below left=1cm and 1.3cm of a] (c) {v$_{5}$};
    \node[vertex, below left=1cm and 0.3cm of b] (d) {v$_{3}$};
    \node[vertex, below right=1cm and 0.3cm of c] (e) {v$_{4}$};
    
    \path[-]
      (b) edge (d)
      (c) edge (e);
      
\end{tikzpicture}}
\caption{Spanning tree (left) and the two (center) and three clusters (right)  partitions generated from it.} \label{part}
\end{figure}

\subsection{On the prior for the number of clusters} \label{ss:ncl_prior}

It is not difficult to see from the proposed model (see Eq.\eqref{eq:partition_prior}--\eqref{eq:rho_prior} of the main manuscript) that the prior distribution of $\rho_{s}$ can induce higher or lower levels of informativity for the prior distribution of ${\bm \pi}_{s}$ based on prior knowledge about the number of clusters. Therefore, it would be more relevant to obtain prior marginal distributions for partition, the probability of removing edges, and the number of clusters. This will help in the prior elicitation of parameters. To do so, let $N_{\mathcal{T}_{s}}$ be the total number of spanning trees associated with a certain graph, $N_{\mathcal{T}_{s}}({\bm \pi}_{s})$ be the total number of spanning trees compatible with the partition ${\bm \pi}_{s}$, and $K_{s}$ be the number of clusters in season $s$. Consider the prior distribution $\mathds{P}[{\bm \pi}_{s} \mid \mathcal{T}_{s}, \rho_{s}]$ defined in Eq.\eqref{eq:partition_prior} of the main manuscript. Then, we have:
\allowdisplaybreaks \begin{align*}
    \mathds{P}[{\bm \pi}_{s} \mid \rho_{s}, u_{s}, \ldots, u_{s-q}] & = \sum_{\mathcal{T}_{s}} \mathds{P}[{\bm \pi}_{s} \mid \mathcal{T}_{s}, \rho_{s}] \; \mathds{P}[\mathcal{T}_{s}] \\
    & = \sum_{\mathcal{T}_{s}} \rho_{s}^{k_{s} - 1} (1 - \rho_{s})^{n - k_{s}} \; \mathds{P}[\mathcal{T}_{s}] \; \mathds{1}_{[{\bm \pi}_{s} \prec \mathcal{T}_{s}]} \\
    & = \rho_{s}^{k_{s} - 1} (1 - \rho_{s})^{n - k_{s}} \frac{N_{\mathcal{T}_{s}}({\bm \pi}_{s})}{N_{\mathcal{T}_{s}}} 
\end{align*}
and 
\allowdisplaybreaks \begin{align*}
\mathds{P}[K_{s} = k_{s} \mid \rho_{s}] & = \sum_{{\bm \pi}_{s}} \mathds{P}[{\bm \pi}_{s} \mid \rho_{s}, u_{s}, \ldots, u_{s-q}] \mathds{1}_{[{\bm \pi}_{s}, k_{s}]} \\
    & = \binom{n - 1}{k_{s} - 1} \rho_{s}^{k_{s} - 1} (1 - \rho_{s})^{n - k_{s}} \frac{N_{\mathcal{T}_{s}}({\bm \pi}_{s})}{N_{\mathcal{T}_{s}}},
\end{align*}
where $\mathds{1}_{[{\bm \pi}_{s}, k_{s}]}$ indicates partitions ${\bm \pi}_{s}$ formed by $k_{s}$ clusters \citep{Teixeira2019}.

Considering the prior distribution $\mathds{P}[\rho_{s} \mid u_{s}, \ldots, u_{s-q}]$ as shown in Eq.\eqref{eq:rho_prior} of the main manuscript, we know from \cite{Jara2013} that $\rho_{s}$ is marginally distributed as Beta$(\upsilon, \kappa)$. To see this result, note that from Eq.\eqref{eq:u_prior} of the main manuscript, it follows that:
\begin{equation*}
    \sum\limits_{l = 0}^{q} u_{s-l} \mid w \sim \; \text{Bin} \left( \sum\limits_{l = 0}^{q} c_{s-l}, w \right),  
\end{equation*}
thus, we have that its marginal distribution is given by:
\allowdisplaybreaks \begin{align*}
   \mathds{P} \left[ \sum\limits_{l = 0}^{q} u_{s-l} \right] & =  \int_{0}^{1} \mathds{P}[u_{s}, \ldots, u_{s-q} \mid w] \; \mathds{P}[w] \; \partial w \\
   & = \binom{\sum\limits_{l = 0}^{q} c_{s-l}}{\sum\limits_{l = 0}^{q} u_{s-l}} \frac{\Gamma(\upsilon + \kappa)}{\Gamma(\upsilon) \Gamma(\kappa)} \int_{0}^{1} w^{\sum\limits_{l = 0}^{q} u_{s-l} + \upsilon - 1} (1 - w)^{\sum\limits_{l = 0}^{q} (c_{s-l} - u_{s-l}) + \kappa - 1} \\
   & = \binom{\sum\limits_{l = 0}^{q} c_{s-l}}{\sum\limits_{l = 0}^{q} u_{s-l}} \frac{\Gamma(\upsilon + \kappa)}{\Gamma(\upsilon) \Gamma(\kappa)} \frac{\Gamma \left( \upsilon + \sum\limits_{l = 0}^{q} u_{s-l} \right) \Gamma \left( \kappa + \sum\limits_{l = 0}^{q} (c_{s-l} - u_{s-l}) \right)}{\Gamma \left( \upsilon + \kappa + \sum\limits_{l = 0}^{q} c_{s-l} \right)},
\end{align*}
which is the density of the BeBin$\left( \upsilon, \kappa, \sum\limits_{l = 0}^{q} c_{s-l} \right)$. Then, we obtain the marginal distribution of $\rho_{s}$ by doing:
\allowdisplaybreaks \begin{align*}
\mathds{P}[\rho_{s}] & = \sum_{u} \mathds{P}[\rho_{s} \mid u_{s}, \ldots, u_{s-q}] \; \mathds{P} \left[ \sum\limits_{l = 0}^{q} u_{s-l} \right] \\
& = \sum_{u} \frac{\Gamma \left( \upsilon + \kappa + \sum\limits_{l = 0}^{q} c_{s-l} \right)}{\Gamma \left( \upsilon + \sum\limits_{l = 0}^{q} u_{s-l} \right) \Gamma \left( \kappa + \sum\limits_{l = 0}^{q} (c_{s-l} - u_{s-l}) \right)} \; \rho_{s}^{\upsilon + \sum\limits_{l = 0}^{q} u_{s-l} - 1} (1 - \rho_{s})^{\kappa + \sum\limits_{l = 0}^{q} (c_{s-l} - u_{s-l}) - 1} \\
& \times \binom{\sum\limits_{l = 0}^{q} c_{s-l}}{\sum\limits_{l = 0}^{q} u_{s-l}} \frac{\Gamma(\upsilon + \kappa)}{\Gamma(\upsilon) \Gamma(\kappa)} \frac{\Gamma \left( \upsilon + \sum\limits_{l = 0}^{q} u_{s-l} \right) \Gamma \left( \kappa + \sum\limits_{l = 0}^{q} (c_{s-l} - u_{s-l}) \right)}{\Gamma \left( \upsilon + \kappa + \sum\limits_{l = 0}^{q} c_{s-l} \right)} \\
& = \frac{\Gamma(\upsilon + \kappa)}{\Gamma(\upsilon) \Gamma(\kappa)} \rho_{s}^{\upsilon - 1} (1 - \rho_{s})^{\kappa - 1} \sum_{u} \binom{\sum\limits_{l = 0}^{q} c_{s-l}}{\sum\limits_{l = 0}^{q} u_{s-l}} \rho_{s}^{\sum\limits_{l = 0}^{q} u_{s-l}} (1 - \rho_{s})^{\sum\limits_{l = 0}^{q} (c_{s-l} - u_{s-l})} \\
& = \frac{\Gamma(\upsilon + \kappa)}{\Gamma(\upsilon) \Gamma(\kappa)} \rho_{s}^{\upsilon - 1} (1 - \rho_{s})^{\kappa - 1},
\end{align*}
which concludes the proof that $\rho_{s}$ is marginally distributed as Beta$(\upsilon, \kappa)$. This result is crucial for determining the marginal distributions of both the partition ${\bm \pi}_{s}$ and the number of clusters $k_{s}$. After taking everything discussed into account, we finally have:
\allowdisplaybreaks \begin{align}
\mathds{P}[{\bm \pi}_{s}] & = \int_{0}^{1} \mathds{P}[{\bm \pi}_{s} \mid \rho_{s}] \; \mathds{P}[\rho_{s}] \; \partial \rho_{s} = \int_{0}^{1} \rho_{s}^{k_{s} - 1} (1 - \rho_{s})^{n - k_{s}} \frac{N_{\mathcal{T}_{s}}({\bm \pi}_{s})}{N_{\mathcal{T}_{s}}} \; \frac{\Gamma(\upsilon + \kappa)}{\Gamma(\upsilon) \Gamma(\kappa)} \rho_{s}^{\upsilon - 1} (1 - \rho_{s})^{\kappa - 1} \; \partial \rho_{s} \nonumber \\
& = \frac{N_{\mathcal{T}_{s}}({\bm \pi}_{s})}{N_{\mathcal{T}_{s}}} \; \frac{\Gamma(\upsilon + \kappa)}{\Gamma(\upsilon) \Gamma(\kappa)} \; \frac{\Gamma(k_{s} + \upsilon - 1) \Gamma(n - k_{s} + \kappa)}{\Gamma(\upsilon - 1 + n + \kappa)} \nonumber \\
\text{and} \hspace{2.2cm} & \nonumber \\
\mathds{P}[K_{s} = k_{s}] & = \int_{0}^{1} \mathds{P}[K_{s} = k_{s} \mid \rho_{s}] \; \mathds{P}[\rho_{s}] \; \partial \rho_{s} \nonumber \\
& = \int_{0}^{1} \binom{n - 1}{k_{s} - 1} \rho_{s}^{k_{s} - 1} (1 - \rho_{s})^{n - k_{s}} \frac{N_{\mathcal{T}_{s}}({\bm \pi}_{s})}{N_{\mathcal{T}_{s}}} \; \frac{\Gamma(\upsilon + \kappa)}{\Gamma(\upsilon) \Gamma(\kappa)} \rho_{s}^{\upsilon - 1} (1 - \rho_{s})^{\kappa - 1} \; \partial \rho_{s} \nonumber \\
& = \binom{n - 1}{k_{s} - 1} \frac{N_{\mathcal{T}_{s}}({\bm \pi}_{s})}{N_{\mathcal{T}_{s}}} \; \frac{\Gamma(\upsilon + \kappa)}{\Gamma(\upsilon) \Gamma(\kappa)} \; \frac{\Gamma(k_{s} + \upsilon - 1) \Gamma(n - k_{s} + \kappa)}{\Gamma(\upsilon - 1 + n + \kappa)}. \label{marg_ncl}
\end{align}

Although the prior distribution of $k_{s}$ depends on the graph topology through $\frac{N_{\mathcal{T}_{s}}({\bm \pi}_{s})}{N_{\mathcal{T}_{s}}}$, given a tree compatible with the partition, the expected number of clusters {\it a priori} is not dependent. Furthermore, note that Eq.\eqref{marg_ncl} corresponds to the beta-binomial density, whose parameters are the same as those used in the marginal distribution of $\rho_{s}$, in addition to the number of edges in the tree, i.e., BeBin$(k_{s} - 1; n - 1, \upsilon, \kappa)$. Having identified this, we can obtain the mean and variance of $k_{s}$ by:
\begin{align}
    \mathds{E}(k_{s} \mid \cdot) & = (n - 1) \frac{\upsilon}{\upsilon + \kappa} + 1, \label{eq:exp_ncl} \\ 
    \mathds{V}(k_{s} \mid \cdot) & = (n - 1) \frac{\upsilon \kappa (\upsilon + \kappa + n - 1)}{(\upsilon + \kappa)^{2} (\upsilon + \kappa + 1)}. \label{eq:var_ncl}
\end{align}

This finding is important for prior elicitation because it shows that $\rho_{s}$ directly influences the number of clusters through its hyperparameters $\upsilon$ and $\kappa$. As an illustration, we calculated the mean and variance of the expected number of clusters by considering the 70 areas of the map. Table~\ref{tab:ncl} displays the values found for different specifications of $\upsilon$ and $\kappa$.
\begin{table}[!t] \centering
\begin{tabular}{c|cccccccc} \hline
\diagbox{$\kappa$}{$\upsilon$} & 0.01 & 1 & 5 & 10 & 15 & 20 & 25 & 30 \\
\hline
0.01& 36 / 1167 & 69 / 24 & 70 / 2 & 70 / 0 & 70 / 0 & 70 / 0 & 70 / 0 & 70 / 0 \\
1   & 2 / 24 & 36 / 408 & 58 / 103 & 64 / 38 & 66 / 20 & 67 / 13 & 67 / 9 & 68 /  7 \\
10  & 1 / 1 & 37 / 38 & 24 / 80 & 36 / 73 & 42 / 60 & 47 / 49 & 50 / 41 & 53 / 34 \\
20  & 1 / 1 & 34 / 13 & 15 / 40 & 24 / 49 & 31 / 49 & 36 / 46 & 39 / 42 & 42 / 39 \\
30  & 1 / 1 & 33 / 7 & 11 / 24 & 18 / 34 & 24 / 38 & 29 / 39 & 32 / 38 & 36 / 36 \\
40  & 1 / 1 & 33 / 4 & 9 / 17 & 15 / 26 & 20 / 30 & 24 / 32 & 28 / 33 & 31 / 33 \\
50  & 1 / 1 & 32 / 3 & 7 / 13 & 12 / 20 & 17 / 25 & 21 / 28 & 24 / 29 & 27 / 30 \\
60  & 1 / 1 & 32 / 2 & 6 / 10 & 11 / 17 & 15 / 21 & 18 / 24 & 21 / 26 & 24 / 27 \\
70  & 1 / 1 & 32 / 2 & 6 / 8 & 10 / 14 & 13 / 18 & 16 / 21 & 19 / 23 & 22 / 24 \\
80  & 1 / 1 & 32 / 2 & 5 / 7 & 9 / 12 & 12 / 16 & 15 / 18 & 17 / 21 & 20 / 22 \\
90  & 1 / 1 & 32 / 1 & 5 / 6 & 8 / 10 & 11 / 14 & 14 / 17 & 16 / 19 & 18 / 20 \\
100 & 1 / 1 & 32 / 1 & 4 / 5 & 7 / 9 & 10 / 12 & 12 / 15 & 15 / 17 & 17 / 19 \\
110 & 1 / 1 & 32 / 1 & 4 / 5 & 7 / 8 & 9 / 11 & 12 / 14 & 14 / 16 & 16 / 17 \\
120 & 1 / 1 & 32 / 1 & 4 / 4 & 6 / 7 & 9 / 10 & 11 / 13 & 13 / 14 & 15 / 16 \\
130 & 1 / 1 & 32 / 1 & 4 / 4 & 6 / 7 & 8 / 9 & 10 / 12 & 12 / 13 & 14 / 15 \\
140 & 1 / 1 & 31 / 1 & 3 / 3 & 6 / 6 & 8 / 9 & 10 / 11 & 11 / 13 & 13 / 14 \\
150 & 1 / 1 & 31 / 1 & 3 / 3 & 5 / 6 & 7 / 8 &  9 / 10 & 11 / 12 & 12 / 13 \\
\hline
\end{tabular} \caption{Number of cluster {\it a priori} (mean/variance) for different specifications of $\upsilon$ and $\kappa$.} \label{tab:ncl} 
\end{table}

\subsection{Autocorrelation function} \label{ss:rho}

\begin{figure} [!h] \centering
 \includegraphics[width=.95\textwidth]{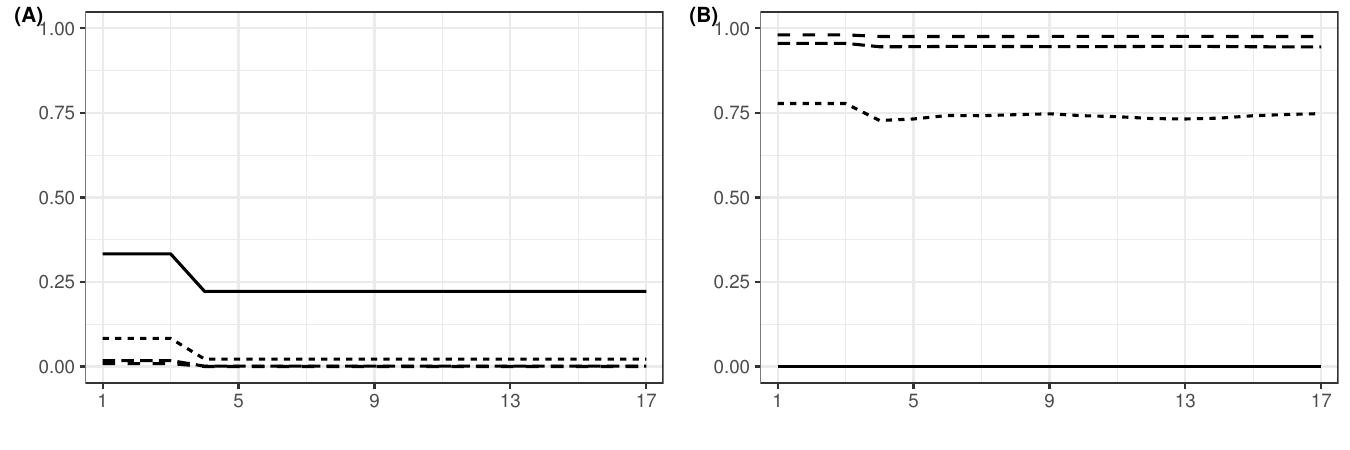}
\caption{Autocorrelation function of $\{ \rho_{s} \}$ for $q = 3$, $s = 1$, and $l = 1, \ldots, 17$. (A) Values obtained by fixing $\{c_{s}\} = 1$ and varying $\upsilon$ and $ \kappa$, with $\upsilon = \kappa = 1$ (solid line), $\upsilon = 1$, $\kappa = 5$ (dotted line), $\upsilon = 5$, $\kappa = 50$ (dashed line), and $\upsilon = 10$, $\kappa = 100$ (longdashed line). (B) Values obtained by fixing $\upsilon = \kappa = 1$ and randomly generating $\{c_{s}\}$ from the prior distribution with $\zeta = 1$ (solid line), $\zeta = 10$ (dotted line), $\zeta = 50$ (dashed line), and $\zeta = 100$ (longdashed line).} \label{fig:cor}
\end{figure}

As discussed in Section~\ref{s:SPPPM} of the main manuscript, one advantage of using this approach lies in the calculation of the temporal autocorrelation of $\{ \rho_{s} \}$. This function depends on the parameters $\upsilon$, $\kappa$, and $\{c_{s}\}$. Thus, any prior knowledge about this autocorrelation may assist in the prior elicitation of these parameters. To set the hyperparameters of $\zeta$, for instance, we computed the autocorrelation function of $\{ \rho_{s} \}$ for some values of $\upsilon$, $\kappa$, and $\{ c_{s} \}$. Figure~\ref{fig:cor}(A) displays autocorrelation values obtained by varying $\upsilon$ and $\kappa$, while $\{ c_{s} \} = 1$ is kept fixed. In contrast, Figure~\ref{fig:cor}(B) illustrates the autocorrelation values obtained by fixing $\upsilon = \kappa = 1$ and varying hyperparameter $\zeta$. In this case, it is evident that the values of $\{c_{s}\}$ have a significant impact on autocorrelation, whereas $\upsilon$ and $\kappa$ do not appear to play a significant role. As noted in the model formulation, the time dependence of the partition sequence is driven by time dependence of ${ \rho_{s} }$. This connection, illustrated in Figure~\ref{fig:model} of the main manuscript, highlights the importance of the correlation among ${ \rho_{s} }$.

Additionally, within each season, the correlation among observations is indirect, as they share common parameters (${\bm \theta}$ and ${\bm z}$). To analyze the data autocorrelation, we begin by generating probability vectors based on prior distributions with different hyperparameter values. Next, using these probabilities, we create random partitions and subsequently produce synthetic data. Finally, we assess the autocorrelation using a standard autocorrelation function. To do so, let us set $\upsilon = 10$ and $\kappa = 100$ so that, according to Table~\ref{tab:ncl}, the {\it a priori} expected number of clusters in the first season is around 10\% of the total number of areas. Then, we randomly generate $\{c_{s}\}$ from its prior distribution with $\zeta = 1, 10, 50, 100$. See Section~\ref{app:sim} for the generation settings of the other model components. Once we have the dataset, we calculate the correlation function using {\tt acf} R function.

\begin{figure} [!h] \centering
 \includegraphics[width=\textwidth]{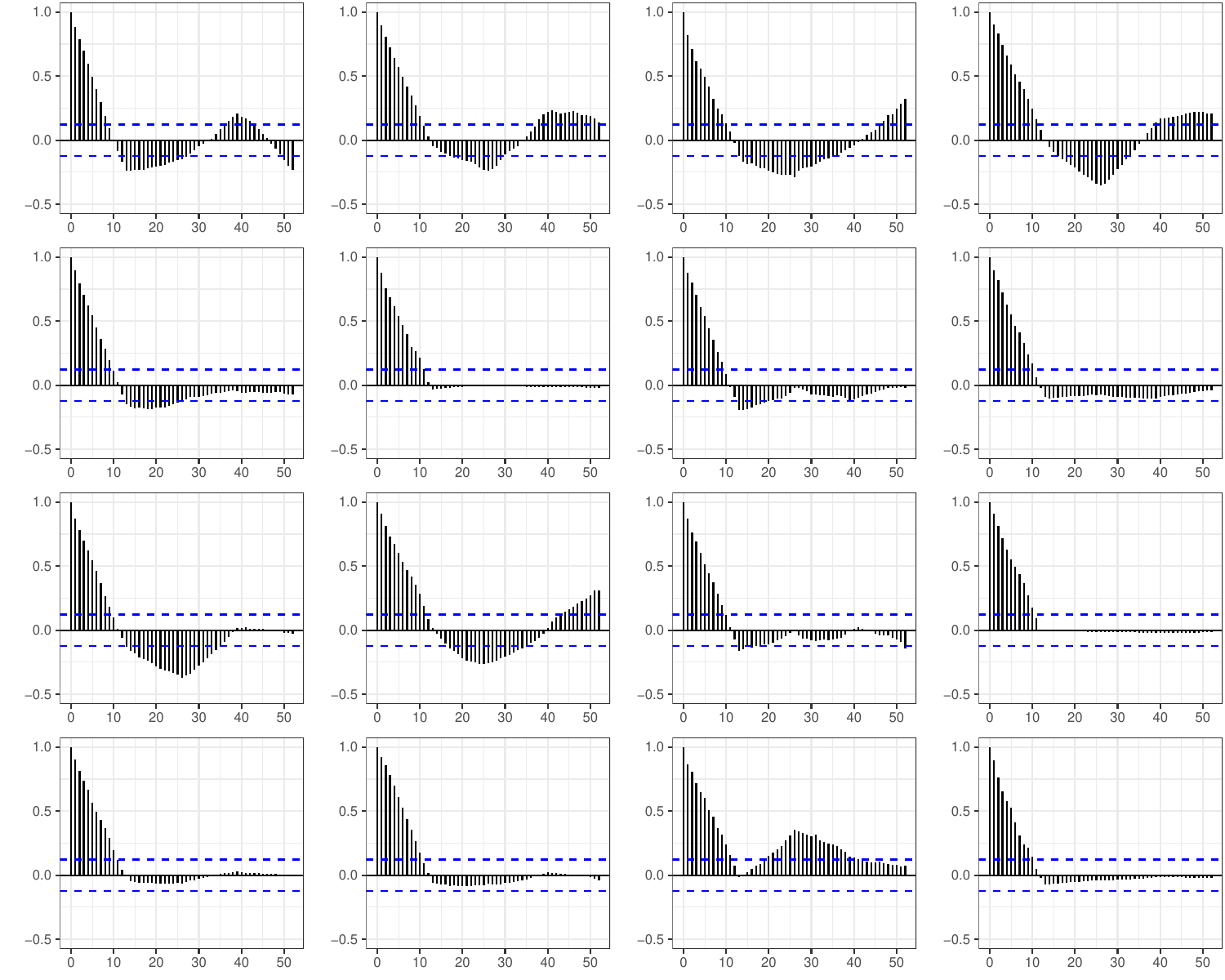}
\caption{Autocorrelation function of $\{ y_{t} \}$ for different areas represented in the rows. Data generated with different $\zeta$ values is represented in the columns (from left to right: $\zeta = 1, 10, 50, 100$). } \label{fig:cor_y}
\end{figure}

Figure~\ref{fig:cor_y} displays the autocorrelation function of $\{ y_{t} \}$ for different areas. It is difficult to determine whether the correlation observed in the data is influenced by the correlation of $\{ \rho_{s} \}$, even though the partition has a direct impact on the data through $\theta$ values. However, it is unquestionable that the data are autocorrelated over time. Furthermore, we can observe a seasonal trend, which can be derived from cluster parameters, dispersion parameters, or covariates. It is also worth noting that, in order to keep the expected number of clusters reasonably small, the values of $\rho$ must remain low, therefore, assuming $\zeta = 50$ or $\zeta = 100$ is unrealistic for data analysis as it encourages a large number of clusters.

\section{Posterior Inference} \label{app:post}

Although posterior inference is analytically intractable for this model, the fact that some full conditionals follow known distributions simplifies the sampling process. The parameters $\theta_{js}^{\star}$, $z_{is}$, $\rho_{s}$, $w$, and $\zeta$ are sampled via the usual Gibbs sampler algorithm, and a Metropolis-within-Gibbs algorithm is used to sample from the posteriors of $\beta$, $\delta$, $u$, $c$, $\upsilon$, and $\kappa$.

The most challenging step is sampling from the posterior of $(\pi_s,\mathcal{T}_s)$. We implement a Gibbs sampler algorithm to explore the space of spanning trees and partitions induced by removing the edges. This section focuses on detailing the last two sampling processes - for the partition and the tree - which are the most time-consuming computations. Nonetheless, the complete pseudocode may be found in Section~\ref{app:MCMC}.

\subsection{Sampling the partition} \label{ss:partition}

Proposed by \cite{Teixeira2015}, the Gibbs algorithm used to sample from the posterior of the random partitions given a compatible spanning tree is based on the strategy of \cite{Barry1993}. Adapted to the graph context, each coordinate of this binary vector represents an edge, and its dimension is the number of edges in the spanning tree, i.e., $n - 1$. If the vector coordinate is $0$, the edge must be removed from the tree to form the partition. Otherwise, the edge must not be removed. Thus, samples from the posterior distribution of the partition can be obtained by sampling from the posterior distribution of the binary vector by means of the usual Gibbs sampler algorithm. Let ${\bm \nu} = ( \nu_{1}, \ldots, \nu_{(n-1)} )$ be the binary representation of a partition ${\bm \pi}$, and let ${\bm \Omega}$ denote the complete parameter vector. The probability of removing an edge can be obtained based on its full conditional probability $\mathds{P}[\nu_{l} \mid {\bm \nu}_{-l}, {\bm \Omega}, {\bm Y}]$, where ${\bm \nu}_{-l} = ( \nu_{1}, \ldots, \nu_{(l-1)}, \nu_{(l+1)}, \ldots, \nu_{(n-1)} )$. Since the variable $\nu_{l}$ can assume only two values, it is sufficient to compute the ratio between them, which is formulated as:
\begin{equation}
R_{l} = \frac{\mathds{P}[\nu_{l} = 1 \mid {\bm \nu}_{-l}, {\bm \Omega}, {\bm Y}]}{\mathds{P}[\nu_{l} = 0 \mid {\bm \nu}_{-l}, {\bm \Omega}, {\bm Y}]} = \frac{\mathds{P}[{\bm \pi}^{(1)} \mid {\bm \Omega}, {\bm Y}]}{\mathds{P}[{\bm \pi}^{(0)} \mid {\bm \Omega}, {\bm Y}]},
\label{ratio} \end{equation}
where ${\bm \pi}^{(1)}$ represents the partition without any changes and ${\bm \pi}^{(0)}$ is the new partition obtained by removing the $l$-th edge. Note that the numerator and denominator differ only at position $l$. By removing edge $l$, the cluster containing the areas connected by this edge is split into two new clusters. These two different partitions induce a different number of clusters, which implies distinct coordinates in ${\bm \theta}^{\star}$ in the numerator and denominator, making it unfeasible to run a standard Gibbs sampler. To solve this problem, we integrate over ${\bm \theta}^{\star}$ in \eqref{ratio}. This calculation is straightforward since the full conditional of ${\bm \theta}^{\star}$ is a well-known distribution. Additionally, all the other clusters remain unchanged. This modified Gibbs sampler can be defined as a partially collapsed Gibbs sampler \citep{Dyk2008}. To achieve faster convergence of the MCMC algorithm, we also integrate $\bm \rho$ out from \eqref{ratio}. By doing this, we reduce instability in the algorithm by eliminating potential divisions by zero.

For the proposed model formulated in Section~\ref{sec:model} of the main manuscript, after integrating out ${\bm \theta}^{\star}$ and $\bm \rho$, the ratio for season $s$ is:
\begin{equation*}
R_{ls} = \frac{ \left( n - k_{s} + \kappa + \sum\limits_{l = 0}^{q} (c_{s-l} - u_{s-l}) - 1 \right)}{\left( k_{s} + \upsilon + \sum\limits_{l = 0}^{q} u_{s-l} - 1 \right)} \frac{\Gamma (a)}{b^{a}} \frac{f^{(1)}_{s}({\bm Y}_{S_{j}})}{f^{(0)}_{s}({\bm Y}^{(1)}_{S_{j}}) f^{(0)}_{s}({\bm Y}^{(2)}_{S_{j}})},
\end{equation*}
where $f_{s}({\bm Y}_{S_{j}})$ denotes the normalizing constant obtained when integrating ${\bm \theta}^{\star}$ out, and is given by 
\begin{equation*}
f_{s}({\bm Y}_{S_{j}}) = \frac{ \Gamma \left( a + \sum\limits_{i \in S_{j}} \sum\limits_{t \in s} y_{it}\right)}{ \left( b + \sum\limits_{i \in S_{j}} z_{is} \sum\limits_{t \in s} O_{it} \exp \{ X_{it}^{\top} \beta \} \right)^{ \left( a + \sum\limits_{i \in S_{j}} \sum\limits_{t \in s} y_{it}\right) }}.
\end{equation*}

Here, function $f^{(1)}_{s}({\bm Y}_{S_{j}})$ is related to the whole cluster, whereas $f^{(0)}_{s}({\bm Y}^{(1)}_{S_{j}})$ and $f^{(0)}_{s}({\bm Y}^{(2)}_{S_{j}})$ are computed from the two sub-clusters formed when the $l$-th edge is removed from the tree. Once $R_{ls}$ is obtained, $\nu_{l}$ can be updated by a Metropolis-within-Gibbs step with a uniformly generated candidate.

\subsection{Sampling the tree}

To sample a tree compatible with the partition, we need its full conditional distribution. By assuming the prior distributions in (\eqref{eq:partition_prior}) and \eqref{eq:tree_prior}, it is not difficult to see that such a distribution is the uniform distribution over the subset of trees compatible with the current partition. Thus, at each step of the Gibbs sampler, it is sufficient to ensure that the new tree and current partition are compatible. Recall that compatibility is assumed when removing some edges from the tree produces the partition. To ensure that this constraint is satisfied, we consider the algorithm proposed by \cite{Teixeira2015}. We first assign weights to the edges in the original graph. Edges that connect vertices belonging to the same group receive a low weight, whereas edges that connect vertices belonging to different groups receive a high weight. Specifically, we assume the Unif$(0, 1)$ distribution to generate low values and the Unif$(10, 20)$ distribution to generate higher values. Note that these values are arbitrarily assigned; the key is that the weights for edges connecting vertices from different groups must be higher than the weights of other edges. Once the weights are assigned, we use a minimum spanning tree (MST) algorithm to sample a tree that is compatible with the current partition. 

In Computer Science literature, there are different algorithms designed to find a minimum spanning tree for a weighted undirected graph. The idea behind these algorithms is to compute the spanning tree with the minimal sum of weights. By using one MST algorithm and following the proposed weight assignments, edges connecting vertices in the same component will be selected before edges that connect vertices in distinct components, ensuring that the resulting spanning tree respects the current partition. In this case, we use the Prim algorithm \citep{Jungnickel2013}, which is preferred when there are a large number of edges in the graph.

\section{Prior elicitation} \label{app:prior_spec}

\begin{figure} [!h] \centering
 \includegraphics[width=0.95\textwidth]{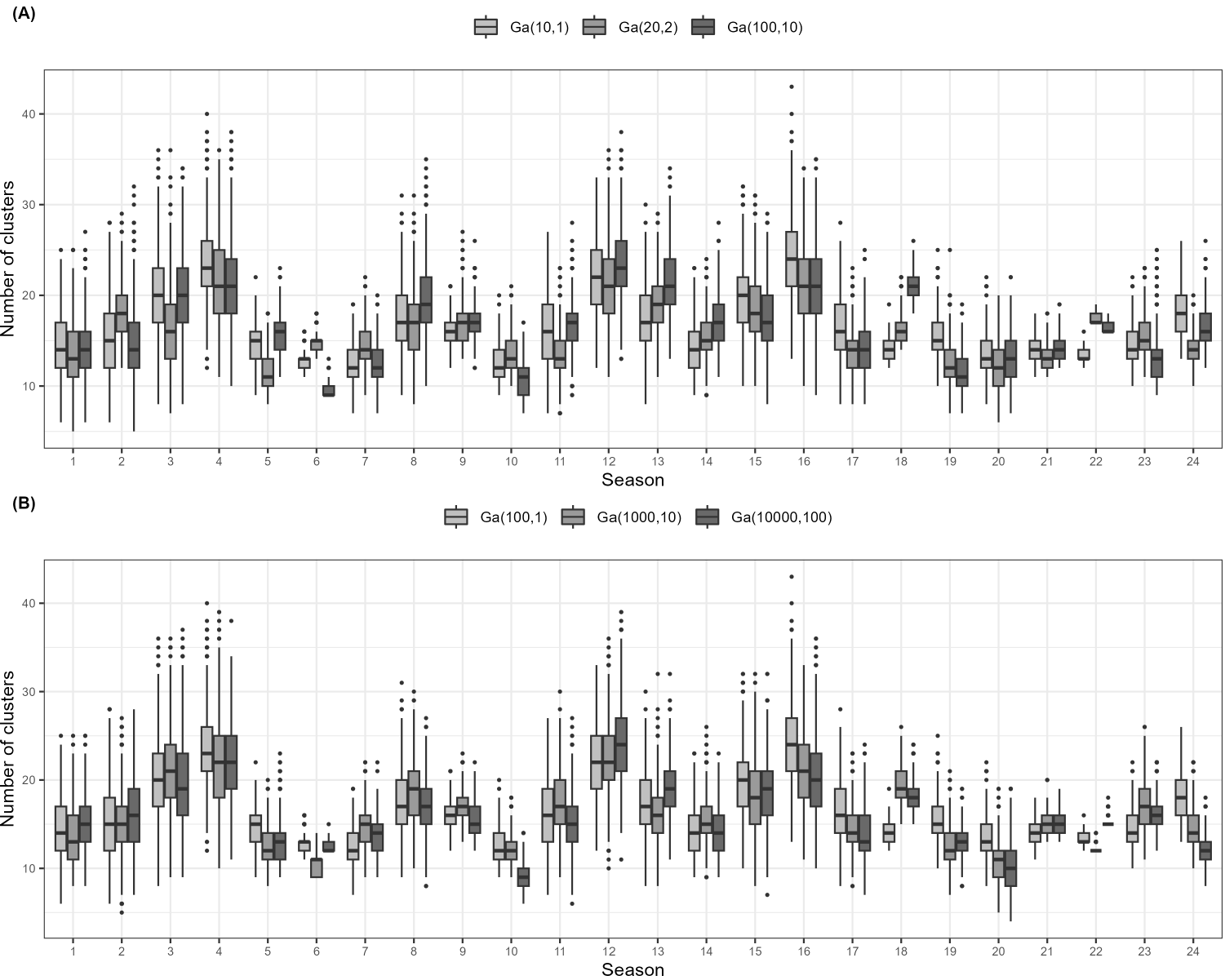}
\caption{Posterior distributions of the number of clusters ($k$) under different prior distributions for: (A) $\upsilon$ and (B) $\kappa$. The results were obtained by fitting the model to the application data.} \label{fig:upsilon_kappa_sensi}
\end{figure}

As presented in the main manuscript, the values of $\upsilon$ and $\kappa$ have a significant impact on the number of clusters $k$. Expressions for the mean \eqref{eq:exp_ncl} and variance \eqref{eq:var_ncl} can guide the setting of the hyperparameters $\upsilon$ and $\kappa$ according to the $k$ that one expects to estimate. In the case of the dengue data used in Section~\ref{sec:dataapp} of the main manuscript, assuming $\upsilon \approx 10$ and $\kappa \approx 100$ implies $k = 14$, which represents approximately 10\% of the total areas. Considering this, we explored three prior settings for $\upsilon$ such that the mean is fixed, $\mathds{E}(\upsilon) = 10$, while the variance varies with $\mathds{V}(\upsilon) = 10, 5, 1$. To do so, we assumed Ga(10,1), Ga(20,2), and Ga(100,10), respectively. Similarly, we set three prior distributions for $\kappa$ with $\mathds{E}(\kappa) = 100$ and $\mathds{V}(\kappa) = 100, 10, 1$. In this case, we used Ga(100,1), Ga(1000,10), and Ga(10000,10), respectively. Figure~\ref{fig:upsilon_kappa_sensi} displays the posterior distributions of $k$ over time under these prior distributions. In Figure~\ref{fig:upsilon_kappa_sensi}(A), the prior distribution of $\upsilon$ varies as mentioned, while $\kappa$ follows a Ga(100,1) distribution. Conversely, Figure~\ref{fig:upsilon_kappa_sensi}(B) presents a scenario where the prior distribution of $\upsilon$ is Ga(10,1) and $\kappa$ varies as previously described. Overall, the {\it a posteriori} number of clusters is robust to the {\it a priori} variability of the $\upsilon$ and $\kappa$ distributions.

\begin{figure} [!t] \centering
 \includegraphics[width=0.95\textwidth]{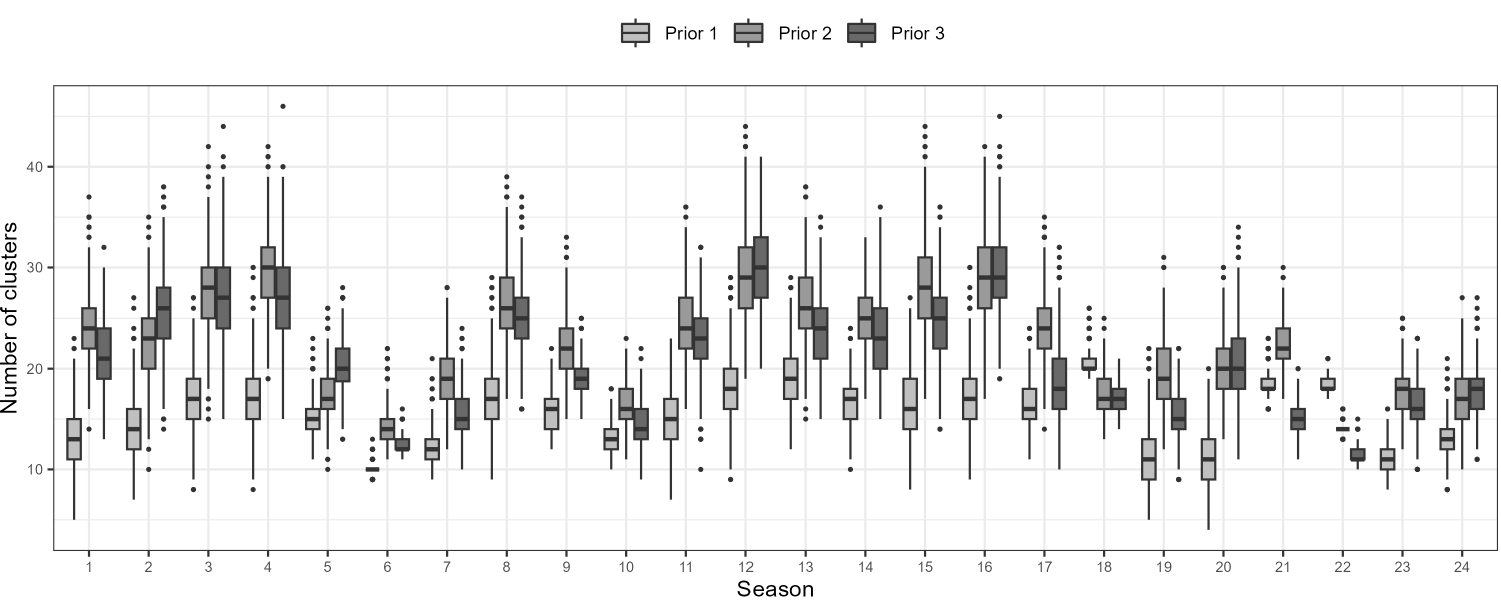}
\caption{Posterior distributions of the number of clusters ($k$) under different prior distributions for $\upsilon$ and $\kappa$. The results were obtained by fitting the model to the application data.} \label{fig:ncl_sensi}
\end{figure}

Another strategy was also considered for exploring the hyperparameter settings. Instead of fixing the means of $\upsilon$ and $\kappa$ in order to estimate a certain number of clusters, we set prior distributions to estimate different values of $k$. Specifically, we set $\upsilon$ and $\kappa$ considering an {\it a priori} mean of $\mathds{E}(k) = 14, 29, 44$, which corresponds to 10\%, 20\%, and 30\% of the total number of areas in the study. Figure~\ref{fig:ncl_sensi} shows the posterior distributions of $k$ under these settings. It can be seen that there was more variation in the posterior distributions of $k$ than previously observed in Figure~\ref{fig:upsilon_kappa_sensi}. This suggests that it is desirable to have an {\it a priori} understanding of the number of clusters one expects to estimate.

\begin{figure} [!t] \centering
 \includegraphics[width=0.95\textwidth]{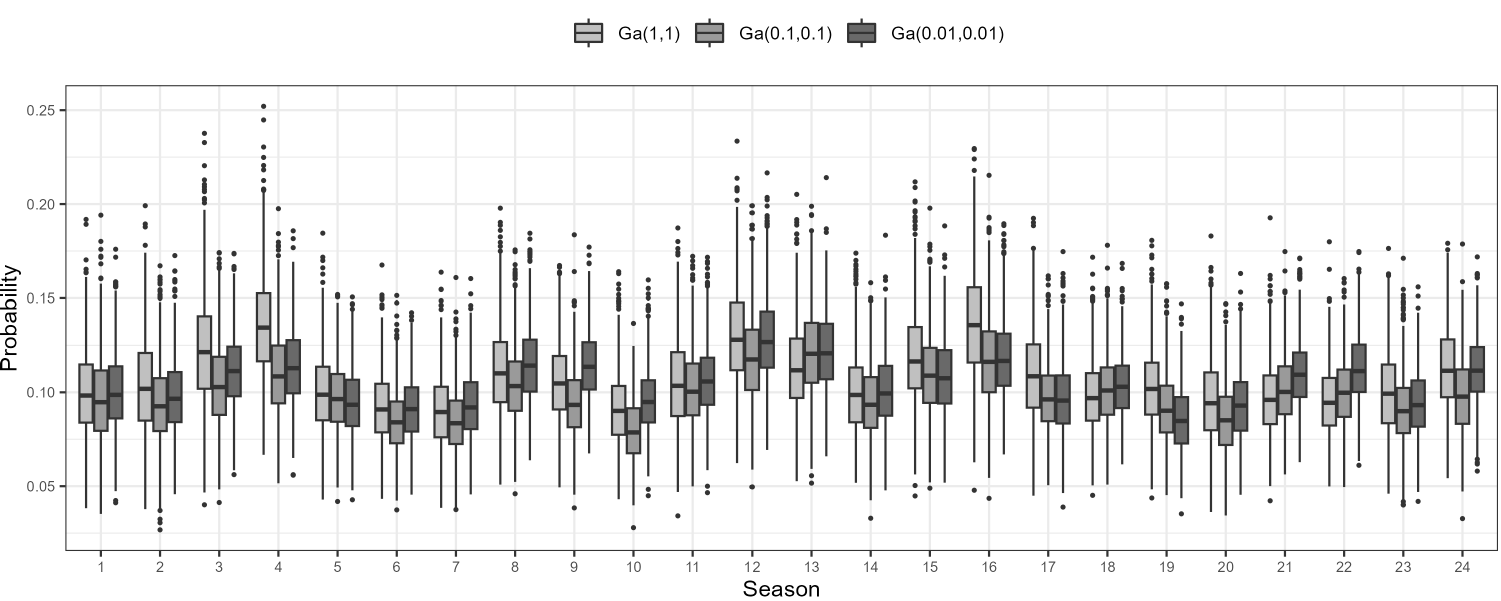}
\caption{Posterior distributions of the probability of removing edges ($\rho$) under different prior distributions for $\zeta$. The results were obtained by fitting the model to the application data.} \label{fig:zeta_sensi}
\end{figure}

Another parameter that needs its distribution defined is $\zeta$. As seen in Section~\ref{s:SPPPM}, $\zeta$ is the hyperparameter of the latent variables $\{ c_{s} \}$ and directly impacts the probability of removing edges, $\rho_{s}$, as well as its autocorrelation function, as shown in Eq.~\eqref{eq:cor_rho} of the main manuscript. To understand the effect of $\zeta$ on ${\bm \rho}$, we explore three settings for the prior distribution of $\zeta$. Figure~\ref{fig:zeta_sensi} presents the posterior distributions of the probability of removing edges under the different prior distributions for $\zeta$. In general, the probabilities are robust to the prior specification, with slight differences in the posterior of $\rho_{s}$.

\section{MCMC algorithm} \label{app:MCMC}

In this section, we describe the Metropolis-within-Gibbs sampler algorithm used to obtain posterior samples from the model presented in Section~\ref{sec:model} of the main manuscript. Recall that $i = 1, \ldots, n$, $t = 1, \ldots, T$, and $s = 1, \ldots, n_{s}$ represent area, epidemiological week, and season, respectively. Additionally, let ${\bm \Omega} = \{ {\bm z}, {\bm \beta},$ ${\bm \theta}, {\bm \delta}, {\mathcal{T}}, {\bm \pi}, {\bm \rho}, {\bm u},$ ${\bm c}, w, \zeta, \upsilon, \kappa \}$ denote the complete parameter vector. The algorithm is then given as follows:

\vspace{0.5cm} \begin{enumerate} \item[1.] Update $\upsilon$: the update of this parameter includes a random walk Metropolis–Hastings step, where the full conditional distribution is given by: \end{enumerate}
\begin{equation*}
    p(\upsilon \mid \cdot) \propto \left[ \prod_{s = 1}^{n_s} \frac{\Gamma \left( \upsilon + \kappa + \sum\limits_{l = 0}^{q} c_{s-l} \right)}{\Gamma \left( \upsilon + \sum\limits_{l = 0}^{q} u_{s-l} \right) } \right] \frac{\Gamma(\upsilon + \kappa)}{\Gamma(\upsilon)} \left( w \prod_{s = 1}^{n_s} \rho_{s} \right)^{\upsilon} \upsilon^{a_{_{\upsilon}}-1} \exp \{ - b_{_{\upsilon}} \upsilon \}.
\end{equation*}

\begin{enumerate} \item[2.] Update $\kappa$: similar to the previous step, the update of this parameter also includes a random walk Metropolis–Hastings step, where the full conditional distribution is given by:\end{enumerate}
\begin{equation*}
    p(\kappa \mid \cdot) \propto \left[ \prod_{s = 1}^{n_s} \frac{\Gamma \left( \upsilon + \kappa + \sum\limits_{l = 0}^{q} c_{s-l} \right)}{\Gamma \left( \kappa + \sum\limits_{l = 0}^{q} (c_{s-l} - u_{s-l}) \right)} \right] \frac{\Gamma(\upsilon + \kappa)}{\Gamma(\kappa)} \left( (1 - w) \prod_{s = 1}^{n_s} (1 - \rho_{s}) \right)^{\kappa} \kappa^{a_{_{\kappa}}-1} \exp \{ - b_{_{\kappa}} \kappa \}.
\end{equation*}

\begin{enumerate} \item[3.] Update $\zeta$: the full conditional of this parameter is a well-known distribution, thus: \end{enumerate}
\begin{equation*}
    (\zeta \mid \cdot) \sim \text{Ga} \left( a_{_\zeta} + \sum_{s = 1}^{n_{s}} c_{s}, \; b_{_\zeta} + n_{s} \right).
\end{equation*}

\begin{enumerate} \item[4.] Update $w$: the full conditional for this parameter is also a widely recognized distribution, hence: \end{enumerate}
\begin{equation*}
    (w \mid \cdot) \sim \text{Be} \left( \upsilon + \sum_{s = 1}^{n_{s}} u_{s}, \; \kappa + \sum_{s = 1}^{n_{s}} (c_{s} - u_{s}) \right).
\end{equation*}

\vspace{0.5cm}
\noindent For $s = 1, \ldots, n_{s}$, do: 
\vspace{0.2cm}

\begin{enumerate} \item[5.] Update $\mathcal{T}_{s}$: the following update is based on Prim's algorithm \citep{Jungnickel2013}. For each pair of edges belonging to the original graph, weights are assigned as follows:
\begin{enumerate}
       \item the edges that connect vertices belonging to the same group receive a low weight,
       \item the edges that connect vertices belonging to different groups receive a high weight.
\end{enumerate}
   
   Once the weights are assigned, the minimum spanning tree is obtained. This new sampled tree is compatible with the current partition.
\end{enumerate}

\begin{enumerate} \item[6.] Update $\pi_{s}$: as described in Section~\ref{ss:partition}, the sampling of a partition given a compatible spanning tree is performed based on \cite{Barry1993}'s strategy of representing a partition as a vector of binary variables. Hence, let ${\bm U}_{s}$ be the binary representation of the partition at season $s$, and $l$ denotes the edge under evaluation.

\vspace{0.2cm} For $l = 1, \ldots, n - 1$, do: \vspace{0.2cm}
\begin{enumerate}[leftmargin=.4in]
       \item[6.1] Calculate 
\begin{equation*}
    R_{ls} = \frac{ \left( n - k_{s} + \kappa + \sum\limits_{l = 0}^{q} (c_{s-l} - u_{s-l}) - 1 \right)}{\left( k_{s} + \upsilon + \sum\limits_{l = 0}^{q} u_{s-l} - 1 \right)} \frac{\Gamma (a)}{b^{a}} \frac{f^{(1)}(S_{j})}{f^{(0)}(S_{j_{1}}) f^{(0)}(S_{j_{2}})},
\end{equation*}
where $f_{s}({\bm Y}_{S_{j}}) = \frac{ \Gamma \left( a + \sum\limits_{i \in S_{j}} \sum\limits_{t \in s} y_{it}\right)}{ \left( b + \sum\limits_{i \in S_{j}} z_{is} \sum\limits_{t \in s} E_{it} \exp \{ {\bm X}_{it}^{\top} {\bm \beta} \} \right)^{ \left( a + \sum\limits_{i \in S_{j}} \sum\limits_{t \in s} y_{it}\right) }}.$
       \vspace{0.2cm} \item[6.2] Sample: $u \sim \text{Unif}(0, 1)$.
       \vspace{0.2cm} \item[6.3] Remove the edge if $R_{ls} < \frac{u}{1-u}$.
\end{enumerate}

\vspace{0.2cm} end of the loop iterating in $l$.  \vspace{0.2cm}
\end{enumerate}

\begin{enumerate} \item[7.] Update $c_{s}$: the update of this parameter includes a random walk Metropolis–Hastings step, where the full conditional distribution is given by:
\end{enumerate}
\begin{equation*}
    p(c_{s} \mid \cdot ) \propto \left[ \prod_{h = 0}^{q} \frac{\Gamma \left( \upsilon + \kappa + \sum\limits_{l = 0}^{q} c_{s-l+h} \right)}{\Gamma \left( \kappa + \sum\limits_{l = 0}^{q} (c_{s-l+h} - u_{s-l+h}) \right)} \right] \frac{1}{(c_{s} - u_{s})!} \left[ \zeta (1 - w) \prod_{h = 0}^{q} (1 - \rho_{s+h}) \right]^{c_{s}} \mathds{1}_{[c_{s} \geq u_{s}]}.
\end{equation*}

\begin{enumerate} \item[8.] Update $u_{s}$: similarly, the update of this parameter includes a random walk Metropolis-Hastings step, where the full conditional distribution is given by:
\end{enumerate}
\allowdisplaybreaks \begin{align*}
    p(u_{s} \mid \cdot) \propto & \left[ \prod_{h = 0}^{q}\frac{1}{\Gamma \left( \upsilon + \sum\limits_{l = 0}^{q} u_{s-l+h} \right) \Gamma \left( \kappa + \sum\limits_{l = 0}^{q} (c_{s-l+h} - u_{s-l+h}) \right)} \right] \binom{c_{s}}{u_{s}} \\ & \times \left[ \frac{w}{(1-w)} \prod_{h = 0}^{q} \frac{\rho_{s+h}}{(1-\rho_{s+h})} \right]^{u_{s}} \mathds{1}_{[u_{s} \leq c_{s}]}.
\end{align*}

\begin{enumerate} \item[9.] Update $\rho_{s}$: the full conditional of this parameter is a well-known distribution, thus: \end{enumerate}
\begin{equation*}
    (\rho_{s} \mid \cdot) \sim \text{Be} \left( k_{s} + \upsilon + \sum_{l = 0}^{q} u_{s-l} - 1, n - k_{s} + \kappa + \sum_{l = 0}^{q} (c_{s-l} - u_{s-l}) \right)
\end{equation*}

\begin{enumerate} \item[10.] Update ${\bm \theta}_{s}$:  this parameter is independently updated for each cluster, where its full conditional is a well-known distribution.

\vspace{0.2cm} For $j = 1, \ldots, k_{s}$ do: \vspace{0.2cm} 
\begin{equation*}   
    (\theta_{js}^{\star} \mid \cdot ) \sim  \text{Ga} \left(a + \sum\limits_{i \in S_{j}} \sum\limits_{t \in s} y_{it}, b + \sum\limits_{i \in S_{j}}  z_{is} \sum\limits_{t \in s} E_{it} \exp \Big\{ {\bm X}_{it}^{\top} {\bm \beta} \Big\} \right) 
\end{equation*}
\vspace{0.2cm} end of the loop iterating in $j$.  \vspace{0.2cm} \end{enumerate}

\begin{enumerate} \item[11.] Update ${\bm z}_{s}$: This parameter is independently updated for each area, where its full conditional is a well-known distribution.

\vspace{0.2cm} For $i = 1, \ldots, n$ do: \vspace{0.2cm} 
\begin{equation*}   
(z_{is} \mid \cdot) \sim \text{GIG} \left( \sum\limits_{t \in s} y_{it} - \frac{1}{2}, \; 2 \theta_{js}^{\star} \sum\limits_{t \in s} E_{it} \exp \{ {\bm X}_{it}^{\top} {\bm \beta} \} + \exp \{ {\bm V}_{is}^{\top} {\bm \delta} \}, \; \exp \{ {\bm V}_{is}^{\top} {\bm \delta} \} \right) 
\end{equation*}

\vspace{0.2cm} end of the loop iterating in $i$.  
\end{enumerate}

\vspace{0.5cm}
\noindent end of the loop iterating in $s$. 
\vspace{0.5cm}

\noindent Steps 12 to 16 are included to generate the $q$ extra values of $\bm u$, $\bm c$, and $\bm \rho$ used in steps 7 and 8. 

\vspace{0.2cm}
\noindent For $s = n_{s} + 1, \ldots, n_{s} + q$, do: 
\vspace{0.2cm}

\begin{enumerate} \item[12.] Update $\mathcal{T}_{s}$: Repeat step 5. \end{enumerate}

\begin{enumerate} \item[13.] Update $\pi_{s}$: analogous to step 6, let ${\bm U}_{s}$ be the binary representation of the partition at season $s$, and $l$ denotes the edge under evaluation. Then the prior distribution defined in Eq.(8) is used to obtain the ratio $R_{ls}$

\vspace{0.2cm} For $l = 1, \ldots, n - 1$, do: \vspace{0.2cm}
\begin{enumerate}[leftmargin=.5in]
       \item[13.1] Calculate $ R_{ls} = \frac{1 - \rho_{s}}{\rho_{s}}$.
       \vspace{0.2cm} \item[13.2] Sample: $u \sim \text{Unif}(0, 1)$.
       \vspace{0.2cm} \item[13.3] Remove the edge if $R_{ls} < \frac{u}{1-u}$.
\end{enumerate}

\vspace{0.2cm} end of the loop iterating in $l$.  \vspace{0.2cm}
\end{enumerate}

\begin{enumerate} \item[14.] Update $c_{s}$: Repeat step 7.
\end{enumerate}

\begin{enumerate} \item[15.] Update $u_{s}$: Repeat step 8.
\end{enumerate}

\begin{enumerate} \item[16.] Update $\rho_{s}$: Repeat step 9.
\end{enumerate}

\vspace{0.2cm}
\noindent end of the loop iterating in $s$. 
\vspace{0.5cm}

\begin{enumerate} \item[17.] Update $\bm \beta$: the following update includes a random walk Metropoli-Hastingss step, where the full conditional distribution is given by:
\end{enumerate}
\allowdisplaybreaks \begin{align*}
    p({\bm \beta} \mid \cdot) \propto & \exp \left\{ \sum_{i=1}^{n} \sum_{t=1}^{T} y_{it} {\bm X}_{it}^{\top} {\bm \beta} \right\} \exp \left\{ - \sum_{s = 1}^{n_{s}} \sum_{t \in s} \sum_{j = 1}^{k_{s}} \sum_{i:c_{i} = j} E_{it} z_{is} \theta_{js}^{\star} \exp \Big\{ {\bm X}_{it}^{\top} {\bm \beta} \Big\} \right\} \\ & \times \exp \left\{ - \frac{1}{2} ({\bm \beta} - {\bm \mu}_{\beta})^{\top} {\bm \Sigma}_{\beta}^{-1}({\bm \beta} - {\bm \mu}_{\beta}) \right\}.
\end{align*}

\begin{enumerate} \item[18.] Update $\bm \delta$: similar to step 17, the update of this parameter includes a random walk Metropoli-Hastingss step, where the full conditional distribution is given by:
\end{enumerate}
\allowdisplaybreaks \begin{align*}
    p({\bm \delta} \mid \cdot) \propto & \exp \left\{ \frac{1}{2} \sum_{i=1}^{n} \sum_{s=1}^{n_{s}} {\bm V}_{is}^{\top} {\bm \delta} \right\} \exp \left\{ - \sum_{s = 1}^{n_{s}} \sum_{t \in s} \sum_{i=1}^{n} \frac{\exp \Big\{ {\bm V}_{is}^{\top} {\bm \delta} \Big\} (z_{is} - 1)^{2}}{2 z_{is}} \right\} \\ & \times \exp \left\{ - \frac{1}{2} ({\bm \delta} - {\bm \mu}_{\delta})^{\top} {\bm \Sigma}_{\delta}^{-1}({\bm \delta} - {\bm \mu}_{\delta}) \right\} .
\end{align*}

\section{Simulation study} \label{app:sim}

In this section, we provide an in-depth explanation of two simulation studies that illustrate various aspects of the model we proposed. We chose the 70 microregions of the Brazilian state of Minas Gerais as our underlying map (i.e., $i = 1, \ldots, 70$), where two regions are treated as neighbors if they share a common geographic boundary. To perform posterior inference, we saved samples of size 1,000 obtained after running 10,000 iterations, discarding the first 70\% as burn-in, and thinning by 3 to avoid correlation. Convergence was monitored graphically. Models were fitted with the following prior specifications: ${\bm \mu}_{\beta} = {\bm \mu}_{\delta} = {\bm 0}$ and ${\bm \Sigma}_{\beta} = {\bm \Sigma}_{\delta} = 10 {\bm I}$, allowing considerable variability for regression coefficients in both mean and dispersion components. We also assumed $a_{\theta} = b_{\theta} = 1$, so that the offset plays an important role in explaining the outcome. As mentioned before, the values of $\upsilon$ and $\kappa$ directly affect the number of clusters. Based on Table~\ref{tab:ncl}, we set $a_{\upsilon} = 10$, $b_{\upsilon} = 1$, $a_{\kappa} = 100$, and $b_{\kappa} = 1$, which implies an {\it a priori} expected number of clusters equal to 7, representing 10\% of the total number of areas. Finally, $\bm c$ is the parameter set affecting the temporal autocorrelation structure of $\bm \rho$. By defining $a_{\zeta} = 1$ and $b_{\zeta} = 1$, we prevent $\bm c$ from assuming very high values. Moreover, we are assuming that the {\it a priori} temporal correlation of the sequence of probabilities is low. We used the Watanabe-Akaike information criterion (WAIC, \cite{Gelman2014}) to check the goodness of fit. To evaluate the accuracy of partition estimates, we employed the method available in the {\tt salso} R package \citep{Dahl2020}, using the variation information (VI) loss function to estimate the partitions, and then we used the Rand Index \citep[RI,][]{Hubert1985} to measure the similarity between the true and estimated partitions.

\subsection{Simulation 1: comparing Poisson-inverse Gaussian and Poisson models} \label{app:sim1}

\begin{figure} [!t] \raggedright
\includegraphics[width=\textwidth]{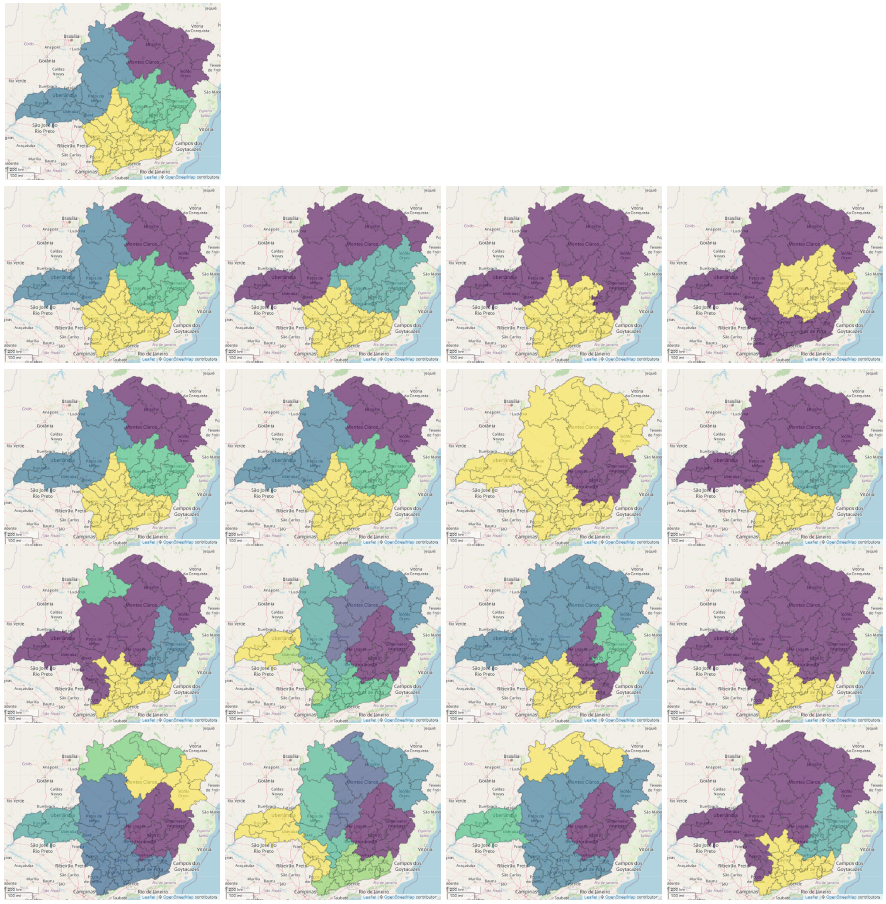}
\caption{ Map of spatial partitions used in Simulation 1. Row 1: Scenario 1 - a constant partition in time. Row 2: Scenario 2 - a different partition for each season that is repeated over the years. Rows 3-5: Scenario 3 - a different partition for each season and year. } \label{fig:map_part_sim1}
\end{figure}

This first simulation study aims to compare the performance of spatio-temporal Poisson-inverse Gaussian and Poisson models when applied to equidispersed and overdispersed data. To produce artificial scenarios that are faithful to reality, we used the observed information as offsets and covariates. The population size (per 100k people) of each area was considered as an offset, whereas temperature and humidity measured for each area and epidemiological week were taken as the covariates forming the design matrix ${\bm X}$ (see Section \ref{sec:motivation} for the covariate descriptions). Regression coefficients were set to ${\bm \beta} = (0.4, 0.1)$ for temperature and humidity, respectively. Figure~\ref{fig:map_part_sim1} displays the maps showing each partition structure considered in this simulation study.

Cluster-specific parameters were set with a two-units difference between clusters as follows:
\small \begin{equation*} \begin{tabular}{ll} 
\multicolumn{1}{c}{\bf Scenario 1} & \multicolumn{1}{c}{\bf Scenario 3} \\
${\bm \theta}^{\star}_{s} = (1, 3, 5, 7) \quad \; \; $ for $s = 1, \ldots, 12$ & \multirow{7}{*}{${\bm \theta}^{\star}_{s} = \begin{cases}
        (1, 3, 5, 7, 9, 11) & \text{for} \, s = 6 \\
        (1, 3, 5, 7, 9, 11) & \text{for} \, s = 10 \\
        (1, 3, 5, 7, 9) & \text{for} \, s = 9 \\
        (1, 3, 5, 7) & \text{for} \, s = 1, 2, 5, 7, 11 \\
        (1, 3, 5) & \text{for} \, s = 4, 12 \\
        (5, 7) & \text{for} \, s = 3 \\
        (1, 3) & \text{for} \, s = 8 \\
\end{cases}$} \\
& \medskip \\
\multicolumn{1}{c}{\bf Scenario 2} & \\
${\bm \theta}^{\star}_{s} = \begin{cases}
        (1, 3, 5, 7) & \text{for} \, s = 1, 5, 9 \\
        (3, 5, 7) & \text{for} \, s = 2, 6, 10 \\
        (3, 5) & \text{for} \, s = 3, 7, 11 \\
        (1, 3) & \text{for} \, s = 4, 8, 12 \\
\end{cases}$ & \\
\end{tabular} \end{equation*} \normalsize

We used information on temperature and humidity to build the design matrix ${\bm V}$. Unlike ${\bm X}$, when constructing ${\bm V}$, the measurements were averaged over all weeks in each season. Furthermore, ${\bm V}$ includes a column of 1s for the intercept term. Regression coefficients were set to ${\bm \delta} = (-0.3, 0.2, -0.4)$ for intercept, temperature, and humidity, respectively. We fit the PIG and Poisson models with $q = 1$ to analyze the synthetic datasets using the MCMC algorithm described in Section~\ref{app:MCMC}.

The first facet of the proposed model to be studied is its ability to recover regression coefficients (shown in the main manuscript Section~ref{s:sim1}). To evaluate the accuracy of partition estimates, we employed the method available in the {\tt salso} R package \citep{Dahl2020} with the VI loss function to estimate the 12 partitions for each synthetic dataset. To measure the similarity between the true and estimated partitions, we used the RI implemented in the same package. The higher the RI value, the closer two partitions are to each other. Table~\ref{tab:sim1} displays RI values averaged over 100 replicates and 12 seasons. Additionally, Table~\ref{tab:sim1_1} presents RI values calculated for each time point. On average, the RI for the PIG and Poisson models when the data are equidispersed is approximately equal. In this case, the accuracy of the partition estimates obtained is high, which aligns with all previously mentioned findings. For overdispersed data, the Poisson model almost completely loses its clustering capacity, while the PIG model provides RI values exceeding 60\% on average. The accuracy of partition estimation using the PIG model is further explored in Section~\ref{app:sim2}.

The conclusion drawn from this simulation study is that a PIG model can be applied to both equidispersed and overdispersed data. When overdispersion is not observed, the model assumes its particular case, which is equivalent to the Poisson model. Therefore, its ability to estimate and cluster is similar. However, the opposite does not occur. When a Poisson model is applied to overdispersed data, the model's clustering capacity decreases drastically. In this case, $\bm \theta$ attempts to explain the interaction between the cluster-specific intercept and the dispersion parameter, making it difficult for areas to co-cluster.
\begin{center} \begin{table} [!t]
 \caption{Rand index over time. Values are averaged over the 100 generated datasets.} \label{tab:sim1_1}
\resizebox{\columnwidth}{!}{
\begin{tabular}{cccccccccccccc}
\toprule
\multicolumn{2}{@{}c}{\multirow{2}{*}{\bf Equidispersed}} & \multicolumn{12}{@{}c}{\small RI}  \\ 
\cmidrule{3-14}
 & & $s = 1$ & $s = 2$ & $s = 3$ & $s = 4$ & $s = 5$ & $s = 6$ & $s = 7$ & $s = 8$ & $s = 9$ & $s = 10$ & $s = 11$ & $s = 12$\\ 
\midrule
 \multirow{2}{*}{Sce 1} & {\small PIG} & 0.99 & 0.99 & 0.99 & 0.99 & 0.99 & 0.99 & 0.98 & 1.00 & 0.99 & 0.99 & 0.99 & 0.99 \\
                        & Poisson & 0.99 & 0.99 & 0.99 & 0.99 & 0.99 & 0.99 & 0.98 & 0.99 & 0.99 & 0.99 & 0.99 & 0.99 \\
 \multirow{2}{*}{Sce 2} & {\small PIG} & 0.99 & 1.00 & 0.96 & 1.00 & 0.99 & 1.00 & 0.95 & 0.99 & 0.99 & 1.00 & 0.96 & 0.73 \\
                        & Poisson & 0.99 & 1.00 & 0.96 & 1.00 & 0.99 & 1.00 & 0.95 & 0.99 & 0.99 & 1.00 & 0.95 & 0.73 \\
 \multirow{2}{*}{Sce 3} & {\small PIG} & 0.99 & 0.99 & 0.98 & 0.99 & 1.00 & 0.98 & 1.00 & 0.99 & 1.00 & 0.99 & 0.99 & 0.88 \\
                        & Poisson & 0.99 & 0.99 & 0.97 & 0.99 & 1.00 & 0.98 & 1.00 & 0.99 & 1.00 & 0.99 & 0.99 & 0.88 \\
 \midrule
 \multicolumn{2}{@{}c}{\bf Overdispersed} & & & & & \\
\cmidrule{1-2}
 \multirow{2}{*}{Sce 1} & {\small PIG} & 0.66 & 0.67 & 0.63 & 0.65 & 0.66 & 0.67 & 0.64 & 0.66 & 0.66 & 0.64 & 0.65 & 0.67 \\
                        & Poisson & 0.04 & 0.04 & 0.04 & 0.04 & 0.04 & 0.03 & 0.04 & 0.04 & 0.03 & 0.04 & 0.05 & 0.05\\
 \multirow{2}{*}{Sce 2} & {\small PIG} & 0.61 & 0.69 & 0.51 & 0.61 & 0.61 & 0.69 & 0.50 & 0.62 & 0.63 & 0.71 & 0.50 & 0.59 \\
                        & Poisson & 0.04 & 0.03 & 0.02 & 0.01 & 0.04 & 0.02 & 0.02 & 0.01 & 0.04 & 0.03 & 0.03 & 0.01 \\
 \multirow{2}{*}{Sce 3} & {\small PIG} & 0.64 & 0.66 & 0.60 & 0.74 & 0.70 & 0.63 & 0.58 & 0.66 & 0.63 & 0.62 & 0.63 & 0.70 \\
                        & Poisson & 0.04 & 0.04 & 0.02 & 0.02 & 0.02 & 0.05 & 0.05 & 0.01 & 0.02 & 0.06 & 0.01 & 0.02 \\
\bottomrule
\end{tabular} } 
\end{table} \end{center}

\begin{table} [!t] \centering
 \caption{95\% credible intervals for some model parameters. Values are averaged over the 100 generated datasets.} \label{tab:sim1_2}
\begin{tabular}{ccccccccccccc}
\toprule
\multicolumn{2}{@{}c}{\bf Equidispersed} & $\rho$ & $\upsilon$ & $\kappa$ & $w$ & $\zeta$ \\ 
\midrule
 \multirow{2}{*}{Sce 1} & {\small PIG} & (0.03, 0.10) & (4.59, 9.35)& (86.07, 125.41)& (0.03, 0.13) & (0.59, 7.99)  \\
                        & Poisson & (0.03, 0.10) & (4.59, 9.39)& (86.31, 124.99)& (0.03, 0.13)& (0.65, 7.99) \\
 \multirow{2}{*}{Sce 2} & {\small PIG} & (0.01, 0.08) & (3.30, 7.38)& (86.79, 126.20)& (0.02, 0.11) & (0.64, 7.60) \\
                        & Poisson & (0.01, 0.08) & (3.34, 7.42)& (86.75, 126.31)& (0.02, 0.11)& (0.61, 7.81) \\
 \multirow{2}{*}{Sce 3} & {\small PIG} & (0.03, 0.11) & (4.79, 9.87)& (85.27, 123.97)& (0.03, 0.13) & (0.70, 8.33) \\
                        & Poisson & (0.03, 0.11) & (4.82, 9.91)& (85.02, 123.79)& (0.03, 0.13)& (0.66, 8.22) \\
\midrule
\multicolumn{2}{@{}c}{\bf Overdispersed} & & & & & \\
\cmidrule{1-2}
 \multirow{2}{*}{Sce 1} & {\small PIG} & (0.02, 0.10) & (3.84, 8.79)& (86.17, 125.31)& (0.02, 0.12)& (0.60, 7.69) \\
                        & Poisson & (0.54, 0.72) & (45.85, 74.08)& (34.64, 52.30)& (0.55, 0.73)& (3.52, 15.07) \\
 \multirow{2}{*}{Sce 2} & {\small PIG} & (0.02, 0.10) & (3.46, 8.55)& (85.57, 124.54)& (0.02, 0.12)& (0.61, 8.01) \\
                        & Poisson & (0.54, 0.73) & (45.73, 74.16)& (34.07, 51.23)& (0.56, 0.74)& (3.55, 15.2) \\
 \multirow{2}{*}{Sce 3} & {\small PIG} & (0.02, 0.11) & (4.22, 9.73)& (85.19, 123.99)& (0.02, 0.13)& (0.66, 8.07) \\
                        & Poisson & (0.56, 0.74) & (46.85, 74.94)& (33.50, 50.39)& (0.57, 0.75)& (3.65, 15.15) \\
\bottomrule
\end{tabular}
\end{table}

\subsection{Simulation 2: exploring temporal dependence orders} \label{app:sim2}

\begin{figure} [!t] \raggedright
\includegraphics[width=\textwidth]{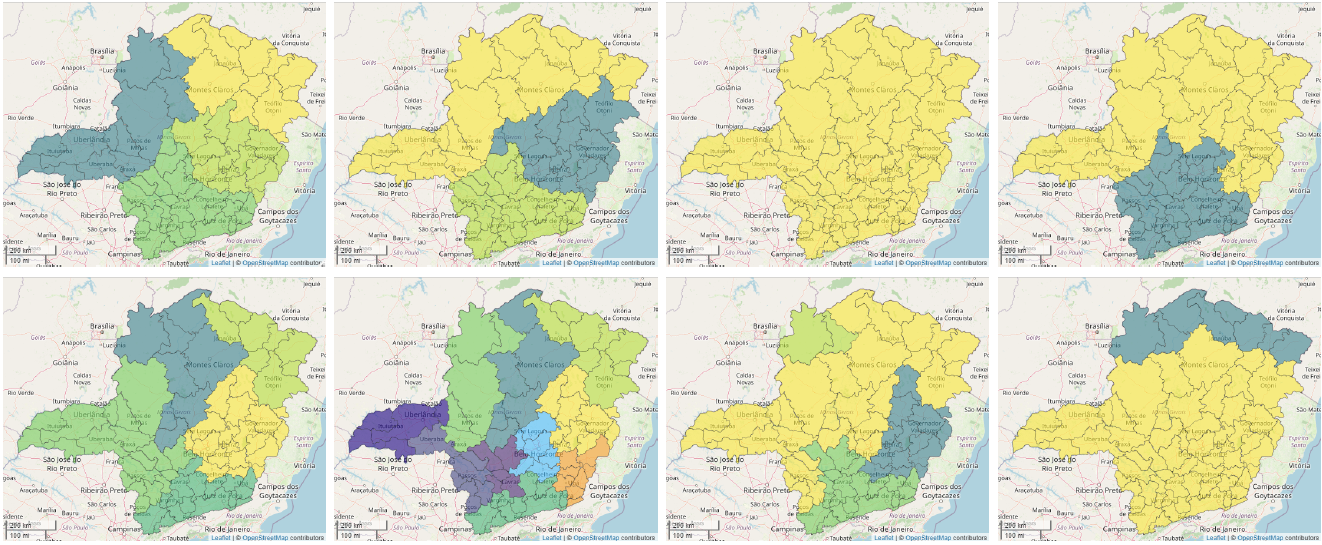}
\caption{Map of spatial partitions used in simulation 2. Scenario where there is a different partition for each season that is repeated over the years. Row 1: partition used in scenarios 1 and 2. Row 2: partition used in scenarios 3 and 4.} \label{fig:map_part_sim2}
\end{figure}

We now turn our attention to the order of temporal dependence and its impact on clustering. In this case, we considered a clustering scenario where each season has a different partition that is repeated over the years. Following this structure, we examine two situations. In the first case, the number of clusters remains consistent, with 4, 3, 1, and 2 clusters for summer, autumn, winter, and spring, respectively. The second case presents greater variation in the number of clusters, with 5, 10, 4, and 2 clusters for the seasons. Figure~\ref{fig:map_part_sim2} displays the maps showing each partition structure considered in this simulation study.

Once the partitions were created, the proposed model (Eq.\eqref{eq:Poi}--\eqref{eq:delta_prior} of the main manuscript) was considered as a data-generating mechanism to produce 100 synthetic datasets with 260 time points representing the epidemiological weeks (i.e., $t = 1, \ldots, 260$). This is equivalent to 5 years, resulting in 20 seasons, each lasting 13 weeks (i.e., $s = 1, \ldots, 20$). The design matrices ${\bm X}$ and ${\bm V}$, as well as the regression coefficients ${\bm \beta}$, were considered the same as in Section~\ref{app:sim1}. Meanwhile, ${\bm \delta} = (3.5, 0.2, -0.4)$, which restricts the dispersion values to be around one. After generating the datasets, we fitted PIG models with $q = 1, 2, \ldots, 5$ to all synthetic datasets using the MCMC algorithm described in Section~\ref{app:MCMC}. In addition, we fit a particular case where partitions were independently sampled. Note that to do this, it is sufficient to fix  ${\bm u} = {\bm c} = {\bm 0}$, thus $\rho_{s}$ is independent and identically distributed as Be$(\upsilon, \kappa)$.

To evaluate the goodness of fit, we calculated the WAIC for each fitted model. Then, we computed the frequency with which each model obtained the lowest WAIC value over the 100 datasets, as displayed in Figure~\ref{fig:sim2}. Overall, adding a temporal structure to the partition prior tends to enhance the fit performance compared to using the independent version. The only case in which the model with independent partitions outperformed the others was scenario 3. However, it was unclear which dependency order yielded the best fit. Looking at the WAIC values averaged over the 100 generated datasets (Table~\ref{tab:sim2_1}), the differences between the models seem imperceptible. This occurred similarly in all four scenarios.

\begin{figure} [!h] \centering
\includegraphics[width=\textwidth]{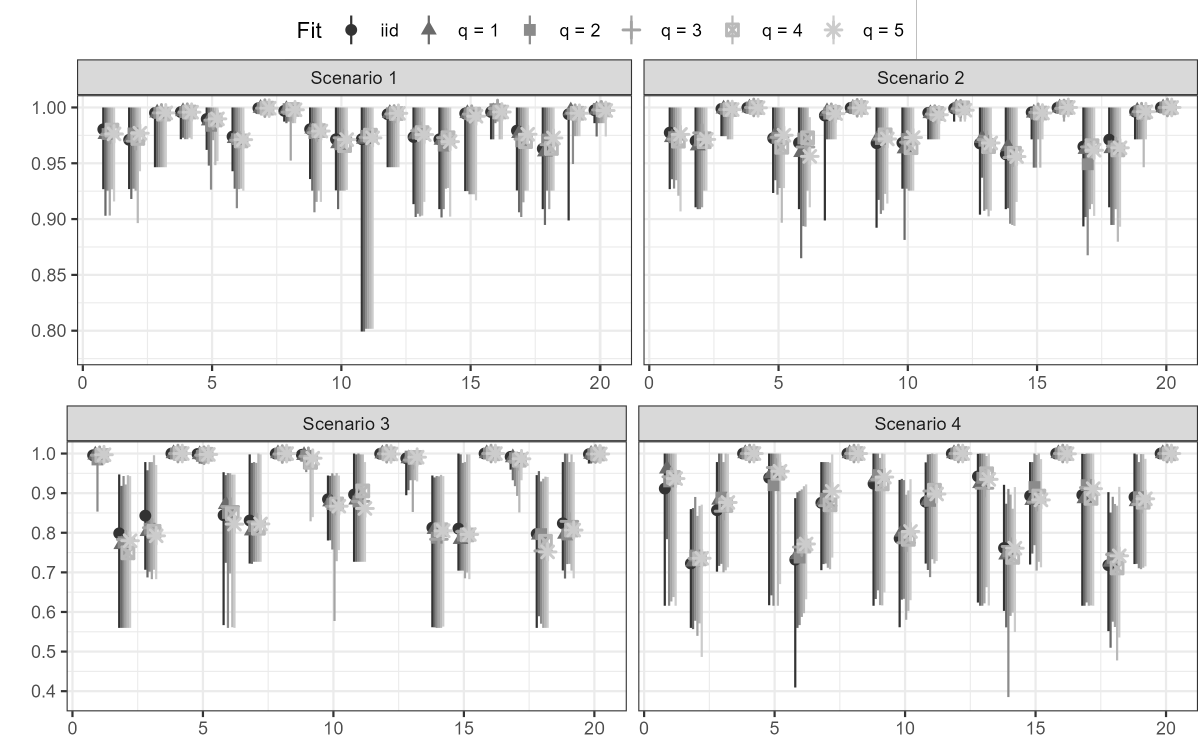}
\caption{RI values to measure similarity between estimated and true partitions over time. Higher RI value indicate higher accuracy of partition estimates.} \label{fig:sim2_1}
\end{figure}

As shown in Figure~\ref{fig:sim2_1} and Table~\ref{tab:sim2_1}, the model provided good accuracy in partition estimation, with RI values averaging over 90\% across all scenarios. Nonetheless, similar to the WAIC, it is challenging to identify the best dependence order. On average, the RI values are comparable among the models (Figure~\ref{fig:sim2_1}). In fact, all parameters used in the temporal structure were estimated similarly among the models, as observed in Table~\ref{tab:sim2_2}.

\begin{center} \begin{table} [!h]
\caption{Model fit performance metrics used to compare different autoregressive order values. Values are averaged over the 100 generated datasets. Lower WAIC values indicate better fit. Higher RI value indicate higher accuracy of partition estimates.} \label{tab:sim2_1}
\resizebox{\columnwidth}{!}{
\begin{tabular}{cccccccc} \toprule
& & \multicolumn{6}{@{}c}{Fit} \\ 
\cmidrule{3-8}
 & & iid & $q = 1$ & $q = 2$ & $q = 3$ & $q = 4$ & $q = 5$ \\ 
\midrule
Sce 1 & {\small WAIC} $|$ {\small RI} & 77,078 $|$ 0.98 & 77,079 $|$ 0.98 & 77,077 $|$ 0.98 & 77,078 $|$ 0.98 & 77,078 $|$ 0.98 & 77,077 $|$ 0.98 \\
Sce 2 & {\small WAIC} $|$ {\small RI} & 87,055 $|$ 0.98 & 87,051 $|$ 0.98 & 87,054 $|$ 0.98 & 87,052 $|$ 0.98 & 87,055 $|$ 0.98 & 87,053 $|$ 0.98 \\
Sce 3 & {\small WAIC} $|$ {\small RI} & 87,497 $|$ 0.92 & 87,505 $|$ 0.91 & 87,504 $|$ 0.91 & 87,505 $|$ 0.91 & 87,506 $|$ 0.91 & 87,504 $|$ 0.90 \\
Sce 4 & {\small WAIC} $|$ {\small RI} & 104,672 $|$ 0.89 & 104,668 $|$ 0.89 & 104,670 $|$ 0.89 & 104,671 $|$ 0.89 & 104,668 $|$ 0.89 & 104,669 $|$ 0.90 \\
\bottomrule 
\end{tabular} } 
\end{table} \end{center} 

\begin{table}[!h] \centering
\caption{Summary for some model parameters. Values are averaged over the 100 generated datasets. Lower WAIC values indicate better fit.} \label{tab:sim2_2}
\begin{tabular}{cccccc}
\toprule
 \multirow{2}{*}{Fit} & \multicolumn{5}{@{}c}{95\% credible intervals} \\ 
\cmidrule{2-6}
 & $\rho$ & $\upsilon$ & $\kappa$ & $w$ & $\zeta$ \\ 
\midrule
iid     & (0.01, 0.08) & (3.12, 6.39) & (88.07, 127.13) & -- & -- \\
$q = 1$ & (0.01, 0.08) & (3.14, 6.57) & (87.90, 127.53) & (0.01, 0.08) & (2.27, 9.81) \\
$q = 2$ & (0.01, 0.08) & (3.13, 6.59) & (87.66, 126.74) & (0.01, 0.08) & (2.35, 9.62) \\  
$q = 3$ & (0.01, 0.08) & (3.16, 6.70) & (87.32, 126.66) & (0.01, 0.08) & (2.30, 9.62) \\ 
$q = 4$ & (0.01, 0.08) & (3.21, 6.80) & (87.13, 126.88) & (0.01, 0.07) & (2.33, 9.55) \\ 
$q = 5$ & (0.01, 0.08) & (3.24, 6.98) & (87.11, 126.87) & (0.01, 0.07) & (2.40, 9.69) \\  
\midrule
iid     & (0.01, 0.07) & (2.93, 6.12) & (87.90, 127.29) & -- & -- \\ 
$q = 1$ & (0.01, 0.07) & (2.99, 6.22) & (87.95, 127.42) & (0.01, 0.08) & (2.33, 9.48) \\
$q = 2$ & (0.01, 0.07) & (2.97, 6.29) & (87.41, 127.21) & (0.01, 0.07) & (2.26, 9.69) \\
$q = 3$ & (0.01, 0.07) & (3.03, 6.42) & (87.25, 127.14) & (0.01, 0.07) & (2.18, 9.48) \\ 
$q = 4$ & (0.01, 0.07) & (3.00, 6.45) & (87.03, 126.73) & (0.01, 0.07) & (2.29, 9.71) \\
$q = 5$ & (0.01, 0.07) & (3.06, 6.53) & (86.95, 126.91) & (0.01, 0.07) & (2.36, 9.60) \\
\midrule
iid     & (0.03, 0.11) & (5.32, 9.79) & (85.63, 123.92) & -- & -- \\
$q = 1$ & (0.03, 0.11) & (5.20, 9.76) & (85.39, 123.77) & (0.03, 0.12) & (2.49, 10.42) \\
$q = 2$ & (0.03, 0.11) & (5.11, 9.75) & (85.55, 123.99) & (0.03, 0.12) & (2.38, 9.92) \\
$q = 3$ & (0.03, 0.11) & (5.14, 9.89) & (85.63, 124.20) & (0.03, 0.11) & (2.43, 9.60) \\
$q = 4$ & (0.03, 0.11) & (5.08, 9.84) & (85.18, 123.75) & (0.03, 0.11) & (2.32, 10.14) \\
$q = 5$ & (0.03, 0.11) & (5.04, 9.94) & (85.40, 123.38) & (0.03, 0.11) & (2.47, 10.15) \\
\midrule
iid     & (0.03, 0.11) & (5.21, 9.56) & (85.95, 124.29) & -- & -- \\
$q = 1$ & (0.03, 0.11) & (5.20, 9.74) & (85.86, 124.33) & (0.03, 0.12) & (2.41, 9.89) \\
$q = 2$ & (0.03, 0.11) & (5.12, 9.70) & (85.88, 124.13) & (0.03, 0.11) & (2.51, 9.99) \\
$q = 3$ & (0.03, 0.11) & (5.21, 9.91) & (85.69, 124.04) & (0.03, 0.11) & (2.40, 10.05) \\ 
$q = 4$ & (0.03, 0.11) & (5.09, 9.84) & (85.98, 123.83) & (0.03, 0.11) & (2.41, 9.86) \\  
$q = 5$ & (0.03, 0.11) & (5.13, 10.06) & (85.21, 123.76) & (0.03, 0.11) & (2.56, 10.12) \\  
\bottomrule
\end{tabular}
\end{table} 

\subsection{Simulation 3} \label{app:sim3}

\begin{figure} [!h] \raggedright
 \includegraphics[width=\textwidth]{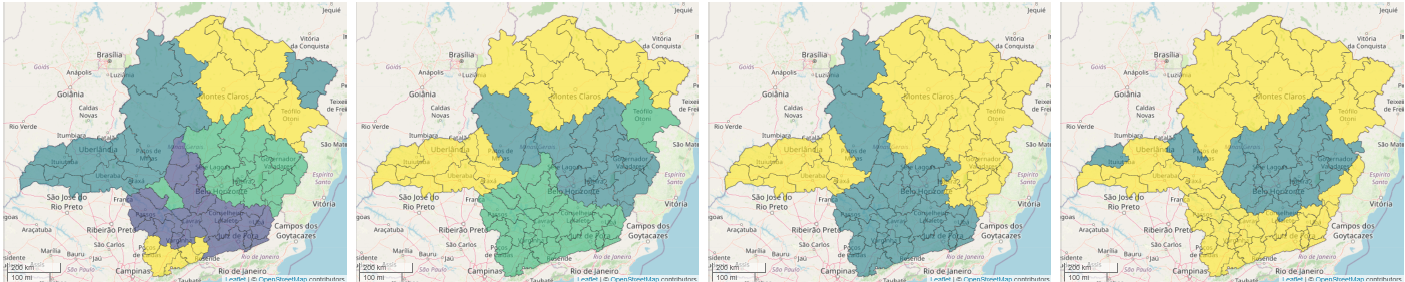}
\caption{Map of spatial partitions used in simulation 3. Scenario where there is a different partition for each season (summer, autumn, winter, and spring) that repeat over the years.} \label{fig:map_part_sim3}
\end{figure}

The incorporation of random spanning trees into PPM for areal clustering ensures that the resulting clusters maintain spatial contiguity, thereby guaranteeing that clustering remains contiguous. To investigate how the model deals with non-contiguous clusters, we have arbitrarily established partitions comprised of non-contiguous clusters. Specifically, we have devised a scenario featuring four distinct partitions representing the seasons, as depicted in Figure~\ref{fig:map_part_sim3}, and we consider that they repeat over the years. The first partition (summer) contains three non-contiguous clusters, each divided into two components. The second partition (autumn) consists of two non-contiguous clusters, also segmented into two components. The only non-contiguous cluster in third and fourth partitions (winter and spring, respectively) is divided into two and three components, respectively. Furthermore, there is a contiguous cluster in every partition.

Cluster-specific parameters were set as follows:
\small \begin{equation*}
{\bm \theta}^{\star}_{s} = \begin{cases}
        (1, 10, 25, 45) & \text{for} \, s = 1,5,9,13,17 \\
        (1, 5, 25) & \text{for} \, s = 2,6,10,14,18 \\
        (5, 20) & \text{for} \, s = 3,7,11,15,19 \\
        (1, 15) & \text{for} \, s = 4,8,12,16,20. \\
\end{cases}
\end{equation*} \normalsize

Upon the establishment of the partitions, the proposed model (Eq.\eqref{eq:Poi}--\eqref{eq:delta_prior} in the main manuscript) was employed as a data-generating mechanism to create 100 datasets, each comprising 260 time points corresponding to epidemiological weeks (i.e., $t = 1, \ldots, 260$). This temporal framework encapsulates five years, thereby delineating 20 seasons, each spanning 13 weeks (i.e., $s = 1, \ldots, 20$). The design matrices ${\bm X}$ and ${\bm V}$, alongside the regression coefficients ${\bm \beta}$ and ${\bm \delta}$, were maintained consistent with those outlined in Section~\ref{app:sim2}. Subsequent to the generation process, PIG model with $q = 1$ was fitted to all datasets utilizing the MCMC algorithm detailed in Section~\ref{app:MCMC}.

\begin{figure} [!h] \raggedright
\hspace{2cm}
\includegraphics[width=0.75\textwidth]{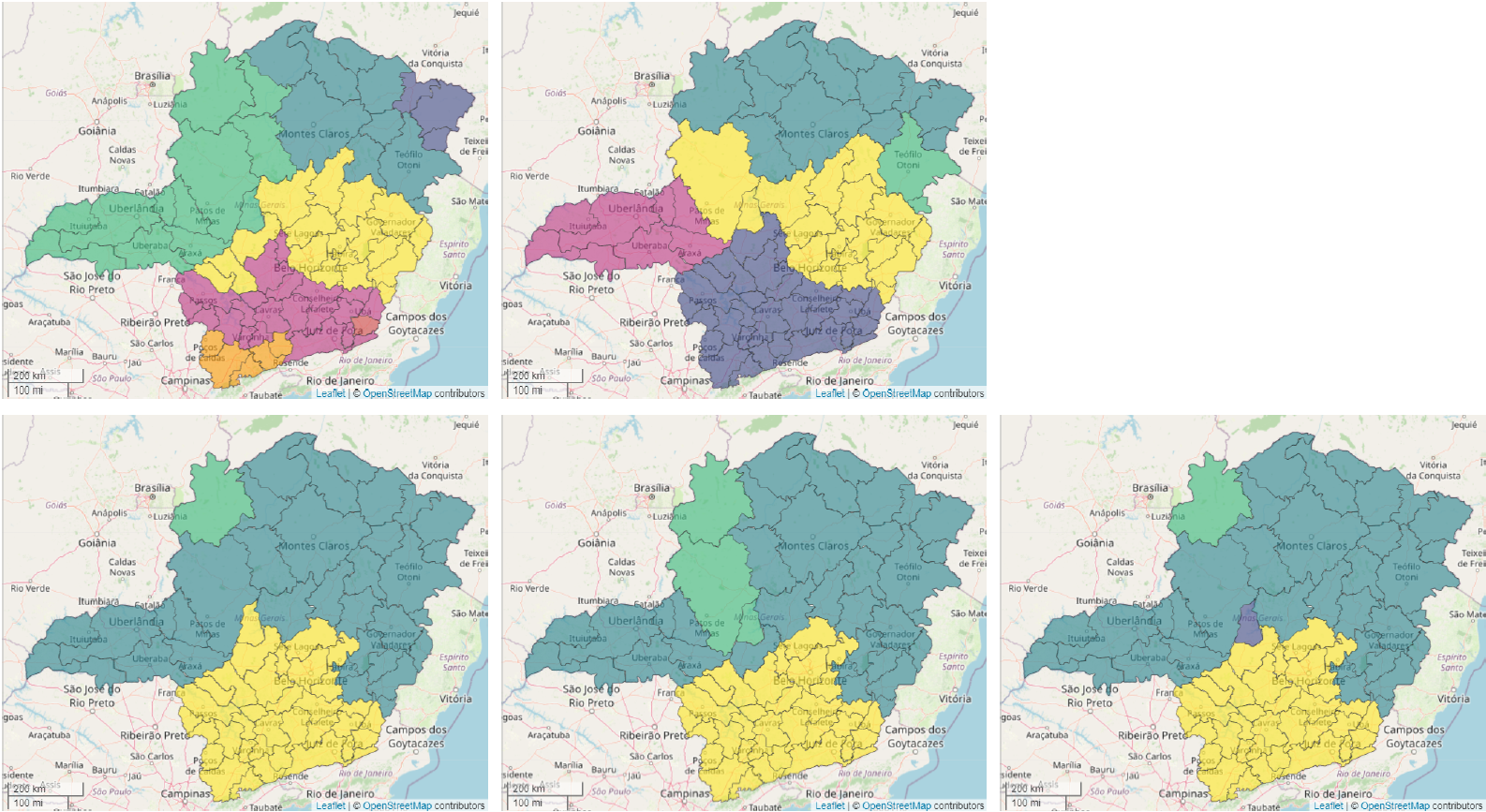}
\caption{Cases in which the model merged the non-contiguous cluster with neighboring regions to make the resulting cluster contiguous. Row 1: estimated partitions for summer. Row 2: estimated partitions for winter.} \label{fig:sim3}
\end{figure}

In accordance with previous simulation studies, we employed the {\tt salso} R package \citep{Dahl2020}, utilizing the VI loss function to estimate the four partitions for each synthetic dataset. Subsequently, we examined the model's behavior in relation to non-contiguous clusters. A summary of the results obtained is displayed in Table~\ref{tab:sim3}. In the majority of instances, the model effectively delineated the non-contiguous cluster based on the quantity of its constituent components. Specifically, when the non-contiguous cluster comprised two components, the model partitioned this cluster into two distinct clusters. Similarly, when the cluster consisted of three components, it was segregated into three separate clusters. However, there were a limited number of occurrences in which the model opted not to subdivide the non-contiguous cluster, instead, it merged it with neighboring regions to render the resulting cluster contiguous. More precisely, this occurred twice (0.4\%) for the summer partition and 22 times (4.4\%) for the winter partition (in which three different partitions were repeatedly estimated). Figure~\ref{fig:sim3} displays the corresponding maps.
\begin{table} [!h] \centering
 \caption{ Frequency in which the model segregated or merged the non-contiguous clusters over the 100 datasets. Indicated colors are based on Figure~\ref{fig:map_part_sim3}.} \label{tab:sim3}
\begin{tabular}{cccccccc}
\toprule
\multirow{2}{*}{Partition} & Non-contiguous & Number of           & \multicolumn{5}{@{}c}{Number of estimated clusters} \\ 
\cmidrule{4-8}
                           & cluster         & original components & 1 & 2 & 3 & 4 & 5 \\ 
\midrule
 \multirow{3}{*}{1} & 1 (yellow) & 2 & --    & 93\%   & 4.2\%  & 2.8\%  & -- \\
                    & 2 (blue)   & 2 & --    & 58\%   & 30.6\% & 11.2\% & 0.2\% \\
										& 3 (green)  & 2 & 0.4\% & 84.6\% & 14.8\% & 0.2\%  & -- \\
\midrule
 \multirow{2}{*}{2} & 1 (yellow) & 2 & --    & 94.6\% & 5.2\%  & 0.2\%  & -- \\
                    & 2 (green)  & 2 & --    & 90\%   & 9.6\%  & 0.4\%  & -- \\
\midrule
                 3  & 1 (yellow) & 2 & 4.4\% & 94\%   & 1.6\%  & --     & -- \\
\midrule
                 4  & 1 (blue)   & 3 & --    & --     & 98.6\% & 1.4\%  & -- \\
\bottomrule
\end{tabular}
\end{table}

\section{Complementary results} \label{app:results}

In this section, we provide complementary results that support Section~\ref{sec:dataapp} of the main manuscript. 

To address time-related trends in the dengue data, we analyzed a range of values for the dependence order parameter by adjusting $q$ from 1 to 12, corresponding to three years. This approach allowed us to evaluate the impact of choosing $q$ on the posterior inference of $\bm \rho$ and, in turn, on the partitioning. We also examined the simplest model where the partitions are treated as independent. Table~\ref{tab:app_1} shows a posterior summary of the parameters related to temporal structure, $\rho, \upsilon, \kappa, w,$ and $\zeta$, as well as the WAIC values used as a criterion for goodness-of-fit to determine the most suitable model. Regarding the estimated parameters, we noticed a significant similarity among the results obtained from different temporal orders, particularly for the probabilities of removing edges. When evaluating the WAIC values, we observed that assuming a first-order autoregressive structure is the best option for fitting the dengue data from the Southeast region of Brazil. It is worth noting that the highest WAIC value was obtained in the case of independent partitions, which corroborates the need to consider temporal correlations between spatial partitions.
\begin{center} \begin{table} [!h]
 \caption{Summary for some model parameters. Lower WAIC values indicate better fit.} \label{tab:app_1}
\resizebox{\columnwidth}{!}{
\begin{tabular}{ccccccc}
\toprule
 \multirow{2}{*}{Fit} & \multicolumn{5}{@{}c}{95\% credible intervals} & \multirow{2}{*}{\small WAIC} \\ 
\cmidrule{2-6}
 & $\rho$ & $\upsilon$ & $\kappa$ & $w$ & $\zeta$ &  \\ 
\midrule
iid & (0.07, 0.18) & (10.64, 16.45) & (84.04, 117.55) & - & - & 878,186 \\
$q = 1$ & (0.07, 0.18) & (9.73, 16.28) & (81.14, 117.73) & (0.08, 0.21) & (3.96, 15.49) & {\bf 814,474} \\ 
$q = 2$ & (0.07, 0.18) & (8.66, 15.50) & (80.20, 117.83) & (0.08, 0.19) & (3.46, 13.05) & 819,334 \\ 
$q = 3$ & (0.07, 0.19) & (8.68, 17.11) & (79.96, 117.60) & (0.08, 0.19) & (3.81, 12.03) & 820,384 \\ 
$q = 4$ & (0.07, 0.17) & (8.86, 15.87) & (82.61, 118.51) & (0.08, 0.18) & (3.33, 11.05) & 827,939 \\ 
$q = 5$ & (0.08, 0.19) & (9.29, 16.87) & (80.66, 116.92) & (0.09, 0.20) & (4.12, 13.69) & 824,301 \\ 
$q = 6$ & (0.07, 0.17) & (8.47, 15.74) & (81.51, 119.62) & (0.08, 0.18) & (4.15, 13.52) & 847,291 \\ 
$q = 7$ & (0.06, 0.16) & (7.70, 15.09) & (80.76, 118.52) & (0.07, 0.16) & (4.53, 13.94) & 818,774 \\ 
$q = 8$ & (0.07, 0.17) & (8.15, 16.57) & (82.01, 119.02) & (0.08, 0.18) & (3.95, 15.83) & 817,737 \\ 
$q = 9$ & (0.07, 0.18) & (9.13, 16.76) & (81.51, 118.52) & (0.08, 0.18) & (3.53, 12.29) & 823,026 \\ 
$q = 10$ & (0.07, 0.17) & (7.93, 15.49) & (81.91, 118.09) & (0.08, 0.17) & (3.40, 12.03) & 825,716 \\ 
$q = 11$ & (0.07, 0.17) & (8.06, 15.30) & (81.74, 118.20) & (0.08, 0.16) & (3.66, 14.52) & 816,214 \\ 
$q = 12$ & (0.07, 0.17) & (8.04, 15.57) & (81.84, 120.34) & (0.08, 0.16) & (3.89, 12.30) & 844,795 \\ 
\bottomrule
\end{tabular} } 
\end{table} \end{center}

\begin{figure} [!h] \centering
\includegraphics[width=\textwidth]{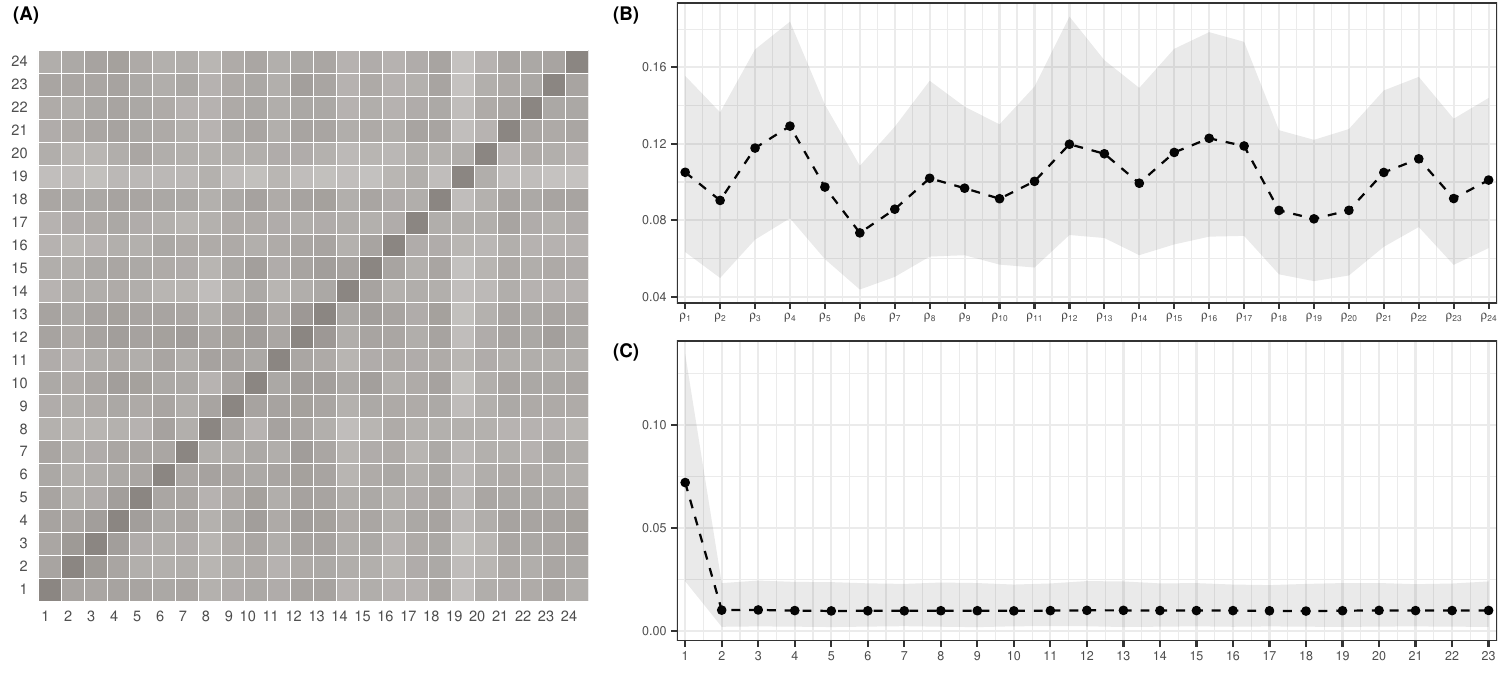}
\caption{Temporal dependence of estimated partitions. (A) Summary of the lagged RI values. (B) Posterior mean and 95\% credible interval of the probability of removing edges at each season $\{\rho_{s}\}$. (C) Autocorrelation function of $\{ \rho_{s} \}$.} \label{fig:corr_plot}
\end{figure}

\begin{figure} [!h] \centering
 \includegraphics[width=\textwidth]{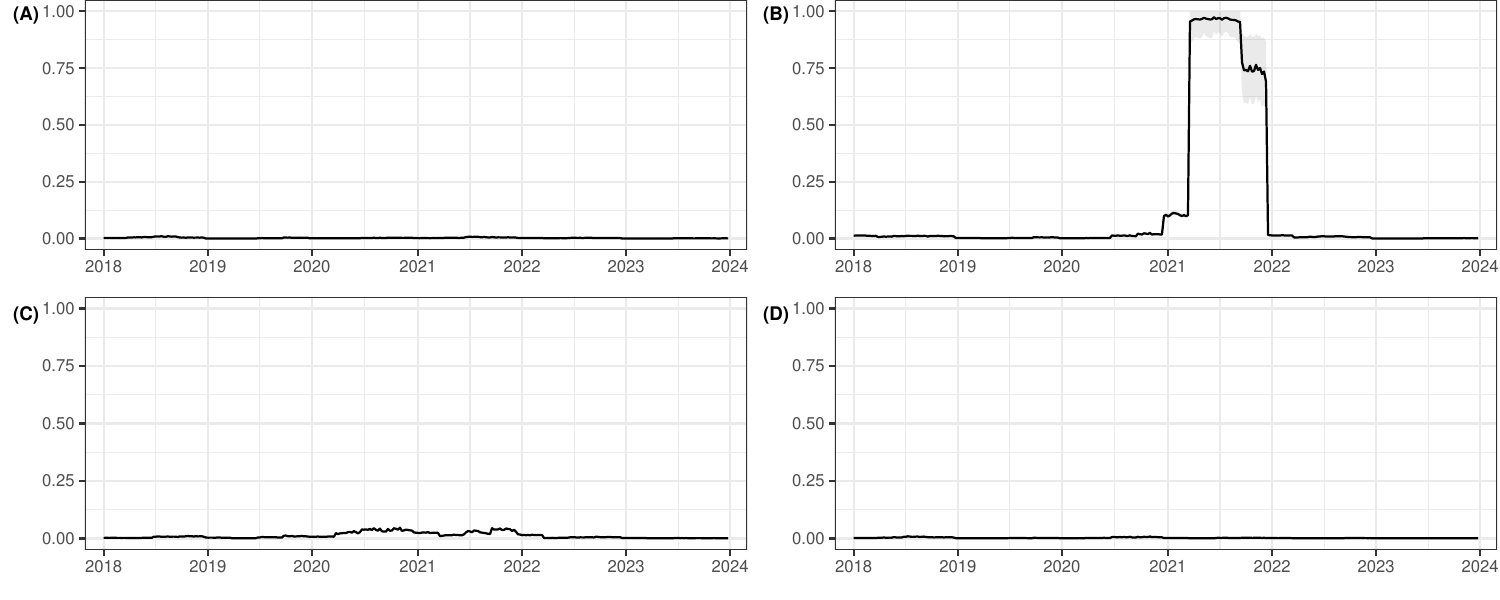}
\caption{$\mathds{E}(Y_{it} \mid \lambda_{it}, \psi_{is}) / \mathds{V}(Y_{it} \mid \lambda_{it}, \psi_{is})$ over time for the capitals of each state. (A) Belo Horizonte - MG, (B) Vit{\'o}ria - ES, (C) Rio de Janeiro - RJ, and S{\~a}o Paulo - SP.} \label{fig:z_plot}
\end{figure}

\begin{figure} [!h] \raggedright
 \includegraphics[width=\textwidth]{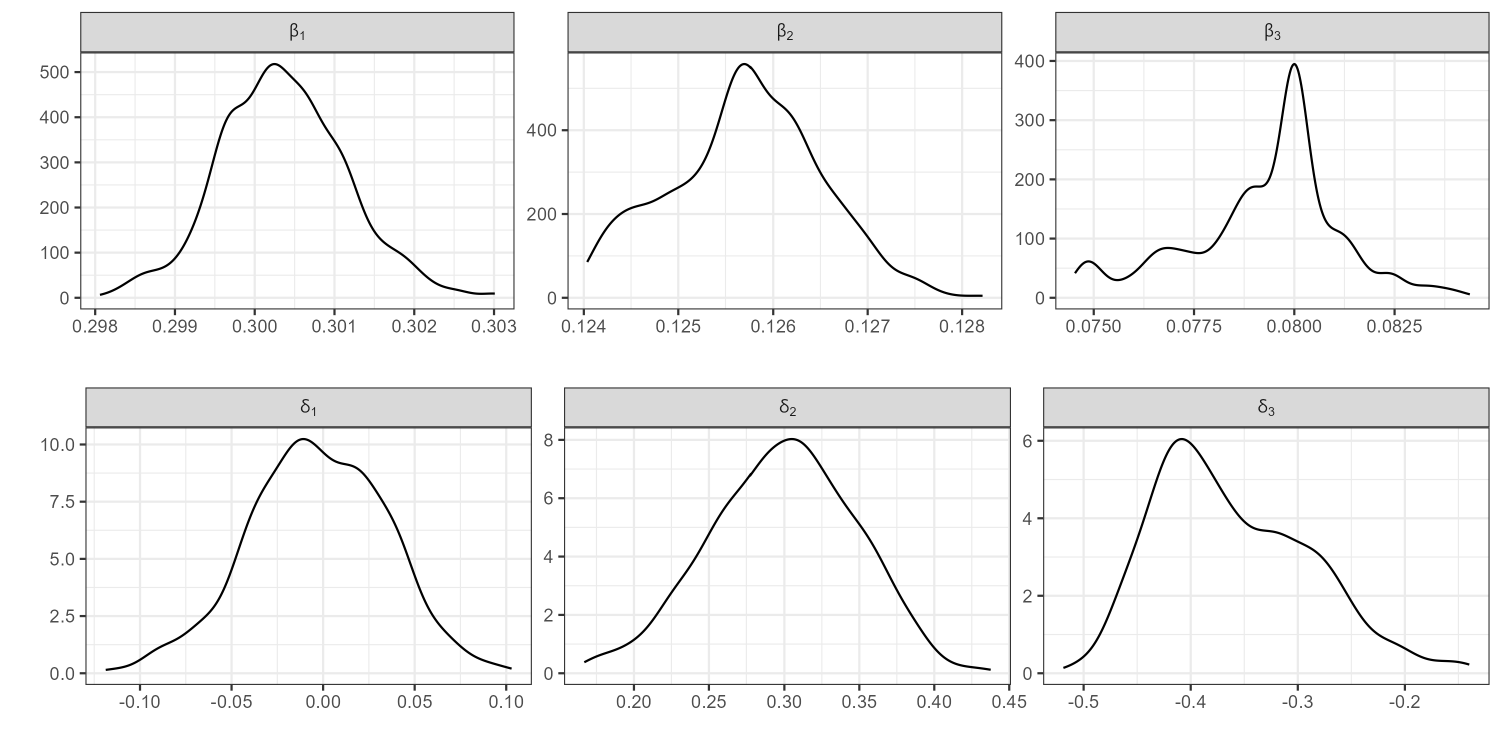}
\caption{Posterior distribution of the regression coefficients. Used in mean struture: $\beta_{1}$, $\beta_{2}$, and $\beta_{3}$ corresponding to temperature, humidity, and HDI. Used in dispersion struture: $\delta_{0}$, $\delta_{1}$, and $\delta_{2}$ corresponding to intercept, temperature, and humidity.} \label{fig:reg_coef}
\end{figure}

Results presented from now on were obtained by fitting the proposed model with a dependence order of $q = 1$, see Section~\ref{sec:dataapp} of the main manuscript for further details. Figure~\ref{fig:corr_plot} illustrates the temporal dependence of the estimated partitions based on various measures, including lagged RI values, the posterior distribution of $\{\rho_{s}\}$, and the autocorrelation function of $\{ \rho_{s} \}$. Figure~\ref{fig:time_est_disp2} complements Figure~\ref{fig:time_est_disp} in the main manuscript by showing the geographic distribution of dispersion indicators for all seasons from 2018 to 2023. To complement Figure~\ref{fig:time_z_lambda} in the main manuscript, we present the ratio $\mathds{E}(Y_{it} \mid \lambda_{it}, \psi_{is}) / \mathds{V}(Y_{it} \mid \lambda_{it}, \psi_{is})$, calculated over time for the capitals of each state. We observe that Belo Horizonte (MG), Rio de Janeiro (RJ), and S{\~a}o Paulo (SP) exhibited overdispersion throughout the entire period, as shown in Figures~\ref{fig:z_plot}(A), (C), and (D), respectively. In contrast, Vit{\'o}ria (ES) demonstrated a period of equidispersion in mid-2021, as illustrated in Figure~\ref{fig:z_plot}(B). A graphical representation of the product $\bm{O z \lambda}$, which indicates Poisson's rate, is shown in Figure~\ref{fig:rate_plot}. In this case, we calculated $\bm{O z \lambda}$ for Janu{\'a}ria (MG) and Campinas (SP) to complement the discussion presented in the main manuscript. Finally, Figure~\ref{fig:reg_coef} illustrates the posterior distribution of the regression coefficients.

\end{document}